\documentclass[letterpaper,twocolumn,10pt]{article}
\usepackage{usenix-2020-09}
\date{} 
\microtypecontext{spacing=nonfrench} 

\usepackage{amsthm}
\usepackage{amsmath}
\usepackage{amsfonts}
\usepackage{amssymb}
\usepackage{mathrsfs}
\usepackage{cancel}
\usepackage[overload]{empheq}

\usepackage{listings}
\usepackage{sty/listings-pseudo}
\usepackage[scaled=0.85]{beramono}

\usepackage{booktabs} 
\usepackage{enumitem}
\usepackage{graphicx}
\usepackage{subfig} 
\usepackage{stackengine}
\usepackage{tabto}
\usepackage{xspace}
\usepackage{xparse}

\usepackage[available,functional]{usenix-badges/usenixbadges}

\NewDocumentCommand\system{}{Chop Chop\xspace}
\NewDocumentCommand\broadcast{}{Atomic Broadcast\xspace}

\NewDocumentCommand\client{}{\chi}
\NewDocumentCommand\broker{}{\beta}
\NewDocumentCommand\server{}{\sigma}

\NewDocumentCommand{\Batchified}{}{Distilled\xspace}

\NewDocumentCommand\Batchification{}{Distillation\xspace}

\NewDocumentCommand\Fullbatchification{}{Full \batchification}

\NewDocumentCommand\batchify{}{distill\xspace}
\NewDocumentCommand\batchified{}{distilled\xspace}
\NewDocumentCommand\fullbatchified{}{fully \batchified}
\NewDocumentCommand\partialbatchified{}{partially \batchified}
\NewDocumentCommand\batchification{}{distillation\xspace}

\NewDocumentCommand\fullbatchification{}{full \batchification}

\NewDocumentCommand\Batches{}{\Batchified batches\xspace}

\NewDocumentCommand\batch{}{\batchified batch\xspace}
\NewDocumentCommand\batches{}{\batchified batches\xspace}
\NewDocumentCommand\fullbatch{}{\fullbatchified batch\xspace}
\NewDocumentCommand\partialbatch{}{\partialbatchified batch\xspace}
\NewDocumentCommand\fullbatches{}{\fullbatchified batches\xspace}

\newcommand{\bftsmart}{BFT-SMaRt\xspace}
\newcommand{\bullshark}{Bullshark\xspace}
\newcommand{\bs}{\bullshark}
\newcommand{\hotstuff}{HotStuff\xspace}
\newcommand{\hs}{\hotstuff}
\newcommand{\narwhal}{Narwhal\xspace}
\newcommand{\nw}{\narwhal}
\newcommand{\nwbs}{Narwhal-Bullshark\xspace}
\newcommand{\nwbssig}{Narwhal-Bullshark-sig\xspace}
\newcommand{\cchs}{Chop Chop-\hs}

\newcommand{\eg}{e.g.,\xspace}

\newcommand{\ie}{i.e.,\xspace}

\usepackage[shortcuts]{glossaries-extra}
\glsdisablehyper

\newabbr{bft}{BFT}{Byzantine fault tolerant}
\newabbr{smr}{SMR}{state machine replication}

\usepackage[capitalize]{cleveref}
\renewcommand{\Cref}[1]{\cref{#1}} 
\renewcommand{\autoref}[1]{\cref{#1}} 

\newcommand{\secref}{$\S\mkern-4mu$}
\crefname{section}{\secref}{$\S\S\mkern-4mu$}
\crefname{appendix}{Appx.}{Apps.}
\crefname{thm}{Theorem}{Theorems}

\usepackage[compact]{titlesec}

\newcommand{\captionspace}{-16pt}

\newcommand{\eqspace}{-5pt}
\AtBeginEnvironment{equation*}{\vspace{\eqspace}}
\AtEndEnvironment{equation*}{\vspace{\eqspace}}
\AtBeginEnvironment{equation}{\vspace{\eqspace}}
\AtEndEnvironment{equation}{\vspace{\eqspace}}

\NewDocumentCommand\cstob{}{Client-Server Total Order Broadcast\xspace}
\NewDocumentCommand\cstobprefix{}{CSTOB\xspace}

\NewDocumentCommand\stob{}{Server Total Order Broadcast\xspace}
\NewDocumentCommand\stobprefix{}{STOB\xspace}

\NewDocumentCommand\cl{}{\cstobprefix~Client\xspace}
\NewDocumentCommand\clab{}{\cstobprefix Client}
\NewDocumentCommand\clin{}{cl}

\NewDocumentCommand\bkab{}{\cstobprefix Broker}
\NewDocumentCommand\bkin{}{bk}

\NewDocumentCommand\sr{}{\cstobprefix~Server\xspace}
\NewDocumentCommand\srab{}{\cstobprefix Server}
\NewDocumentCommand\srin{}{sr}
\NewDocumentCommand\stobi{}{stob}

\NewDocumentCommand\process{}{\pi}

\NewDocumentCommand\brokers{}{B}

\NewDocumentCommand\servers{}{\Sigma}

\NewDocumentCommand\dir{}{directory\xspace}

\NewDocumentCommand\diral{}{Rank\xspace}

\NewDocumentCommand\dirab{}{Directory}
\NewDocumentCommand\dirin{}{dir}

\NewDocumentCommand\dirsrab{}{DirectoryServer}
\NewDocumentCommand\dirsrin{}{dsr}

\NewDocumentCommand\rp{m}{{\left(#1\right)}}
\NewDocumentCommand\qp{m}{{\left[#1\right]}}

\NewDocumentCommand\ap{m}{{\left\langle#1\right\rangle}}

\NewDocumentCommand\ceil{m}{{\left\lceil #1 \right\rceil}}

\NewDocumentCommand\sign{o}{
    \IfValueTF{#1}
    {{\text{sign}\rp{#1}}}
    {{\text{sign}}}
}

\NewDocumentCommand\identity{o}{
	\IfValueTF{#1}
	{{\text{Id}\rp{#1}}}
	{{\text{Id}}}
}

\NewDocumentCommand\indicator{o}{
	\IfValueTF{#1}
	{{\text{I}\qp{#1}}}
	{{\text{I}}}
}

\stackMath
\NewDocumentCommand\pagoda{m}{
    \savestack{\tmpbox}{\stretchto{
        \scaleto{
            \scalerel*[\widthof{\ensuremath{#1}}]{\kern+1.5pt\bigwedge\kern+1.5pt}
            {\rule[-\textheight/2]{1ex}{\textheight}}
        }{\textheight}
    }{0.5ex}}
    \stackon[1pt]{#1}{\tmpbox}
}

\NewDocumentCommand\powerset{om}{
    \IfValueTF{#1}
    {{\mathbb{P}^{#1}\rp{#2}}}
    {{\mathbb{P}\rp{#2}}}
}

\NewDocumentCommand\choice{o}{
	\IfValueTF{#1}
	{{\mathfrak{C}\rp{#1}}}
	{\mathfrak{C}}
}

\theoremstyle{definition}
    \newtheorem{theorem}{Theorem}
\theoremstyle{definition}
    \newtheorem{lemma}{Lemma}
    \newtheorem{definition}{Definition}
    \newtheorem{notation}{Notation}

\theoremstyle{definition}
    \newtheorem{assumption}{Assumption}

\makeatletter
\newcommand{\assumptiontag}[1]{
    \let\oldtheassumption\theassumption
    \renewcommand{\theassumption}{#1}
    \g@addto@macro\endassumption{
        \addtocounter{assumption}{-1}
        \global\let\theassumption\oldtheassumption
    }
}
\makeatother

\definecolor{mygray}{rgb}{0.5,0.5,0.5}

\let\origthelstnumber\thelstnumber
\makeatletter
\newcommand*\StartMultiline{%
  \lst@AddToHook{OnNewLine}{%
    \let\thelstnumber\relax%
     \advance\c@lstnumber-\@ne\relax%
    }%
}

\newcommand*\EndMultiline{%
  \lst@AddToHook{OnNewLine}{%
   \let\thelstnumber\origthelstnumber%
   \advance\c@lstnumber\@ne\relax}%
}

\NewDocumentCommand{\event}{mmo}{
    \IfValueTF{#3}
    {{\ap{#1.\textrm{#2} \mid #3}}}
    {{\ap{#1.\textrm{#2}}}}
}

\title{Chop Chop: Byzantine Atomic Broadcast to the Network Limit}

\author{\rm{Martina Camaioni} \quad \rm{Rachid Guerraoui} \quad \rm{Matteo Monti} \\ \rm{Pierre-Louis Roman} \quad \rm{Manuel Vidigueira} \quad \rm{Gauthier Voron}\\
Ecole Polytechnique F\'{e}d\'{e}rale de Lausanne (EPFL)
}

\hypersetup{pdftitle={Chop Chop: Byzantine Atomic Broadcast to the Network Limit}}
\hypersetup{pdfauthor={Martina Camaioni, Rachid Guerraoui, Matteo Monti, Pierre-Louis Roman, Manuel Vidigueira, Gauthier Voron}}

\begin{document}

\maketitle

\begin{abstract}
At the heart of state machine replication, the celebrated technique enabling decentralized and secure universal computation, lies \broadcast, a fundamental communication primitive that orders, authenticates, and deduplicates messages. 
This paper presents \system, a Byzantine \broadcast system that uses a novel authenticated memory pool to amortize the cost of ordering, authenticating and deduplicating messages, achieving ``line rate'' (\ie closely matching the complexity of a protocol that does not ensure any ordering, authentication or Byzantine resilience) even when processing messages as small as 8 bytes.
\system attains this performance by means of a new form of batching we call \emph{\batchification}.
A \batch is a set of messages that are fast to authenticate, deduplicate, and order.
Batches are \batchified using a novel interactive protocol involving \emph{brokers}, an untrusted layer of facilitating processes between clients and servers.
In a geo-distributed deployment of 64 medium-sized servers, \system processes 43,600,000 messages per second with an average latency of 3.6 seconds.
Under the same conditions, state-of-the-art alternatives offer two orders of magnitude less throughput for the same latency.
We showcase three simple \system applications: a Payment system, an Auction house and a ``Pixel war'' game, respectively achieving 32, 2.3 and 35 million operations per second.
\end{abstract}

\section{Introduction}
\label{sec:introduction}

Is an Internet computer feasible?
A computer that is highly-available, decentralized, secure, universal and shared by all?
Theory says yes: \ac{smr}~\cite{schneider-smr-csur90,tobcast-csur04} enables decentralized universal computation in the face of arbitrary failures~\cite{byz-generals-toplas82,bft-fire-nsdi08}.
In practice, however, \ac{smr}'s inefficiency still makes for limited throughput.
At the heart of \ac{smr} lies \broadcast~\cite{byz-atomic-broadcast-ic95}, a powerful consensus-equivalent primitive that comes with fundamental bounds~\cite{dr85} and constraints~\cite{flp85}, hindering its real-world performance despite decades of extensive research~\cite{pbft-osdi99,bft-separate-agreement-execution-sosp03,oracles-constantinople-jcrypt05,zyzzyva-sosp07,upright-sosp09,bft-smart-dsn14,raynal-podc14,abstract-tocs15,xft-osdi16,hotstuff-podc19,pompe-osdi20} and attention from industry, where \ac{smr} powers a myriad of blockchains and ledgers~\cite{book-wattenhofer-blockchain,ethereum-yellow-paper-2014, stellar-whitepaper-2016,algorand-sosp17,hyperledger-fabric-eurosys18,omniledger-sp18,teechain-sosp19,diembft-2021-08,dfinity-whitepaper-2022,aptos-whitepaper-2022,sui-whitepaper-2022}.

\begin{figure}[t]
    \centering
    \includegraphics[width=\columnwidth]{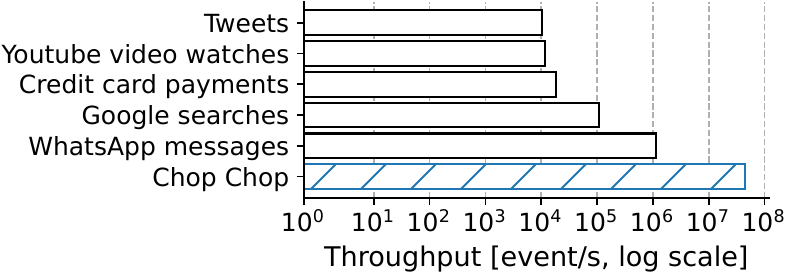}
    \vspace{\captionspace}
    \caption{\textbf{Throughput of Internet-scale services.}
    }
    \label{fig:motivation}
\end{figure}

When deployed globally, seminal \broadcast implementations, such as \bftsmart~\cite{bft-smart-dsn14} and \hotstuff~\cite{hotstuff-podc19}, can deliver a few thousand messages per second, three orders of magnitude short of the millions of requests per second collectively handled by the Internet's largest, centralized services (\cref{fig:motivation}).
Pushing \broadcast's throughput into the tens of millions of messages per second seems a necessary stepping stone towards achieving an Internet-scale computer.

\paragraph{Towards line rate.}

While slow and expensive, ordering messages in \broadcast is amenable to \emph{batching}~\cite{pbft-osdi99}: order once, deliver in bulk.
This observation motivated the development of \emph{memory pool} (mempool) protocols~\cite{narwhal-tusk-eurosys22,bullshark-ccs22,stratus-icde23}, as initiated by \narwhal~\cite{narwhal-tusk-eurosys22}, designed to amortize ordering.
This strategy proved effective, \eg \bullshark~\cite{bullshark-ccs22} delivers in the order of 380,000 messages per second when accelerated by \narwhal.
Despite this improvement, however, state-of-the-art batching still falls short of achieving \emph{line rate}, \ie matching the communication complexity of a protocol that does not ensure any ordering, authentication, or Byzantine resilience.
In such a simplified setting, a server could simply deliver a sequence of application messages as it receives them from the network: $b$ bits received, $b$ bits delivered.
Modern connections have enough bandwidth to receive tens of millions of application messages per second:\footnote{
    Payloads as small as 12 bytes can have real-world applications (see \cref{subsection:broadcastsbottlenecks}).
    A 5 Gbit/s link can receive 52 millions such payloads per second.
}
2.5 orders of magnitude of gap still exist between \broadcast and unordered, unauthenticated dissemination.
It is natural to ask if such a large gap is inherent to atomicity's unavoidable cost of ordering, authenticating and deduplicating messages.
This paper answers in the negative, accelerating \broadcast by a further two orders of magnitude with a system that performs close to optimal efficiency, \ie within 8\% of line-rate, even when handling 40 million requests per second.

\paragraph{\system.}

We present \system, a Byzantine \broadcast system using a novel authenticated mempool.
Mempools amortize the cost of ordering by having an underlying instance of \broadcast order batches.
Classic methods of batching, however, fail to also amortize authentication and deduplication: each payload in a batch still carries an individual public key, signature and sequence number.

\system addresses this shortcoming with a new form of batches: \emph{\batches}.
Unlike a classic batch, a \batch contains condensed information that allows to authenticate and deduplicate its messages in bulk, much faster than in existing schemes.
\Batches leverage the strong ordering of \broadcast to minimize redundant information.

\paragraph{Trustless brokers.}

\system produces \batches using a novel interactive protocol involving \emph{brokers}, a layer of facilitating processes between clients and servers.
\Batches are faster for servers to receive and process, but expensive for brokers to produce: \batchification is interactive and relies on expensive cryptographic operations for brokers.

Importantly, however, incorrectly \batchified batches are visibly malformed.
As such, brokers can be \emph{untrusted}: good brokers take load off the servers; bad ones cannot compromise the system's safety.
Servers are exposed to every message in the system, bottleneck easily, and only a threshold of them can be compromised before the system loses safety.
Brokers, instead, can be spun up by anyone, outside of \system's security perimeter, to meet the load produced by clients.

\paragraph{Evaluation.}

We evaluate \system in a cross-cloud, geo-distributed environment including 320 medium-sized AWS EC2 machines and 64 OVH machines.
We simulate up to 257 million clients and consider 12 experimental environments. 
Setting up each environment requires the installation of 13~TB of synthetic workload.
A naive installation using \texttt{scp} from a single machine would take 68 hours.
We designed \texttt{silk}, a one-to-many peer-to-peer file transfer tool optimized for high latency connections, to install the files in 30 minutes instead.

We compare \system's throughput and end-to-end latency against its baselines in multiple real-world scenarios including server failures, adverse network conditions, and applications running.
In all scenarios, \system's throughput outperforms its closest competitor by up to two orders of magnitude, with no penalty in terms of latency.
When put under stress, \system orders, authenticates and deduplicates upwards of 43,600,000 messages per second with a mean latency of 3.6 seconds.
Except under the most adverse network conditions and proportions of faulty clients, \system still achieves millions of operations per second.

\paragraph{Applications.}

Unlike most \broadcast implementations~\cite{hotstuff-podc19, bft-smart-dsn14, narwhal-tusk-eurosys22, bullshark-ccs22}, \system does not offload authentication and deduplication to the application.
This allows \system-based applications to focus entirely on their core logic without ever engaging in expensive, and easy to get wrong, cryptography.
To showcase this, we implement three simple applications to evaluate on top of \system: a Payment system, an Auction house and an instance of the game ``Pixel war''.
These three simple applications (300 lines of logic) work effectively with messages as small as 8 bytes, further underlying the communication overhead represented by public keys, signatures and sequence numbers in non-\batchified systems.
Both Payments and Pixel war inherit \system's throughput, respectively processing over 32 and 35 million operations per second.
Even the Auction house, which is single-threaded, achieves 2.3 million operations per second.
(These applications are meant as examples, and further optimization is beyond the scope of this paper.)

\paragraph{Contributions.}

We identify authentication and deduplication as the main bottlenecks of batched \broadcast; we introduce \batches to extend the amortizing properties of batching to authentication and deduplication; we present \batchification, an interactive protocol to produce \batches, and identify the opportunity to offload it to an untrusted set of brokers; we implement \system, a Byzantine \broadcast system that takes advantage of \batchification through an authenticated mempool; we thoroughly evaluate \system, improving state-of-the-art \broadcast throughput by two orders of magnitude, maintaining near line-rate performance up to 40 million requests per second; we showcase \system through a Payment system, an Auction house and an instance of the ``Pixel war'' game, respectively achieving 32, 2.3 and 35 million operations per second.

\paragraph{Roadmap.}

\Cref{section:broadcast} introduces \broadcast, discusses classic batching mechanisms and highlights the cost of authenticating and deduplicating messages in the resulting batches.
\Cref{section:batches} presents \batches and introduces a simplified failure-free version of \system's protocol.
\Cref{section:system} describes \system's fault-tolerant protocol in detail.
\Cref{sec:implementation} discusses \system's implementation.
\Cref{sec:evaluation} discusses \system's empirical evaluation, highlighting the challenges of such large scale experiments.
We summarize related work in \Cref{section:relatedwork} and future work in \Cref{section:conclusions}.
\Cref{appendix:artifact} describes \system's artifact.
\Cref{appendix:chop} contains the full correctness proof of \system.

\section{Atomic Broadcast}
\label{section:broadcast}

In an \broadcast system~\cite{ct96}, \emph{clients} broadcast messages that are delivered by \emph{servers}. 

\paragraph{Properties~\cite{book-cachin-guerraoui-rodrigues}.}

Correct servers deliver the same messages in the same order (\emph{agreement}). Messages from correct clients are eventually delivered (\emph{validity}). Spurious messages cannot be attributed to correct clients (\emph{integrity}). No message is delivered more than once (\emph{no duplication}).

\subsection{Cost of Atomic Broadcast}
\label{subsection:broadcastsbottlenecks}

Informally, \broadcast's most distinctive property, agreement, is also the most challenging to satisfy.
Correct servers must coordinate to \emph{order} messages without compromising liveness.
A great deal of research effort has been put in developing ordering techniques, optimizing for latency~\cite{fast-bft-consensus-dsn05,zyzzyva-sosp07} or communication complexity~\cite{proba-quorum-ic01,kauri-sosp21}.

Integrity and no duplication, instead, allow for simple solutions. 
Clients can ensure integrity by \emph{authenticating} their messages using digital signatures: servers simply ignore incorrectly authenticated messages.
For no duplication, clients can tag each message with a strictly increasing \emph{sequence number}: after ordering, servers discard old messages as replays.

Both techniques---we call them \emph{classic authentication} and \emph{classic sequencing}---are non-interactive, easy to implement, and agnostic of the protocol employed to order messages.
Arguably due to the simplicity and effectiveness of classic authentication and sequencing, most \broadcast implementations overlook integrity and no duplication entirely: they offload authentication and sequencing to the application, focusing on the more challenging task of ordering.

\paragraph{Batching for ordering.}

Lacking an efficient technique to minimize its complexity, ordering could be \broadcast's main bottleneck.\footnote{
    Byzantine \broadcast among $n$ participants cannot be achieved with a bit complexity smaller than $\Theta(n^2)$~\cite{dr85}.
}
The well-known strategy of \emph{batching}, however, is both general and effective at amortizing the agreement cost of an \broadcast implementation~\cite{pbft-osdi99,bft-fire-nsdi08}.

Broadly speaking, batching is orchestrated by a \emph{broker} as follows~\cite{narwhal-tusk-eurosys22}.\footnote{
    In the literature, servers usually play the role of brokers. 
    As we discuss in \cref{section:system}, however, \system minimizes its load on the servers by offloading brokerage to a separate, trustless set of processes.
}
Over a small window of time, the broker collects multiple client-issued messages in a batch, which it disseminates to the servers;
the broker then submits to an underlying instance of \broadcast a cryptographic hash of the batch it collected;
upon delivering the hash of a batch from \broadcast, a server retrieves the batch, and delivers to the application all the messages it contains.
Because the size of a hash is constant, the cost of ordering a batch does not depend on its size: as batches become larger, the cost of ordering each message goes to zero.
In practice, batching can effectively eliminate the cost of ordering in any real-world implementation of \broadcast. 

\paragraph{Cost of integrity and no duplication.}

Batching does not efficiently uphold integrity and no duplication.
Regardless of how many messages are batched together, the cost of classic authentication and sequencing stays constant: one public key, one signature and one sequence number for each message.

In practice, these costs dominate the computation and communication budget of a batched \broadcast system (see \cref{subsection:thepowerofbatchification}).
On the one hand, signatures are among the most CPU-intensive items in the standard cryptographic toolbox, dwarfing in particular symmetric primitives such as hashes and ciphers.
On the other, public keys, signatures and sequence numbers can easily account for the majority of a batch's size.

To illustrate these costs, consider the example of a payment system.
A payment operation requires three fields: sender, recipient, and amount.
Sender and recipient fit in 4~B each if the system serves less than 4 billion users.
Amount needs 4~B for payments between 1 cent and 40 millions.
Hence, a payment can be encoded in just 12~B.
Using public keys to identify sender and recipient (2$\times$32~B using Ed25519~\cite{curve25519-pkc06,rfc8032-ed25519}) and attaching a signature (64~B) and a sequence number (8~B) to each message inflates payloads to 140~B.
For payments, \emph{91\% of the bandwidth is spent on integrity and no duplication}.

\subsection{Existing Mitigations}
\label{subsection:twostepsforward}

\system integrates the two following techniques to reduce the bandwidth and CPU cost of authentication.

\paragraph{Short identifiers.}

Repeated public keys consume a significant slice of a server's communication budget.
A workaround is to have servers store public keys in an indexed \emph{directory}~\cite{directory-protocol-beatcs10}.

\system uses \diral as its directory (see \cref{appendix:diral}).
In \diral, upon first joining the system, a client announces its public key via \broadcast to \emph{sign up}.
Upon delivering a sign-up message, a server appends the new public key to its directory.
The same public key appears at the same position in the directory of all correct servers thanks to \broadcast's agreement.
Having signed up, a client uses its position in the directory as identifier instead of its public key.

In the previous example of a payment system, using such identifiers reduces a payment size by 40\%, from 140~B to 84~B.
However, a signature per payment must still be transmitted.

\paragraph{Pooled signature verification.}

Authenticating a batch by verifying its signatures is a computationally intensive task for a server~\cite{pbft-osdi99,mirbft-jsys22}.
However, Red Belly\cite{red-belly-sp21} and Mir~\cite{mirbft-jsys22} showed that not all servers need to authenticate all batches.
Indeed, assuming at most $f$ faulty servers, a broker optimistically asks only $f + 1$ servers to authenticate a batch to be certain to reach at least one correct server.
If $f + 1$ servers do not reply by a timeout, the broker extends its request to $f$ additional servers, thus reaching at least $f + 1$ correct servers.

A correct server that authenticates a batch sends back to the broker a \emph{witness shard}, \ie a signed statement that the batch is correctly signed.
The broker aggregates $f + 1$ identical shards into a \emph{witness}, which it sends to the other $2f$ servers.
Because every witness contains at least one correct shard, the servers can trust the witness instead of verifying the batch.

Assuming $3f + 1$ servers, this technique shaves up to two-thirds off the system's authentication complexity.

\section{Distilled Batches}
\label{section:batches}

\system's main contribution is \emph{\batchification}, a set of techniques aimed at extending the amortizing properties of batches to authentication and sequencing.

\paragraph{Background: multi-signatures.}

\system makes use of multi-signature schemes~\cite{multisig-report83} to authenticate batches.
Secret keys produce signatures that can be verified against the corresponding public keys.
Public keys and signatures, however, can be \emph{aggregated}.
Let $(p_1, r_1), \ldots, (p_n, r_n)$ be distinct key pairs, and $s_1, \ldots, s_n$ be signatures produced by $r_1, \ldots, r_n$ on the \emph{same} message $m$: $p_1, \ldots, p_n$ (resp., $s_1, \ldots, s_n$) can be aggregated into a 
constant-sized aggregate public key $p$ (resp., aggregate signature $s$).

Remarkably, $s$ can be verified in constant time against $p$ and $m$~\cite{bls-multisig-asiacrypt18,musig-schnorr-dcc19}.
\system uses BLS multi-signatures~\cite{bls-multisig-asiacrypt18} which can be aggregated cheaply and non-interactively: even a non-signing process can compute $p$ (resp., $s$) once provided with $p_1, \ldots, p_n$ (resp., $s_1, \ldots, s_n$) by computing a single multiplication over an elliptic curve.

\subsection{Distillation at a Glance}
\label{subsection:batchificationataglance}

\begin{figure}[tb]
    \centering
    \includegraphics[width=\columnwidth]{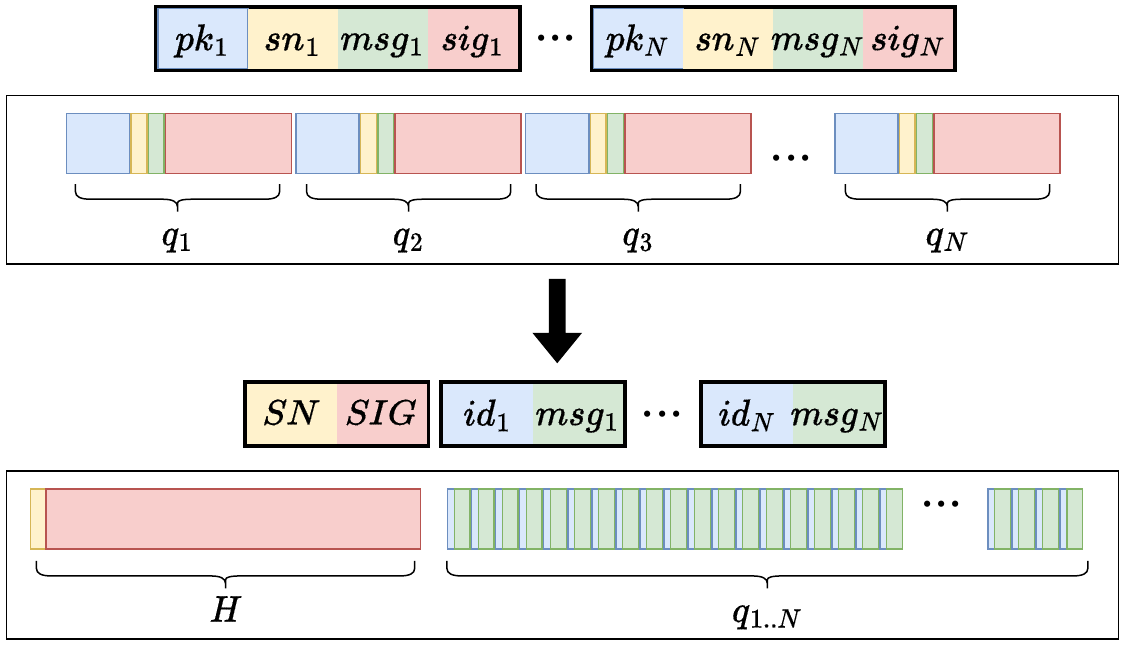}
    \vspace{\captionspace}
    \caption{\textbf{\Fullbatchification in action.} With classic authentication and sequencing, each payload $q_i$ contains a public key $pk_i$, a sequence number $sn_i$, a message $msg_i$ and a signature $sig_i$. In the \fullbatchified case, each $q_i$ reduces to just $id_i$ and $msg_i$: one header $H$, composed of one aggregate sequence number $SN$ and one aggregate signature $SIG$, is sufficient for the entire batch.
    Bars are to scale if small messages are broadcast using Ed25519 for signatures and BLS12-381 for uncompressed multi-signatures: $sn_i$ and $SN$ are 8~B, $msg_i$ is 8~B, $pk_i$ is 32~B, $sig_i$ is 64~B, $SIG$ is 192~B.
    }
    \label{figure:fulldistillationinaction}
\end{figure}

In brief, \batchification aims to produce \emph{\batches}.
A \batch has some of its signatures (resp., sequence numbers) replaced by an \emph{aggregate signature} (resp., \emph{aggregate sequence number}).
When maximally successful, \batchification produces a \emph{\fullbatch}, where all signatures (resp., sequence numbers) have been replaced by a \emph{single} aggregate signature (resp., sequence number). 
As we discuss below, \batches are vastly cheaper for servers to receive and process.
\Cref{figure:fulldistillationinaction} depicts the effect of \batchification on a batch.

\paragraph{\Fullbatchification (failure-free).}

For pedagogical purposes, we introduce \batchification under the assumption that all processes are correct.
We detail \system's fault-tolerant \batchification techniques in \cref{subsection:batchification}, optimized and adapted to the Byzantine setting.
As in the classic batching case, a set $\client_1, \ldots, \client_b$ of clients submit their messages $m_1, \ldots, m_b$ to a broker $\broker$. 
Each $\client_i$ selects for its message $m_i$ a sequence number $k_i$ (greater than any sequence number it previously used), then sends $(k_i, m_i)$ to $\broker$. 
Upon receiving all $(k_i, m_i)$-s, $\broker$ computes the aggregate sequence number
\begin{equation*}
    k = \max_i k_i
\end{equation*}
then builds the \emph{batch proposal}
\begin{equation*}
    B = \left[ \left( x_1, k, m_1 \right), \ldots, \left( x_b, k, m_b \right) \right]
\end{equation*}
where $x_i$ is $\client_i$'s numerical identifier in the system (see \cref{subsection:twostepsforward}).
$\broker$ then sends $B$ back to every $\client_i$.
Upon receiving $B$, $\client_i$ produces a multi-signature $s_i$ for the hash $H(B)$ of $B$, which it sends back to $\broker$.
Having collected all multi-signatures, $\broker$ computes the aggregate signature
\begin{equation*}
    s = \prod_i s_i
\end{equation*}
In doing so, $\broker$ obtains the \fullbatch
\begin{equation*}
    \tilde B = \left[ s, k, \left( \left( x_1, m_1 \right), \ldots, \left( x_b, m_b \right) \right) \right]
\end{equation*}
Upon receiving $\tilde B$, any server now can: compute $B$ by inserting $k$ between each $(x_i, m_i)$; compute $H(B)$; use each $x_i$ to retrieve $\client_i$'s public key $p_i$ from its directory; compute the aggregate public key
\begin{equation*}
    p = \prod_i p_i
\end{equation*}
and finally verify $s$ against $p$ and $H(B)$. 

\paragraph{Distillation outcome.}

Having engaged with $\broker$ to \batchify the batch, every $\client_i$ multi-signs the \emph{same} message $H(B)$ and updates its sequence number to the \emph{same} $k$.
This allows $\broker$ to authenticate and sequence all of $\tilde B$ using $s$ and $k$ only.

\paragraph{Distillation safety.}

The proposed \batchification protocol has no safety drawback.
First, because $(x_i, k, m_i)$ appears in $B$, $\client_i$ still gets to authenticate $m_i$.
Intuitively, $\client_i$'s multi-signature on $H(B)$ publicly authenticates \emph{whatever message in $B$ is attributed to $\client_i$}, $m_i$ in this case.
Second, because $k \geq k_i$, $k$ is still a valid sequence number for $m_i$.
Sequence number \batchification might cause $\client_i$ to skip some sequence numbers whenever any $\client_j$ issues some $k_j > k_i$.
Contiguity of sequence numbers, however, is not a requirement for deduplication.
As with classic sequencing, $\client_i$ produces---and servers deliver---messages with strictly increasing sequence numbers; servers disregard all other messages as replays.

\subsection{Distillation Microbenchmark}
\label{subsection:thepowerofbatchification}

Having discussed how \batches are produced, we now estimate the significance of their effect by means of a back-of-the-envelope calculation and a simple microbenchmark on AWS.
Consider a setting where 100 million clients broadcast 8-byte messages, \eg to issue payments (see \cref{subsection:broadcastsbottlenecks}).
We compare \emph{classic authentication and sequencing}, where clients are identified by their public keys, messages are individually signed and sequenced, against \emph{\fullbatches} where clients are identified by a numerical identifier and each batch contains only one aggregated signature and sequence number.
We use Ed25519~\cite{rfc8032-ed25519} for signatures (32~B public keys, 64~B signatures) and BLS12-381~\cite{rfc-bls-wip2020-09} for multi-signatures (192~B uncompressed signatures).
We use uncompressed BLS multi-signatures to save computation time at the cost of storage space (96~B compressed vs. 192~B uncompressed).

\begin{figure}[tb]
    \centering
    \includegraphics[width=\columnwidth]{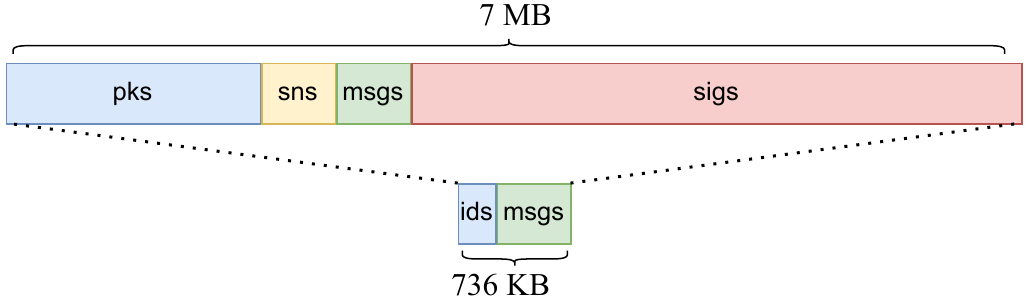}
    \vspace{\captionspace}
    \caption{
    \textbf{\Fullbatchification of a batch of 65,536 payloads (sizes to scale).}
    The aggregate signature and aggregate sequence number do not appear as a result of their small size.
    }
    \label{figure:sizeofabatch}
\end{figure}

\paragraph{Communication complexity.}

Payloads are 112~B per message in the classic case (32~B of public key, 8~B of sequence number, 8~B of message, 64~B of signature) vs. 11.5~B in the \fullbatchified case (28~bits~=~3.5~B of identifier to represent 257M clients, 8~B of message).
Assuming batches of 65,536 messages (\cref{figure:sizeofabatch}), classic batches are exactly 7~MB long, while \fullbatches are 736~KB long including aggregate signature and sequence number.

\paragraph{Computation complexity.}

Running at maximum load, an Amazon EC2 c6i.8xlarge instance authenticates $16.2 \pm 0.4$ classic batches per second using Ed25519's batch verification for 65,536 signatures.
The same machine authenticates $457.1 \pm 0.3$ \fullbatches per second: each authentication requires the aggregation of 65,536 BLS12-381 public keys and the verification of one BLS12-381 multi-signature.

\paragraph{Summary.}

By the order-of-magnitude calculations above, \fullbatches hold the promise to reduce the costs of authentication and sequencing by a factor 9.7 for network bandwidth, and 28.2 for CPU.
\system aims to deliver on that promise for a real-world fault-tolerant system.

\section{Chop Chop}
\label{section:system}

This section overviews \system's architecture, \system's protocol, and provides arguments for its correctness.

\paragraph{Overview.}

\system involves three types of processes (\cref{figure:system-architecture}): broadcasting clients, delivering servers and a layer of broadcast-facilitating brokers between them.
Servers run an \broadcast instance among themselves, to which brokers submit messages.
\system is agnostic to the implementation of \broadcast used by the servers.
On top of the provided broker-to-server \broadcast, Chop Chop implements a much faster client-to-server \broadcast: clients submit messages to the servers, aided by brokers.

\system's protocol unfolds in two phases: \emph{\batchification} (\cref{subsection:batchification}) and \emph{submission} (\cref{subsection:submission}).
In the \batchification phase, clients interact with a broker to gather their messages in a \batch (see \cref{section:batches}).
In the submission phase, the broker disseminates the \batch to the servers and submits the batch's hash to the server-run instance of \broadcast.
Upon delivering its hash from \broadcast, servers retrieve the batch and deliver its messages.
\system's contributions mainly focus on the \batchification phase.
\system's submission strategy closely resembles prior batch-based \broadcast implementations~\cite{narwhal-tusk-eurosys22,bullshark-ccs22,stratus-icde23}.

\subsection{Architecture and Model}
\label{subsection:system}

\begin{figure}[t]
    \centering
    \includegraphics[width=\columnwidth]{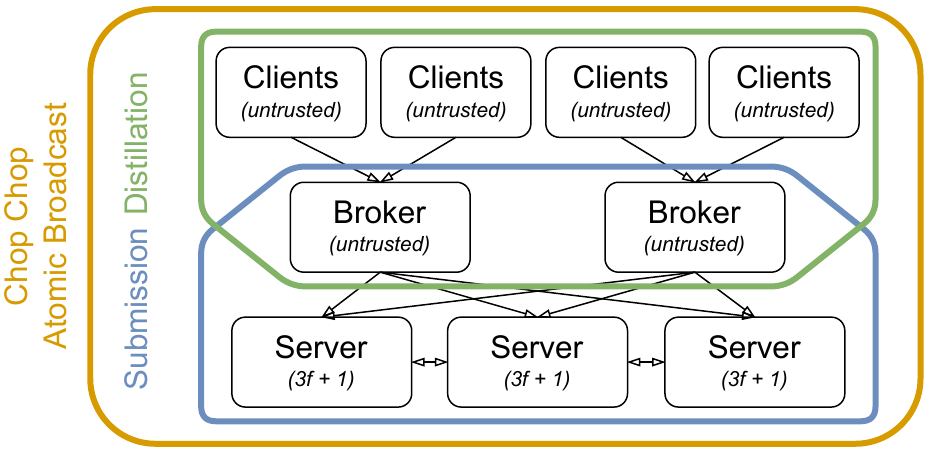}
    \vspace{-20pt}
    \caption{\textbf{\system architecture.}}
    \label{figure:system-architecture}
\end{figure}

\system augments the architecture of a classic \broadcast, as described in \cref{section:broadcast}, with novel brokers.

\paragraph{Clients and servers.}

\emph{Clients} broadcast messages to a (distinct) set of \emph{servers}.
We assume that less than one third of servers can be faulty and behave in an arbitrary manner, \ie be Byzantine~\cite{byz-generals-toplas82}, while all clients can be faulty.
For simplicity, servers form a fixed set that is known by all correct processes at system startup.
\system can be extended for reconfiguration thanks to its modular use of \broadcast~\cite{bft-smart-dsn14,reconf-smr-news10} (\cref{figure:system-architecture}).
Clients issue messages after broadcasting their public keys to the system (see \cref{subsection:twostepsforward}).

\paragraph{Brokers.}
 
We discussed in \cref{section:batches} how both classic and \batches are assembled by a broker.
The role of brokers is traditionally taken by servers.
Given the additional strain put on brokers by \system's interactive \batchification protocol, however, having servers be brokers would result in a waste of scarce, trusted resources.
Importantly, however, \batchification is \emph{trustless}.
On the one hand, agreement rests entirely on \system's underlying \broadcast instance, for which brokers are only clients. 
On the other hand, as we argue in \cref{subsection:batchification,subsubsection:safety}, a faulty broker cannot compromise integrity or no duplication: \batches are publicly authenticated, and correct clients cannot be tricked into using stale sequence numbers.
Hence, \emph{brokers need no trust}: a broker either does its job correctly or produces \batches that are visibly malformed, and easily discarded by all correct servers.

This observation is of paramount importance to the performance of \system: \emph{because \batchification is heavy but trustless, brokers should be distinct from servers}.
Along with clients and servers, we thus assume a third, \emph{independent set of brokers}, sitting between clients and servers, to accelerate \broadcast by assembling client messages in \batches.
We assume that at least one broker is correct; the system loses liveness but not safety if all brokers are faulty.

\paragraph{Network.}

\system guarantees that the batches collected and submitted to servers by correct brokers are well-formed even in asynchrony, but achieves \fullbatchification when the network is synchronous (see \cref{subsection:batchification}).
\system inherits the network requirements of its underlying \broadcast.

\subsection{Distillation Phase}
\label{subsection:batchification}

\begin{figure*}[t]
    \centering
    \includegraphics[height=47mm]{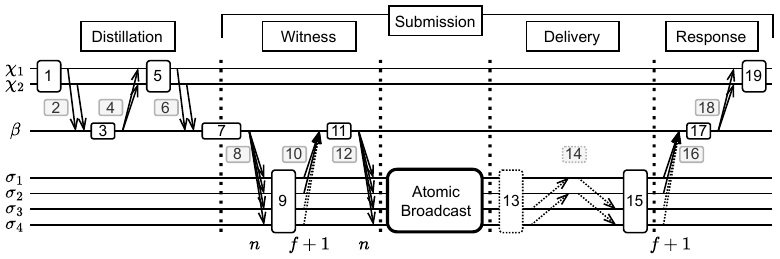}
    \vspace{-5pt}
    \caption{\textbf{Overview of the \system protocol between two clients ($\client_1$, $\client_2$), a broker ($\broker$) and four servers ($\server_1$--$\server_4$).}
    The protocol is comprised of 19 steps (\#1--\#19) and of an underlying instance of \broadcast such as \bftsmart or \hotstuff.
    }
    \label{figure:system-protocol}
\end{figure*}

We introduced in \cref{section:batches} a simplified, failure-free \batchification protocol.
This section describes how \system renders \batchification tolerant to arbitrary failures and improves its performance via a sequence of improvements, each addressing a shortcoming of the simplified protocol.
The complete fault-tolerant protocol of \system is depicted in \cref{figure:system-protocol}.

In the failure-free \batchification protocol: clients $\client_1, \ldots, \client_b$ send their messages $m_1, \ldots, m_b$, with sequence numbers $k_1, \ldots, k_b$ (\#2) to a broker $\broker$ (\cref{figure:system-protocol}, \#1);
$\broker$ identifies the maximum submitted sequence number $k$ and builds a batch proposal $B = [(x_1, k, m_1), \ldots, (x_b, k, m_b)]$ (\#3); 
$\broker$ disseminates $B$ to $\client_1, \ldots, \client_b$ (\#4);
each $\client_i$ produces a multi-signature $s_i$ on $H(B)$ (\#5), which it sends to back $\broker$ (\#6);
$\broker$ aggregates $s_1, \ldots, s_n$ into an aggregate $s$, thus producing a \fullbatch $\tilde B = [s, k, ((x_1, m_1), \ldots, (x_b, m_b))]$ (\#7).

\paragraph{Background: Merkle trees.}

\system uses Merkle trees~\cite{merkle-tree-crypto87} to hash batches.
An $l$-element vector $z_1, \ldots, z_l$ is hashed into a \emph{root} $r$, used as commitment.
For each $i$, $z_i$'s value can be proved by means of a proof of inclusion $p_i$, verifiable against $r$ and $z_i$.
Proofs of inclusions are $O(\log l)$ in size and are verified in $O(\log l)$ time.

\paragraph{What if a broker forges messages?}

A faulty $\broker$ could try to falsely attribute to some $\client_i$ a message $m'_i \neq m_i$.
$\broker$ could do so by replacing $m_i$ with $m'_i$ in $B$, then having $\client_i$ sign $H(B)$, thus implicitly authenticating $m'_i$.
This is easily fixed by having $\client_i$ check that $m_i$ correctly appears in $B$ before signing $H(B)$.

\paragraph{Can a broker avoid sending the entire batch?}

A clear inefficiency of the simplified protocol is that $\broker$ has to convey all of $B$ back to each $\client_i$.
This is fixed using Merkle trees.
Upon assembling $B$, $\broker$ computes the Merkle root $r$ of $B$, along with the Merkle proof $p_i$ for each $(x_i, k, m_i)$ in $B$.
Instead of sending $B$ to all clients, $\broker$ just sends $r$, $k$ and $p_i$ to each $\client_i$.
Upon receiving $r$, $k$ and $p_i$, $\client_i$ checks $p_i$ against $r$ and $(x_i, k, m_i)$, producing $s_i$ on $r$ only if the check succeeds.
If $\client_i$ signs $r$, then $(x_i, k, m_i)$ is necessarily an element of $B$.
Importantly, however, $\broker$ could inject $(x_i, k, m'_i \neq m_i)$ somewhere else in $B$, while still providing $\client_i$ only with the proof for $(x_i, k, m_i)$.
This is solved by having servers ignore every \batch where two or more messages are attributed to the same client.
This way, if $\client_i$ signs $r$, then either $m_i$ is the only message in $B$ attributed to $\client_i$, or $\tilde B$ is rejected by all servers as malformed: either way, integrity is upheld.

\paragraph{What if a client does not multi-sign?}

Under the assumption that $\client_1, \ldots, \client_b$ are correct, $\broker$ can safely wait until it collects all $s_1, \ldots, s_b$. 
This policy is clearly flawed in the Byzantine setting: a single crashed client can prevent $\broker$ from ever aggregating $s$.
Furthermore, lacking an assumption of synchrony, $\broker$ cannot exclude from $\tilde B$ those clients that do not sign $r$ by some timeout: consistently slow clients would always be excluded, and validity would be lost.
This issue is fixed by the fallback mechanism introduced in the following.

\paragraph{Fault-tolerant \batchification.}

Upon first sending $(k_i, m_i)$ to $\broker$ (\#2), $\client_i$ also sends an individual, non-aggregable signature $t_i$ for $(x_i, k_i, m_i)$, which $\broker$ stores.
$\broker$ then waits for $s_i$-s on $r$ until either all $s_i$-s are collected, or a timeout expires.
For every $s_i$ that ends up missing, due to $\client_i$ being crashed or delayed, $\broker$ attaches $(k_i, t_i)$ to $\tilde B$.
Upon receiving $\tilde B$, a server first checks each individual signature $t_i$ against the corresponding $(x_i, k_i, m_i)$.
The server then checks $s$ against the public keys of the clients for which an individual signature $t_i$ was not given, \ie the public keys of all clients that signed $r$ in time.

In summary: fast, correct clients who successfully produce their $s_i$-s in time authenticate their message by multi-signing $r$; slow or crashed clients still get their messages through, individually authenticated by the $t_i$-s that they originally produced.
\Fullbatchification is achieved whenever the network is synchronous and all clients are correct, which we argue is the case in practice for the majority of a system's lifetime.
When the network is asynchronous, however, a fraction of clients might fail to produce their $s_i$ in time, resulting in a \emph{\partialbatch}.
At the limit where all clients fail to sign $r$ in time, $\tilde B$ reduces to a classic batch, degrading server-side performance to pre-\batchification levels.
We underline that safety and liveness are preserved regardless of synchrony.

\paragraph{What if a broker replays messages?}

A problem introduced by the last fix is that $\client_i$ authenticates both $k_i$ and $k$ as sequence numbers for $m_i$, allowing a faulty $\broker$ to play $m_i$ twice, hence breaking \broadcast's no duplication.
This is fixed by having each client engage in the broadcast of only one message at a time.
This way, while $\broker$ can indeed replay $m_i$, it can only do so consecutively: all sequence numbers $\client_i$ authenticates for $m_i$ belong to a range that does not contain sequence numbers for any other message $m_{i' \neq i}$ issued by $\client_i$.

This observation is key to the following fix: along with the last sequence number $\bar k_\client$ each client $\client$ used, a correct server $\server$ stores the last message $\bar m_\client$ that $\client$ broadcast; upon ordering a message $m$ with sequence number $k$ from $\client$, $\server$ delivers $m$ if and only if $k > \bar k_\client$ and $m \neq \bar m_\client$.
In doing so, $\server$ discards all consecutive replays of $\bar m_\client$, thus preventing replays in general.

\paragraph{What if a client broadcasts too frequently?}

The last fix relies on clients broadcasting one message at a time.
Depending on latency, a client broadcasting too frequently might accrue an ever-growing queue of pending messages.
This issue is fixed by flushing application messages in bursts, akin to Nagle's buffering algorithm for TCP.

\paragraph{What if a client uses the largest possible sequence number?}

Assuming that a finite number of bits (\eg $64$) are allocated to representing sequence numbers, a faulty client $\client_m$ could set its $k_m$ to the largest possible sequence number $k_{max}$ (\eg $2^{64} - 1$). 
In doing so, $\client_m$ would force all other $\client_i$-s to update their sequence number to $k_{max}$.
Since correct clients only use strictly increasing sequence numbers, no $\client_i$ could ever broadcast again: sequence numbers would run out.
Proving the \emph{legitimacy} of sequence numbers fixes this issue.

\paragraph{Legitimate sequence numbers.}

By the rule that we established, no more than one message from the same client can appear in the same batch.
Moreover, correct clients always tag their messages with the smallest sequence number they have not yet used, \ie the largest they have used plus one.
By induction, we then have that unless some client misbehaves, no client ever needs to use a sequence number larger than the number of batches ever delivered by the servers: the largest sequence number any client submits to the very first batch is $0$, therefore no client submits a sequence number larger than $1$ to the second batch, and so on. 
This observation allows us to define as \emph{legitimate} any sequence number smaller than the number of batches servers have delivered at any given time.

\paragraph{Legitimacy proofs.}

This definition of legitimacy allows for the generation of \emph{legitimacy proofs}: upon delivering the $n$-th batch, a server publicly states so with a signature.
By collecting $f + 1$ server signatures stating that the $n$-th batch was delivered into a certificate $l_n$, any process can publicly prove that any sequence number smaller than $n$ is legitimate.

Upon initially submitting $k_i$ (\#2), $\client_i$ also sends to $\broker$ a certificate $l_n$, for any $n > k_i$;
$\broker$ ignores client submissions that lack such certificate, except when $k_i = 0$ since no certificate is needed.
Upon sending $k$ back to all $\client_i$-s (\#4), $\broker$ attaches the highest $l_{\hat n}$ it collected: $l_{\hat n}$ proves that $k$ is legitimate since $\hat n > k$.
$\client_i$ signs $r$ (\#5) only if $k$ is proved legitimate by $l_{\hat n}$.

This technique ensures correct clients always use legitimate sequence numbers.
Since legitimate sequence numbers grow only with the number of delivered batches, no correct client is forced to skip too far ahead, compromising its own liveness. 

\paragraph{What if a broker crashes?}

If $\broker$ fails to engage in the protocol, each $\client_i$ can submit its message to any other broker.

\subsection{Submission Phase}
\label{subsection:submission}

The submission phase ensures that all servers efficiently deliver a \batch, and that all broadcasting clients receive a proof that their messages were delivered.

\paragraph{Witness.}

Having gathered a \batch $\tilde B$ (\#7), $\broker$ moves on to have $f + 1$ servers signs a \emph{witness shard} for $\tilde B$.
In signing a witness shard for $\tilde B$, a server $\server$ simultaneously makes two statements.
First, $\tilde B$ is \emph{well-formed}: $\server$ successfully verified $\tilde B$'s signatures and found all messages in $\tilde B$ to have a different sender.
Second, $\tilde B$ is \emph{retrievable}: $\server$ stores $\tilde B$ and makes it available for retrieval, should any other server need it.
We call a \emph{witness} for $\tilde B$ the aggregation of $f + 1$ witness shards for $\tilde B$.
Because any set of $f + 1$ processes includes a correct process, when presented with a witness for $\tilde B$ any server can trust $\tilde B$ to be well-formed and retrievable.

As discussed in \cref{subsection:twostepsforward}, witnesses optimize server-side computation.
Only $f + 1$ servers need to engage in the expensive checks required to safely witness $\tilde B$.
All other servers can trust $\tilde B$'s witness, saving trusted CPU resources.

In order to collect a witness for $\tilde B$, $\broker$ sends $\tilde B$ to all servers (\#8).
Optimistically, $\broker$ asks only $f + 1$ servers to sign a witness shard for $\tilde B$, progressively extending its request to $2f + 1$ servers upon expiration of suitable timeouts.
Upon receiving $\tilde B$ (\#9), a correct server $\server$ stores $\tilde B$.
If asked to witness $\tilde B$, $\server$ checks that $\tilde B$ is well-formed and sends back to $\broker$ its witness shard for $\tilde B$ (\#10).
$\broker$ collects and aggregates $f + 1$ shards into a witness for $\tilde B$ (\#11), then submits $\tilde B$'s hash and witness to the server-run \broadcast (\#12).

\paragraph{Delivery.}

Upon delivering $\tilde B$'s hash and witness from \broadcast (\#13), a correct server $\server$ retrieves $\tilde B$, either from its local storage (if it directly received $\tilde B$ from $\broker$ at \#8) or from another server (\#14).
Because $\tilde B$ is retrievable, $\server$ is guaranteed to eventually find a server to pull $\tilde B$ from.
Having retrieved $\tilde B$ (\#15), $\server$ delivers all non-duplicate messages in $\tilde B$ (see \cref{subsection:batchification} for how $\server$ detects duplicates).

\paragraph{Response.}

Finally, $\server$ signs a \emph{delivery certificate}, listing the messages in $\tilde B$ that $\server$ delivered.
$\server$ sends its signature back to $\broker$ (\#16).
By agreement of \broadcast, all correct servers deliver the same subset of messages in $\tilde B$.
As such, $\broker$ is guaranteed to eventually collect $f + 1$ signatures on the same delivery certificate (\#17).
Upon doing so, $\broker$ distributes a copy of $\tilde B$'s delivery certificate to $\client_1, \ldots, \client_b$ (\#18).
Armed with $\tilde B$'s delivery certificate, a correct $\client_i$ 
can publicly prove the delivery of $m_i$ (\#19) and safely broadcast its next message.

\subsection{Correctness}
\label{sec:protocol-correctness}

This section summarizes \system's correctness analysis.
We prove \system's correctness to the fullest extent of formal detail in \cref{appendix:chop}.
(The agreement property defined in \cref{section:broadcast} is split into agreement and total order in \cref{appendix:chop}.)

\subsubsection{Safety}
\label{subsubsection:safety}

The safety of \system is given by its agreement, integrity and no duplication properties (see \cref{section:broadcast}).

\paragraph{Agreement.}

\system inherits agreement from its underlying, server-run instance of \broadcast.
A correct server delivers messages only upon delivering the hash of a batch from the server-run \broadcast.
Upon doing so, a correct server retrieves the full batch, checks its hash, and delivers all its messages in order of appearance.
All correct servers deliver the same messages in the same order assuming cryptographic hashes are collision-resistant.

\paragraph{Integrity.}

A correct server only delivers messages included in a batch witnessed by $f + 1$ servers, \ie by at least one correct server.
A correct server witnesses a batch only if: no more than one message in the batch is attributed to the same client; every client in the batch authenticates its message with a signature or the root of the batch's Merkle tree with a multi-signature.
A correct client multi-signs the root of a batch's Merkle tree only upon receiving a proof of the inclusion of its message in the batch.
As such, if a correct client multi-signs the root of a batch's Merkle tree, either the batch contains only the client's intended message or it is not witnessed.
In summary, a correct server delivers a message $m$ from a correct client $\client$ only if $\client$ broadcast $m$.

\paragraph{No duplication.}

A correct client only broadcasts one message at a time.
As such, while the client might attach multiple sequence numbers to the same message (different brokers may propose different aggregate sequence numbers for the client to authenticate) the sequence numbers the client attaches to each message belong to distinct ranges.
A correct server delivers client messages only in increasing order of sequence number, and ignores repeated messages.
This means that a correct server delivers at most one message from each sequence number range.
In summary, no server delivers a correct client's message more than once.

\subsubsection{Liveness}
\label{subsubsection:liveness}

The liveness of \system is given by its validity property.

\paragraph{Validity.}

If a correct client submits its message to a correct broker, the message is guaranteed to eventually be delivered by all correct servers: even if the client fails to engage in \batchification in a timely manner, its message is still included in a batch which gets disseminated, witnessed and delivered by all correct servers.
Faulty brokers can clearly refuse to service (specific) clients.
Upon expiration of a suitable timeout, however, a correct client submits its message to a different broker.
As we assume that at least one broker is correct, all correct clients are eventually guaranteed to find a correct broker and get their messages delivered by all correct servers.

\subsubsection{Other Attacks}
\label{subsubsection:other-attacks}

As we outlined in \cref{subsubsection:liveness,subsubsection:safety}, \system satisfies all properties of \broadcast.
In this section, we consider other attacks an adversary might deal to impair \broadcast's performance and fairness~\cite{fair-bft-crypto20} in \system.

\paragraph{Denial of service.}

A faulty broker may refuse to service clients, thus forcing them to fall back on other brokers, increasing latency.
A faulty broker may also submit deliberately non-\batchified batches to servers to force them to waste trusted resources to receive and verify individual signatures.
While handling DoS is beyond the scope of this paper, \system is amenable to accountability mechanisms~\cite{peerreview-sosp07}.
Brokers could be asked to stake resource to join the system.
Correct, high-performance brokers could be rewarded, akin to gas fees in Ethereum~\cite{ethereum-yellow-paper-2014}.
Brokers that accrue a reputation of misbehavior or slowness could be banned and lose their initial stake.

\paragraph{Front-running.}

A faulty broker might impact fairness by front-running messages of interest~\cite{flashboys-sp20,sandwich-attacks-sp21}.
While front-running resistance is beyond the scope of this paper, \system is compatible as-is with existing mechanisms to mitigate or prevent front-running, most notably schemes that have clients submit encrypted messages whose content is revealed only after delivery~\cite{f3b-icdcsw22,fairblock-securecomm22}.
Importantly, these encrypt-order-reveal schemes could be selectively employed only for those messages that are vulnerable to front-runs, \eg messages used for stock trading~\cite{mev-measures-sp22}.
Maintaining \system's throughput while providing quorum-enforced fairness for every message~\cite{pompe-osdi20} opens a valuable future avenue of research.

\section{Implementation Details}
\label{sec:implementation}

A straightforward implementation of the protocol we presented in \cref{section:system} would not achieve the throughput and latency we observe in \cref{sec:evaluation}.
In this section, we discuss some of the techniques and optimizations required on the way to practically achieving \system's full potential.
(Many optimizations are however left out due to space constraints).

\paragraph{Code.}

\system is implemented in Rust, totaling 8,900 lines of code.
The main libraries \system depends on are:
\texttt{tokio} 1.12 for an asynchronous, event-based runtime;
\texttt{rayon} 1.5 for worker-based parallel computation;
\texttt{serde} 1.0 for serialization and deserialization;
\texttt{blake3} 1.0 for cryptographic hashes;
\texttt{ed25519-dalek} 1.2 for EdDSA signatures on Curve25519~\cite{rfc8032-ed25519}; 
\texttt{blst} 0.3.5 for multi-signatures on the BLS12-381 curve~\cite{rfc-bls-wip2020-09}.
\system also depends on in-house libraries:
\texttt{talk} (9,800 lines of code) for basic distributed computing and high-level networking and cryptography;
\texttt{zebra} (7,100 lines of code) for Merkle-tree based data structures.

\subsection{Broker}
\label{sec:implementation-broker}

The goal of a \system broker is to produce batches as \batchified as possible (to minimize server load), as large as possible (to amortize ordering), and as quickly as possible (to minimize latency).
Our target is for a broker to assemble one \fullbatch of 65,536 messages ($\sim$ 736~KB, see \cref{figure:sizeofabatch}) per second, with a 1 second \batchification timeout.

\paragraph{Reliable UDP.}

Short-lived TCP connections between broker and clients are easier to work with, but unfeasible for the broker to handle.
Assuming an end-to-end broadcast time of up to 10 seconds, the broker would need to maintain upwards of 600,000 simultaneous TCP connections, which preliminary tests immediately proved unfeasible on the hardware we have access to.
This makes UDP the only option for client-broker communication.
However, UDP lacks the reliability properties of TCP, and tests showed non-negligible packet loss even within the same AWS EC2 availability zone.
As we discussed in \cref{subsection:batchification}, message loss immediately translates to partial \batchification.
We address this issue by means of an in-house, ACK-based, message retransmission protocol based on UDP that also smoothens the rate of outgoing packets.

\paragraph{EdDSA batch verification.}

To avoid spoofing, all client messages are authenticated with signatures. 
At the target rate, however, individually verifying each signature is unfeasible for a broker.
Luckily, \texttt{ed25519-dalek} allows for more efficient batched verification.
A broker buffers the client messages it receives and authenticates them in batches.

\paragraph{Tree-search invalid multi-signatures.}

Clients contributing to the same batch produce matching multi-signatures for the batch's root.
At the target rate the broker cannot independently verify each multi-signature.
We tackle this problem by gathering multiple matching multi-signatures on the leaves of a binary tree: internal nodes aggregate their children.
For each tree, the broker verifies the root multi-signature, recurring only on the children of an invalid parent.
This allows to identify invalid multi-signatures in logarithmic time while enabling batched verification in the good case.

\paragraph{Caching legitimacy proofs.}

Clients justify their sequence numbers with legitimacy proofs. 
Again, the broker cannot verify each proof in time. 
We address this problem by having the broker verify a legitimacy proof only if higher than the highest it previously observed.
As a result, a faulty client might get away with submitting an invalid legitimacy proof but, importantly, not an illegitimate sequence number.

\subsection{Server}

The goal of a \system server is to process \batches as quickly as possible without overflowing its memory.

\paragraph{Batch garbage collection.}

Servers update each other on which batches they delivered.
A server garbage-collects a batch, both messages and metadata, as soon as it is delivered by all other servers.
We underline that, even if a single server fails to deliver a batch, the others cannot garbage-collect it as the slow server might be correct.
This is an inherent limitation of \broadcast: agreement without synchrony can be ensured only in the infinite-memory model.

\paragraph{Identifier-sorted batching.}

No two messages from the same client must appear in the same batch.
To simplify processing, brokers sort the messages in a batch by client identifier.
Servers reject batches whose identifiers are not strictly increasing, thus verifying that all identifiers are distinct in constant size and in linear time.
Sorting messages by identifier also enables parallel deduplication: messages are split by identifier range, chunks are deduplicated independently.

\section{Evaluation}
\label{sec:evaluation}

We evaluate \system focusing on the following research questions (RQs):
What workload can \system sustain (\cref{sec:eval-rq1})?
What are the benefits of \system's \batchification (\cref{sec:eval-rq2})?
How does \system scale to different numbers of servers (\cref{sec:eval-rq3})?
How efficiently does \system use resources overall (\cref{sec:eval-rq4})?
How does \system perform under adverse conditions, such as server failures (\cref{sec:eval-rq5})?
What performance can applications achieve using \system (\cref{sec:eval-rq6})?

\subsection{Baselines}
\label{sec:eval-baselines}

We compare \system against four baselines:
\begin{itemize}[nosep]
    \item \hotstuff~\cite{hotstuff-podc19}: an \broadcast protocol designed for high-throughput (written in C++);
    \item \bftsmart~\cite{bft-smart-dsn14}: an \broadcast protocol, similar to PBFT~\cite{pbft-osdi99}, designed for low-latency (written in Java);
    \item \nwbs: the DAG-based \broadcast protocol \bs~\cite{bullshark-ccs22} with the state-of-the-art high-throughput mempool \nw~\cite{narwhal-tusk-eurosys22} (written in Rust);
    \item \nwbssig: akin to \nwbs but with \nw modified to authenticate messages, thus matching \system's guarantees.
\end{itemize}
We deploy \system with two distinct underlying \broadcast protocols (\cref{figure:system-protocol}): \hotstuff and \bftsmart.

\paragraph{\hotstuff and \bftsmart.}

Evaluating \hotstuff and \bftsmart allows us to assess the base performance of an \broadcast protocol and determine how much acceleration \system provides.
We evaluate \system on top of the same implementations of \hotstuff~\cite{code-hotstuff} and \bftsmart~\cite{code-bftsmart} we benchmark against.
These implementations are production-ready and do not use state-of-the-art mempool protocols, only some basic form of batching.
When evaluated stand-alone, each message in these systems includes 80~B of header composed of a client identifier (8~B), a sequence number (8~B), and a signature (64~B) verified by the servers.
Both systems use batches of 400 messages, \ie of 34.4~KB.

\paragraph{\nwbs.}

As a state-of-the-art mempool, \nw is a close point of comparison for \system.
Servers in \nw scale out following a primary-workers model: each server is paired with one or several workers into a server group.
Similarly to \system, \nw greatly accelerates its underlying \broadcast (here, \bullshark).
Unlike \system, however, \nw leaves the responsibility of authenticating and deduplicating messages to the application.

\paragraph{\nwbssig.}

For a better comparison, we also benchmark \nwbssig: \nwbs where messages are authenticated by \nw in a state-of-the-art way, \ie using batched, multi-core Ed25519 signature verification.
Each message includes an 80~B header as for \hs and \bftsmart.
As for \nwbs, the remaining parameters are the default ones, \eg 500~KB batches.

\begin{figure}[tb]
    \centering
    \includegraphics[width=\columnwidth]{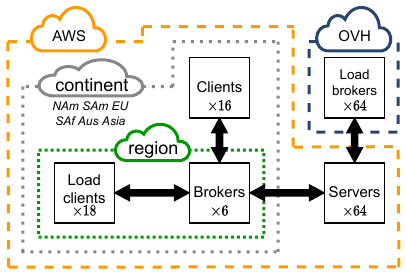}
    \caption{\textbf{Cross-cloud deployment summary.}
    }
    \label{fig:deployment}
\end{figure}

\subsection{Setup}
\label{sec:eval-setup}

Unless otherwise specified---in \cref{sec:eval-rq3,sec:eval-rq4}---the \system benchmarks involve 64 c6i.8xlarge AWS servers, of 32 Intel vCPUs each, geo-distributed across 14 regions.
Brokers assemble, and servers process, batches of 65,536 messages. 
Each message is 8~B in length, resulting in 736~KB batches (\cref{figure:sizeofabatch}).
Baselines always use the same set of server machines as their \system counterpart.
All experiments run with maximum resilience, \eg the system survives 21 faulty servers out of 64.
\cref{fig:deployment} overviews the used deployment.

\paragraph{Matching trusted and total resources.}

Unlike its baselines, \system leverages \emph{untrusted} resources, brokers, to boost its performance.
Lacking a well-defined conversion between trusted and untrusted resources, two extremes can be taken to compare \system with its baselines: we can either match trusted resources, \eg same number of \system servers as \nw workers, or match total resources, \eg same number of servers and brokers in \system as workers in \nw.

Intuitively, the first approach considers untrusted resources to be free while the second considers untrusted resources to be as costly as trusted resources.
We use the first approach in \cref{sec:eval-rq1,sec:eval-rq2,sec:eval-rq3,sec:eval-rq5,sec:eval-rq6} to stress \system, provisioning the system with enough brokers to bottleneck servers.
We use the second approach in \cref{sec:eval-rq4} to assess how efficiently \system uses its hardware resources, trusted or not.

\paragraph{Load clients and load brokers.}

We show in \cref{sec:eval-rq1} that \system servers handle up to 43.6 million operations per second with an average latency of 3.6 seconds.
To produce this level of workload, a real-world deployment would require over 700 brokers, each handling around 200,000 clients broadcasting back-to-back thus totaling hundreds of millions of machines.
As we cannot experiment at such a scale, we introduce two new actors: \emph{load clients} and \emph{load brokers}. 
(In the rest of this section, ``brokers'' and ``clients'' denote real brokers and real clients; the term ``load'' is always used explicitly.)

Load clients connect to brokers and simulate thousands of concurrent client requests.
Most system evaluation typically use this approach to stress the system and measure latency.
However, we explicitly separate clients from load clients in this evaluation.
Clients run on very small machines---less powerful than most smartphones---to provide more accurate end-to-end latency measurements.
We similarly split clients from load clients in all baseline runs.

Load brokers are unique to \system.
Even using load clients, we could not deploy enough brokers to bottleneck \system's servers.
Load brokers work around this limitation, submitting batches of pre-generated messages directly to the servers.
Free from interactions with clients and expensive cryptography, a load broker puts on the servers a load equivalent to that of tens of brokers working at full capacity.

Using load clients and load brokers, we manage to show that brokers can quickly generate large batches of messages, and servers can process large numbers of batches.

\paragraph{Cross-cloud deployment.}

All servers are deployed on AWS, balanced across 14 regions: Cape Town, São Paulo, Bahrain, Canada, Frankfurt, Northern Virginia, Northern California, Stockholm, Ohio, Milan, Oregon, Ireland, London, and Paris. 
For system sizes of 8 in \cref{sec:eval-rq3}, we distribute servers across the first 8 regions from the list, which constitute the most adversarial setup with the highest pairwise latency.

Load brokers are placed in a separate cloud provider, OVH, for two purposes.
First, it provides a better representation of Internet load than a single-cloud deployment.
AWS operates under its own AS so any AS peering bottlenecks would be bypassed by an AWS-only deployment.
Second, OVH is one of the few cloud providers with enough peering with AWS to stress \system without charging for egress bandwidth, saving us from using AWS' costly bandwidth.
The final cost amounted to 25,000 USD in AWS credits.
Using OVH saved us more than 70,000 USD since each of \system's data point on a figure would have cost 1,700 USD in AWS egress bandwidth---21~TB at 0.08~USD per GB $\approx$ 1,700 USD.

For all experiments, we deploy one broker in each continent (Cape Town, São Paulo, Tokyo, Sydney, Frankfurt, and Northern Virginia) and one client in each of the 14 regions above, plus Tokyo and Sydney.
Clients connect to their nearest broker.
We configure the network for geo-distribution and high load, \eg TCP buffer sizes~\cite{tcp-tuning} and UDP parameters.

All baselines run on the same parameters.
For \nwbs, we collocate each server with one of the workers in its server group.
We reproduced \nwbs's original experiments~\cite{bullshark-ccs22} and matched the results.

\paragraph{Hardware.}

All servers, brokers and load clients run on c6i.8xlarge machines with an Intel Xeon Platinum 8375C (32 virtual CPUs, 16 physical cores, 2.9~GHz baseline, 3.5~GHz turbo), 64~GB of memory and 12.5~Gb/s of bandwidth.
We selected these machines since they provide good performance and are in the same ``commodity'' price range as those chosen initially for \system's main baseline: \nwbs.
Clients run on t3.small machines: 2 vCPUs, 1 physical core, 2~GB of memory, and up to 5~Gb/s bandwidth---of which they use less than 1~KB/s.
All machines run Linux Ubuntu 20.04 LTS on the AWS patched version of the Linux kernel 5.15.0, except for the load brokers on OVH which run on Linux kernel 5.4.0---the same kernel was not available.

\paragraph{Challenges.}

The most significant evaluation challenges arose from the scale of the targeted deployment.
The setup and orchestration alone required simultaneous handling of up to 
320 machines across two different cloud providers and 25 regions, as well as transferring 13~TB of files---mostly public keys and pre-generated batches---for each of the 12 setups.
To handle this, we developed a new command-line tool to efficiently deploy distributed systems: \texttt{silk}. 
Among other things, we use \texttt{silk} for peer-to-peer-style file transfer over aggregated TCP connections, as well as for grouped process control.
With \texttt{silk}, transferring all files from a single machine takes around 30 minutes, compared to 68 hours with \texttt{scp}.
The code for \texttt{silk} is available online (see \cref{appendix:artifact}).

Additional challenges came from the real-world nature of the targeted deployment.
First, the connection between OVH and AWS's Asia and Pacific regions was particularly unstable at certain times of day especially when close to saturation. 
For example, Tokyo's connection was frequently degraded between 3pm and 5pm UTC.
Second, the performance of some machines sometimes deviated from their specifications.
As an example, in a setup size of 64, we observed around 2 machines operating with a 10\% lower CPU turbo clock rate than specified.
Considering these variations, we increased the number of servers a broker initially asks for witness shards (see \cref{subsection:submission}) by a margin, \eg $f + 5$ instead of $f + 1$.
This improves system stability---\ie lower latency variability---while slightly reducing maximum throughput.
Unless otherwise specified, we set the margin to 4 in all experiments, \ie $f + 5$.

\paragraph{Plots.}

Every data point is the mean of 5 runs of 2 minutes each (after excluding warmup and cooldown, the relevant cross-section is at least 1 minute).
All plots further depict one standard deviation from the mean using either colored shaded areas or black error bars (which may be too small to notice).
Experimental data can be found online (see \cref{appendix:artifact}).

\begin{figure*}[tbh]
    \centering
    \includegraphics[width=\textwidth]{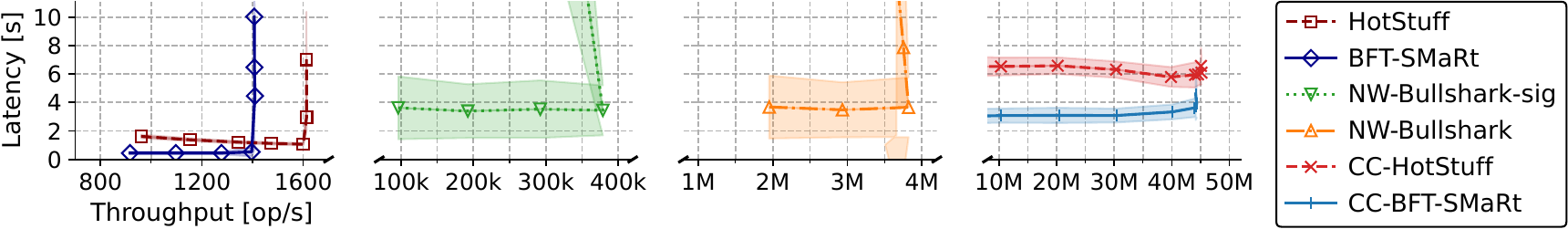}
    \vspace{\captionspace}
    \caption{\textbf{Throughput-latency of \system and of notable \broadcast systems under various input rates.}
    }
    \label{fig:eval-load-handling}
\end{figure*}

\subsection{RQ1 -- Load Handling}
\label{sec:eval-rq1}

\Cref{fig:eval-load-handling} shows the latency and throughput of \system and all its baselines for various input rates of 8~B messages.
The variability is represented using shaded areas.

\paragraph{Baselines.}

Both \bftsmart and \hotstuff showcase stable performances under low loads, respectively achieving around 1,400 and 1,600 operations per second.
\bftsmart's latency is consistently better than \hotstuff's up to its inflection point (0.45--0.53~s vs. 1.2--1.6~s).
We measure up to 3.8M~op/s for \nwbs and up to 382k~op/s for \nwbssig.
The difference in respective throughputs highlights the cost of authentication for servers: verifying signatures reduces the throughput of \nwbs by one order of magnitude.
We observe a latency of around 3.6~s for both \nwbs and \nwbssig.

\paragraph{\system.}

\system achieves close to 44M~op/s while running on top of both \hotstuff and \bftsmart.
\system's latency range is 3.0--3.6~s with \bftsmart and 5.8--6.5~s with \hotstuff.
Notably, the latency of \cchs decreases under high load.
This is due to the internal batching mechanism of the \hotstuff implementation: buffers fill faster under higher load, thus avoiding timeouts.
This has an immediate impact on \system, which feeds \hotstuff at a low rate: \hotstuff alone accounts for over 60\% of \cchs's overall latency.
\bftsmart makes a better fit for \system, as its throughput is sufficient for \system's needs, and its latency is lower than \hotstuff's.

\paragraph{Mempools' trade-off.}

In comparison to \bftsmart and \hotstuff, \system trades latency in favor of throughput.
This trade-off is mostly explained by batching and distillation.
When assembling a batch, a broker has to wait twice: once to collect enough messages to fill a batch, and once to collect all multi-signatures from clients engaging in distillation.
We set both waits' timeout to 1~second.
Notably, \nwbs seems to incur a similar latency cost, as \system's latency approximately matches that of \nwbs, even though \system needs an extra round trip between clients and broker (\cref{figure:system-protocol}, \#4--\#6).

\begin{figure}[tb]
    \centering
    \includegraphics[width=\columnwidth]{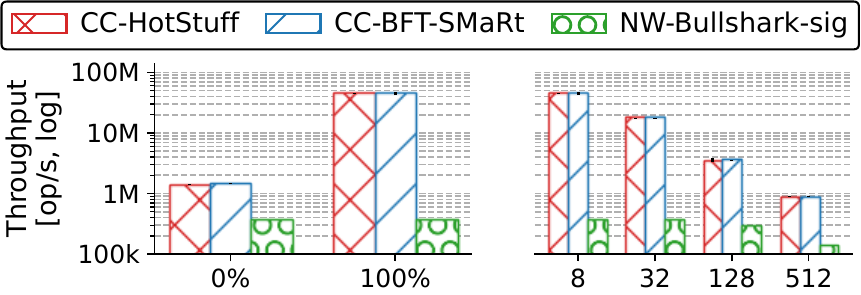}
    \begin{tabular}{@{\hspace{18mm}}c@{\hspace{17mm}}c@{\hspace{18mm}}}
        \begin{minipage}{.25\columnwidth}
            \subfloat[Distillation ratio]{
                \mbox{\hspace{\linewidth}}
                \label{fig:eval-distillation}
            }
        \end{minipage}
        &
        \begin{minipage}{.25\columnwidth}
            \subfloat[Message sizes {[B]}]{
                \mbox{\hspace{\linewidth}}
                \label{fig:eval-payloads}
            }
        \end{minipage}
    \end{tabular}
    \vspace{-10pt}
    \caption{\textbf{Throughput of \system and authenticated \nw with \bs (log scale) when (a) \system has no \batchification and with (b) varying message size.}
    }
    \label{fig:eval-merged-distillation-payloads}
\end{figure}

\subsection{RQ2 -- Distillation Benefits}
\label{sec:eval-rq2}

We showcase the benefits of \batchification by: evaluating throughput with and without \batchification, evaluating \batchification for messages of different sizes, and observing the impact of \batchification on network bandwidth to achieve line rate.

\paragraph{\Batchification vs. mitigations.}

Along with \batchification, \system makes use of two techniques available in the literature to mitigate the cost of \broadcast's authentication: short identifiers and pooled signature verification (see \cref{subsection:twostepsforward}).

\cref{fig:eval-distillation} breaks down \system's throughput, measuring how significantly \batchification alone contributes to \system's performance.
When no message is \batchified, \system's servers bottleneck at 1.5M~op/s, 3.9$\times$ higher than \nwbssig.
This result is in line with both systems bottlenecking on server CPU, as the technique employed by \system to mitigate authentication complexity has only one third of the servers verify each client signature. 
(We conjecture that the additional 1.3 factor may be owed to engineering differences.)
When batches are fully \batchified, \system's throughput grows to 44M~op/s, accounting for the additional 29-fold boost to \system's performance.

\paragraph{\Batchification for larger messages.}

\Cref{fig:eval-payloads} illustrates \system's maximum throughput for message sizes of 8~B to 512~B which may be relevant to applications that cannot work around smaller message sizes, \eg many smart contracts.
\system's throughput is similar with \bftsmart and \hs, decreasing at an approximately 1-to-1 ratio as the message size increases: 44.3M~op/s for 8~B, 17.6M~op/s for 32~B, 3.5M~op/s for 128~B and 890k~op/s for 512~B.

This is in line with expectations.
As we discuss in \cref{subsection:thepowerofbatchification}, a server should receive $\sim b$~bytes in order to deliver a $b$-bytes message in a large, \fullbatchified batch, as \fullbatchification amortizes to zero the communication cost of authenticating and sequencing each message.
For 8~B messages, servers encounter a CPU bottleneck slightly before the link between load brokers and servers is saturated.
This explains why the throughput decreases only 2.52$\times$ when messages grow to 32~B: all remaining server-bound bandwidth is used to convey messages (as messages are larger) while the load on server CPUs is reduced (as less messages are delivered overall).
The system remains communication-bottlenecked as the size of the messages increases, and throughput starts decreasing linearly with message size, \eg \system's throughput for 512~B messages is 4.00$\times$ smaller than for 128~B.

By contrast, \nwbssig bottlenecks on server CPUs longer, due to signature verification, maintaining a stable throughput until 512~B messages finally fill server links.
Overall, \nwbssig's throughput only decreases from 382k~op/s for 8~B messages to 142k~op/s for 512~B messages, which matches their non-authenticated evaluation with 512~B messages.
The gap between \system and \nwbssig at 512~B messages can be mostly attributed to \system's more efficient use of server bandwidth: unlike \narwhal, \system offloads the dissemination of batches to external brokers.
\narwhal's use of worker-to-worker communication in its common path also makes it more prone to be affected by AWS's various upload limitations, \eg AWS upload bandwidth is half the stated download bandwidth, and there are network credit limits for ``burst'' uploading.

\begin{figure}[tb]
    \centering
    \includegraphics[width=\columnwidth]{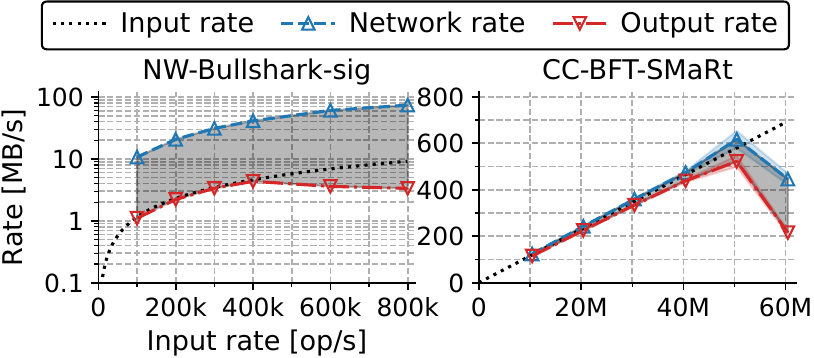}
    \vspace{\captionspace}
    \caption{\textbf{Throughput efficiency of authenticated \narwhal with \bullshark (left, log scale) and \system with \bftsmart (right, linear scale) with various input rates.}
    }
    \label{fig:eval-input-goodput}
\end{figure}

\paragraph{Line rate.}

\Cref{fig:eval-input-goodput} illustrates \system's near line-rate network use by depicting its input, network and output rates:
\begin{itemize}[nosep]
\item Input rate measures the total bytes of useful information---\ie client identifiers and messages---that clients, load clients and load brokers all broadcast per time unit;
\item Network rate measures the ingress bandwidth of servers at their network interface, \ie useful information captured by the input rate as well as the \broadcast's overhead for ordering, authentication and deduplication;
\item Output rate, or ``goodput'', measures the total bytes of useful information that each server delivers per time unit.
\end{itemize}
A system with perfect line rate would match all three rates: input rate would match output rate as messages can be delivered in a timely fashion with no backlogging, and output rate would match network rate as a server would only receive useful information, with no overhead due to \broadcast.
The gray-shaded areas in \cref{fig:eval-input-goodput} highlight this overhead, \ie the difference between network and input rates.
Network and output rates are averaged over all servers.

In this experiment, each of the 257M simulated clients broadcast 8~B messages.
This results in 11.5~B of useful information per broadcast as 28~bits~=~3.5~B are sufficient to represent every identifier.
This conversion is captured by the dotted line which converts the input rate from op/s, represented on the x-axis, to B/s, represented on the y-axis.

For authenticated \nwbs, the output rate closely matches the input rate until signature verification becomes the bottleneck at 378k~op/s, shown by the plateauing output rate.
The gap between \nwbssig's network and input rates is evident, differing by one order of magnitude (notably in line with our back-of-the-envelope calculation in \cref{subsection:thepowerofbatchification}).
In contrast, thanks to \batchification, \system practically achieves line-rate up to its maximum throughput.
Before its inflection point at 40M~op/s, the overhead of \system is less than 8\%.
The drop in output and network rates at 60M~op/s is due to servers surpassing their computational capacity: broadcasts stall, server witness verification gets backlogged and brokers, suspecting server faults, ask for more batch witnesses, further stressing servers' CPUs.

\subsection{RQ3 -- Number of Servers}
\label{sec:eval-rq3}

\Cref{fig:eval-system-sizes} illustrates the maximum throughput for systems of 8 ($f=2$), 16 ($f=5$), 32 ($f=10$) and 64 ($f=21$) servers.
For \system, we adjust the witnessing margin as the system grows by 0, 1, 2, and 4 for 8, 16, 32 and 64 servers respectively (see \cref{sec:eval-setup}).
Both \system and authenticated \nwbs scale well to 64 servers.
Note that, unless trust assumptions are modified, \nwbssig only scales vertically: if a \narwhal server or any of its workers are faulty, the entire server group is compromised.
\system, instead, scales horizontally with the number of brokers.

\subsection{RQ4 -- Overall Efficiency}
\label{sec:eval-rq4}

\begin{figure}[t]
    \centering
    \includegraphics[width=\columnwidth]{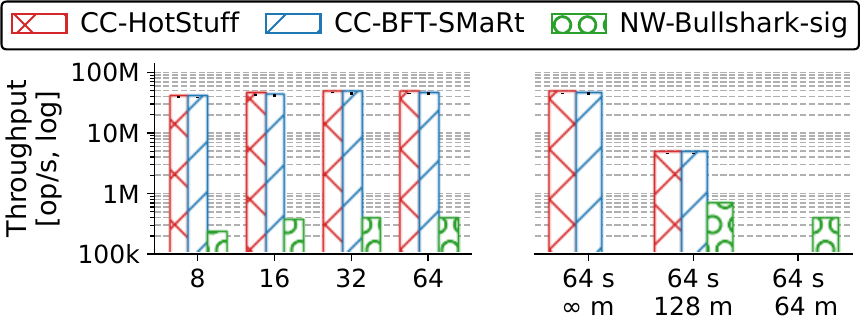}
    \begin{tabular}{@{\hspace{18mm}}c@{\hspace{13mm}}c@{\hspace{18mm}}}
        \begin{minipage}{.25\columnwidth}
            \subfloat[System sizes]{
                \mbox{\hspace{\linewidth}}
                \label{fig:eval-system-sizes}
            }
        \end{minipage}
        &
        \begin{minipage}{.35\columnwidth}
            \subfloat[Matching resources]{
                \mbox{\hspace{\linewidth}}
                \label{fig:eval-matching-trusted-and-untrusted}
            }
        \end{minipage}
    \end{tabular}
    \vspace{-10pt}
    \caption{\textbf{Throughput of \system and authenticated \nw with \bs (log scale) when (a) varying system size, and when (b) varying the number of overall machines (``m'') with 64 servers (``s'').}
    Load brokers in \system simulate tens of brokers, hence are noted ``$\infty$ m''.}
    \label{fig:eval-merged-system-sizes-matching-trusted-resources}
\end{figure}

The center cluster of bars in \cref{fig:eval-matching-trusted-and-untrusted} compares \system's throughput with that of authenticated \nwbs when overall hardware resources are matched.
In this setting, both systems have 128 machines at their disposal.
\system is provided with 64 servers, 64 brokers and 0 load brokers.
Since a load broker uses pre-generated synthetic data to simulate tens of brokers (see \cref{sec:eval-setup}), involving load brokers in this experiment would give an unfair advantage to \system.
\nwbssig is provided with 128 workers, to match \system's total machines, balanced across 64 server groups, to match \system's servers.
The left and right clusters of bars depict \system using load brokers and \nwbssig with 64 server groups containing 1 worker each, respectively, as in the other experiments.

We observe 4.6M~op/s for \system, with servers reporting around 5\% CPU usage.
We observe 679k~op/s for \nwbssig.
\system's higher throughput is in line with expectations.
In \nwbssig, workers are trusted, and as such a worker can only contribute to its own server group.
Instead, since \system brokers are untrusted, a broker's work is useful to all servers.

\begin{figure}[tb]
    \centering
    \includegraphics[width=\columnwidth]{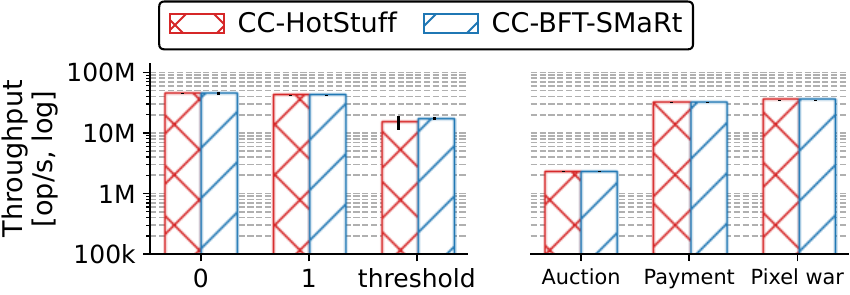}
    \begin{tabular}{@{\hspace{18mm}}c@{\hspace{13mm}}c@{\hspace{18mm}}}
        \begin{minipage}{.25\columnwidth}
            \subfloat[Server failures]{
                \mbox{\hspace{\linewidth}}
                \label{fig:eval-server-failures}
            }
        \end{minipage}
        &
        \begin{minipage}{.35\columnwidth}
            \subfloat[Applications]{
                \mbox{\hspace{\linewidth}}
                \label{fig:eval-applications}
            }
        \end{minipage}
    \end{tabular}
    \vspace{-10pt}
    \caption{
    \textbf{Throughput of \system (log scale) with (a) various server failures and for (b) different applications.}
    }
    \label{fig:eval-merged-server-faults-applications*}
\end{figure}

\subsection{RQ5 -- Chop Chop Under Failures}
\label{sec:eval-rq5}

\Cref{fig:eval-server-failures} depicts \system's throughput when some servers crash 30 seconds into the run.
Performance drops marginally (from 44M~op/s to 43M~op/s) with one crash and by 66\% (down to 15M~op/s) when one-third of the servers crash, resulting in less CPU globally available to witness batches.

\Cref{fig:eval-distillation} captures \system's performance hit when clients fail to engage in \batchification. 
This could be caused by clients being slow or crashed, or brokers being malicious.
Under the most extreme conditions, where no client engages in \batchification, the throughput drops from 44M~op/s to 1.5M~op/s.

\subsection{RQ6 -- Application Use Cases}
\label{sec:eval-rq6}

\Cref{fig:eval-applications} depicts the maximal stable throughput for various application use cases.
In the Auction app, a client can bid an amount on a token it does not own, or take the highest offer it received for an item it owns.
The highest amount bid on each token is locked and cannot be used to bid elsewhere.
Money bid is transferred when the owner of the token takes the offer, or refunded when the bid is raised by another client.
The Auction app is single-threaded and many clients bid on the same token to approximate a real auction.
In the Payments app, clients choose a recipient and an amount to transfer.
In Pixel war, clients choose a pixel and an RGB color to paint on a 2,048 by 2,048 board.
Operations are generated at random.

We observe 2.3M~op/s for the Auction, 32M~op/s for Payments and 35M~op/s for Pixel war.
The bottleneck is the application in all cases, thus \system has sufficient capacity for high, single-application throughput.
\system can also support many separate high-throughput applications simultaneously, making it a fitting \broadcast candidate to power a universal \ac{smr} system, \ie an Internet computer.

\section{Related Work}
\label{section:relatedwork}

We overview below the state-of-the-art most relevant to \system, namely \broadcast systems with high-throughput and efficient signature aggregation schemes.

\paragraph{High-throughput \broadcast.}

Narwhal~\cite{narwhal-tusk-eurosys22} is a mempool protocol that separates the reliable distribution of payloads from the communication-expensive ordering in order to accelerate DAG-based \broadcast~\cite{aleph-aft19,dagrider-podc21,bullshark-ccs22}.
Narwhal utilizes trusted workers to increase throughput while \system relies on \emph{trustless} brokers, for the same effect, and scales out more efficiently.
To circumvent the bottleneck associated with the broadcast leader, approaches using multiple leaders have been developed---both for crash~\cite{epaxos-sosp13,tempo-eurosys21} and arbitrary~\cite{fnf-bft-sirocco23,bigbft-ipccc21,mirbft-jsys22,iss-eurosys22} faults---to scale the broadcast throughput linearly with the number of leaders.
Dissemination trees~\cite{byzcoin-sec16,kauri-sosp21} have also been employed to reduce communication cost and maximize network bandwidth utility, while sharded~\cite{omniledger-sp18,monoxide-nsdi19} and federated~\cite{stellar-consensus-sosp19} approaches reduce communication cost by promoting local communication in geo-distributed setups.
In comparison, \system shows that an optimal distillation mechanism for batches achieves better performances without adding complexity to the \broadcast protocol itself.

Other approaches have shown that the underlying hardware of servers can also be exploited for higher throughput, such as FPGA~\cite{cheapbft-eurosys12,consensus-box-nsdi16} and Intel SGX enclaves~\cite{hybster-eurosys17}.
In comparison, \system uniquely boosts throughput by exploiting \emph{trustless hardware} via brokers.
\broadcast can also be accelerated in data centers by using the topology of the network~\cite{momcast-specpaxos-nsdi15,uomcast-nopaxos-osdi16} or even by running within the network itself using P4-programmable switches~\cite{p4xos-ton20,hovercraft-eurosys20}.
In such low latency environments, the processing overhead incurred by the operating system kernel can be bypassed to further increase the throughput of \broadcast~\cite{apus-paxos-rdma-socc17,hovercraft-eurosys20,ubft-asplos23}.

\paragraph{Signature aggregation.}

Aggregate signatures were first proposed to save space by compacting a large number of signatures into just one~\cite{schnorr-jcrypt91,agg-bls-sig-eurocrypt03}.
Up until recently, aggregation could also save verification time but only in certain cases: either when the signatures are generated by the same signer~\cite[\secref~5.1]{camenisch-batch-jcrypt12}, or when the signatures are on the same message, \ie multi-signatures~\cite{multisig-report83}.
In the latter case, aggregation mechanisms have been proposed to achieve constant-time verification of aggregated multi-signatures for both BLS~\cite{bls-multisig-asiacrypt18} and Schnorr~\cite{musig-schnorr-dcc19} signature schemes.
In particular, multi-signatures are used in cryptocurrencies to have many servers sign the same batch of payloads~\cite{byzcoin-sec16,pixel-sec20}.
Servers in \system use rapidly-verifiable BLS multi-signatures~\cite{bls-multisig-asiacrypt18} for that very purpose.
In addition to aggregating server signatures on batches, \system's distillation mechanism also aggregates all client signatures in a batch in a way that provides constant-time verification.
The theoretical scheme Draft~\cite{oracular-brb-disc22} proposed signature aggregation with similar verification performances but is tailored to Reliable Broadcast.
It is however unclear how Draft could be implemented as a real-world system without compromising liveness.
Indeed, Draft assumes infinite memory to prevent message replay attacks which would rapidly exhaust servers' memory if deployed to match \system's maximum 
throughput (see \cref{sec:eval-setup}).
\system also aggregates client sequence numbers to significantly reduce bandwidth consumption when small messages are broadcast (\cref{figure:fulldistillationinaction}).
\system aggregates sequence numbers thanks to the ordering of and thanks to novel legitimacy proofs (see \cref{subsection:batchification}).

\section{Concluding Remarks}
\label{section:conclusions}

\system's performance comes with two limitations.
First, \system's high throughput makes memory management a challenge: servers fill their memory quickly if unable to garbage-collect under heavy load.
Second, all servers in \system are known at startup and it is unclear if its performance would be maintained when deployed on thousands of servers.
Interesting avenues of future research include sharding to achieve even higher throughput by running multiple, independent, coordinated instances of \system, 
and offloading more tasks to the brokers, such as public key aggregation.

\section*{Acknowledgments}

We thank the OSDI '23 reviewers for their continuous involvement in the revision process.
We further thank Vasileios Trigonakis for his early feedback.
This work has been supported in part by AWS Cloud Credit for Research, the Hasler Foundation (\#21084), and Innosuisse (\#46752.1 IP-ICT).

\clearpage 
\bibliographystyle{plain}
\bibliography{bibliography.bib}

\clearpage
\tableofcontents

\vfill
\appendix
\section{Artifact Appendix}
\label{appendix:artifact}

\subsection*{Abstract}

This artifact is a full implementation of \system, an end-to-end client-server Byzantine \broadcast system leveraging a third-party---brokers---to optimize network usage and vastly reduce the computational load of authenticating client requests on the server-side while preserving safety.

\subsection*{Scope}

This artifact can be used to validate the following claims (\emph{provided the setup is the same}):
\begin{itemize}[nosep]
    \item End-to-end performance (\Cref{fig:eval-load-handling,fig:eval-merged-system-sizes-matching-trusted-resources}).
    \item Distillation benefits (\Cref{fig:eval-merged-distillation-payloads}) and throughput efficiency (\Cref{fig:eval-input-goodput}).
    \item Performance under faults (\Cref{fig:eval-merged-server-faults-applications*}).
\end{itemize}

Note that reproducing the same plots (including error bands) can be prohibitively costly given the scale of the evaluation (number and hourly price of machines, see \cref{sec:eval-setup}).

\subsection*{Contents}

The artifact contains all the source code used to implement \system.
We provide a Dockerfile to set up experiments.
We also provide automated scripts to extract and interpret the data as well as generate plots.
Please refer to the README.md file in the repository for more details.

\subsection*{Hosting}

The repository can be accessed through GitHub:
\url{https://github.com/Distributed-EPFL/chop-chop-osdi24}.

\subsection*{Requirements}

There are no special hardware or software requirements beyond a recent enough version of Rust (\Cref{sec:implementation}) and the desired network layout to evaluate.
If you wish to reproduce the reported results exactly, please see \Cref{sec:eval-setup} for the setup used.

\clearpage
\onecolumn
\section{Chop Chop}
\label{appendix:chop}

This section contains the formal definition (\cref{subsection:chop.interface}), full pseudocode (\cref{subsection:chop.pseudocode}) and full correctness proofs (\cref{subsection:chop.correctness}) of \system.
In the following, \system is presented as a \cstob (\cstobprefix) primitive, for completeness, while it is presented as an \broadcast in the main body of the paper, for simplicity.
The agreement property of \broadcast described in \cref{section:broadcast} is further split in two properties for \cstobprefix: agreement and total order.
The underlying \broadcast of \system---the \broadcast box in \cref{figure:system-protocol}---is referred to as a \stob (\stobprefix) primitive in the following since it only involves servers.
To complete the formalization of \system, \cref{appendix:diral} presents in detail the \dir primitive \diral used in \system to assign identifiers to clients in a dense manner, as briefly mentioned in \cref{subsection:twostepsforward}.

\subsection{TOB Interfaces, Properties and Rules}
\label{subsection:chop.interface}

\paragraph{\cstobprefix interface.}
A \cstob (\cstobprefix) system offers two interfaces, \cl (instance $\clin$) and \sr (instance $\srin$), exposing the following events:
\begin{itemize}
    \item \textbf{Request} $\event{\clin}{Broadcast}[m]$: Broadcasts the message $m$ to all servers.
    \item \textbf{Indication} $\event{\srin}{Deliver}[\client, m]$: Delivers the message $m$ broadcast by client $\client$.
\end{itemize}

\paragraph{\cstobprefix properties.}
A \cstobprefix system satisfies the following properties:
\begin{itemize}
    \item \textbf{No duplication}: No message broadcast by a correct client is delivered more than once by a correct server.
    \item \textbf{Integrity}: If a correct server delivers a message $m$ from a correct client $\client$, then $\client$ previously broadcast $m$.
    \item\textbf{Validity}: If a correct client $\client$ broadcasts a message $m$, then eventually a correct server delivers $m$ from $\client$.
    \item\textbf{Agreement}: If a correct server delivers a message $m$ from a client $\client$, then eventually every correct server delivers $m$ from $\client$.
    \item\textbf{Total order}: Let $m$ and $m'$ be any two different messages. If a correct server delivers $m$ without having delivered $m'$, then no correct server delivers $m'$ before $m$.
\end{itemize}

\paragraph{\stobprefix interface.}
A \stob (\stobprefix) system offers the \stobprefix Server (instance $\stobi$) interface exposing the following events:
\begin{itemize}
    \item \textbf{Request} $\event{\stobi}{Broadcast}[m]$: Broadcasts the message $m$ to all servers.
    \item \textbf{Indication} $\event{\stobi}{Deliver}[m]$: Delivers the message $m$.
\end{itemize}

\paragraph{\stobprefix properties.}
An \stobprefix system satisfies the following properties:
\begin{itemize}
    \item \textbf{No duplication}: No message sent by a correct server is delivered more than once by a correct server.
    \item \textbf{Integrity}: If a correct server delivers a message $m$, then some server has previously broadcast $m$.
    \item \textbf{Validity}: If a correct server broadcasts a message $m$, then eventually a correct server delivers $m$.
    \item \textbf{Agreement}: If a correct server delivers a message $m$, then eventually every correct server delivers $m$.
    \item \textbf{Total order}: Let $m$ and $m'$ be any two different messages. If a correct server delivers $m$ without having delivered $m'$, then no correct server delivers $m'$ before $m$.
\end{itemize}

\paragraph{Client rules.}\label{rules:Rule}
A correct client $\client$ satisfies the following rules:
\begin{enumerate}
[label=\textbf{CR\arabic*}:,ref=CR\arabic*,itemindent=7mm]
    \item \label{rule:client:waitcompletion}
    $\client$ can trigger $\event{\clin}{Broadcast}[m']$ only once it receives the completion certificate for its previously broadcast $m$.
    \item \label{rule:client:noduplicate}
    $\client$ never triggers $\event{\clin}{Broadcast}[m]$ twice in a row with the same message $m$.
\end{enumerate}

\paragraph{Broker rules.}
A correct broker $\broker$ satisfies the following rule:
\begin{enumerate}
[label=\textbf{BR\arabic*}:,ref=BR\arabic*,itemindent=7mm]
    \item \label{rule:broker:waitdelivery}
    $\broker$ can send a message $m'$ only once its previously sent message $m$ has been delivered by \stobprefix.
\end{enumerate}

\clearpage
\subsection{Pseudocode}
\label{subsection:chop.pseudocode}

This section contains the complete code of our \cstobprefix implementation \system.
\system's client is implemented in \cref{subsection:pseudocode.csbalclient}, \system's broker in \cref{subsection:pseudocode.csbalbroker}, and \system's server in \cref{subsection:pseudocode.csbalbroker}.

\subsubsection{Chop Chop Client}
\label{subsection:pseudocode.csbalclient}
\begin{lstlisting}
implements:
    %\clab%, instance %\clin%


uses:
    %\dirab%, instance %\dirin%
    %AuthenticatedPointToPointLinks%, instance %al%


struct Evidence:
    proof: Multisignature,
    number: Integer,
    root: Root


upon <%\clin%.Init>:
    trigger <%\dirin%.Signup>; %\label{line:csignup}%
    await <%\dirin%.SignupComplete | id>; %\label{line:csignupcomplete}%
    submissions: {Message: [(Broker, {Integer})]} = {}; %\label{line:csubmissionsinit}%
    sequence_number = 0; %\label{line:cseqmunbinit}%
    evidence: Evidence = {};


upon <%\clin%.Broadcast | message>: %\label{line:cbrodcastmessage}%
    submissions[message] = {}; %\label{line:cupdatesumissions}%
    submit(message); %\label{line:csubmitinvoke}%


procedure submit(message): %\label{line:cproceduresubmit}%
    %\StartMultiline%if let submission = submissions[message] and let %$\broker$% in (%$\brokers$% \ submission):%\EndMultiline% %\label{line:csubmissioncheck}%
        assignment = dir.export(id); %\label{line:cdirexport}%
        signature = sign([Message, message, sequence_number]); %\label{line:csignaturecompute}%
        %\StartMultiline%trigger <al.Send | %$\broker$%, 
          [Submission, assignment, message, sequence_number, signature, evidence]>;%\EndMultiline% %\label{line:csendsubmit}%
        
        submission.add((%$\broker$%, sequence_number)); %\label{line:caddinitsequncenumbertosub}%
        trigger <timer.Set | [Submit, message], 13 + >; %\label{line:ctimerset}%


upon <timer.Ring | [Submit, message]>: %\label{line:ctimerring}%
    submit(message); %\label{line:cinvokesubmitring}%


upon <al.Deliver | %$\broker$%, [Inclusion, message, root, proof, max_sequence_number, evidence]>: 
    if let submission = submissions[message][%$\broker$%]: %\label{line:cexistsubmissionforinclusion}%
        if !verify(root, proof, (id, max_sequence_number, message)): return;

        result = check(sequence_number, max_sequence_number, evidence);
        sequence_number = max(submission[1]);

        if result not %$\bot$%:
            submission[1].add(result);%\label{line:cupdatesumbission2}%
            multisignature = multisign([Reduction, root]); %\label{line:cmultisignreduction}%
            trigger <al.Send | %$\broker$%, [Reduction, root, multisignature]>; %\label{line:csendreduction}%


procedure check(sequence_number, max_sequence_number, evidence):
    %\StartMultiline%if verify_plurality(evidence.proof, [Completion, evidence.root, evidence.number])
      and max_sequence_number <= evidence.number
      and max_sequence_number >= sequence_number: %\EndMultiline% %\label{line:cchcekcondition}%
        sequence_number = max_sequence_number; %\label{line:cseqnumupdate}%
        evidence = evidence;
        return sequence_number; %\label{line:creturnsequencenumber}%
    else:
        return %$\bot$%;


%\StartMultiline%upon <al.Deliver | %$\broker$%, [Completion, delivery_root, message, sequence_number,
 proof, delivery_number, certificate]>:%\EndMultiline% %\label{line:crecivecompletioncertifcate}%
    %\StartMultiline%if !verify_plurality(certificate, [Completion, delivery_root, delivery_number]): return;%\EndMultiline% %\label{line:cverifyplurality}%

    if evidence.number < delivery_number:
        evidence.number = delivery_number;
        evidence.proof = certificate;
        evidence.root = delivery_root;

    if let submission = submissions[message] and sequence_number in submission[%$\broker$%]:
        if verify(delivery_root, proof, (id, sequence_number, message)): %\label{line:cproverrdim}%
            sequence_number = sequence_number + 1;%\label{line:csequpadte}%
            submissions.remove(message); %\label{line:sremoveelementfromsumissions}%
\end{lstlisting}

\subsubsection{Chop Chop Broker}
\label{subsection:pseudocode.csbalbroker}
\begin{lstlisting}
implements:
    %\bkab%, instance %\bkin%


uses:
    %\dirab%, instance %\dirin%
    %AuthenticatedPointToPointLinks%, instance %al%


struct Evidence:
    proof: Multisignature,
    number: Integer,
    root: Root


struct Submission:
    sequence_number: Integer,
    message: Message,
    signature: Signature,
    evidence: Evidence


enum Batch:
    max_sequence_number: Integer,
    payloads: {Id: (Integer, Message)},
    signatures: {Id: Signature},
    reductions: {Id: MultiSignature},
    straggler: {id: Integer}

    variant Reducing:
        (empty)

    variant Witnessing:
        batch_number: Integer,
        witnesses: {Server: MultiSignature}

    variant Completing:
        completion: {(Root, Integer, {Id: Integer}): {MultiSignature}}


upon <%\bkin%.Init>:
    pending: {Id: [Submission]} (default []) = {}; %\label{line:bpandingdefinition}%
    submissions: {Id: Submission}= {}; %\label{line:binitsubmissions}%
    pool: {Id: Submission} = {}; %\label{line:bpoolinit}%
    collecting: Boolean = false; %\label{line:bcollectinginit}%
    batches: {Root: Batch} = {}; %\label{line:bemptybatches}%
    batch_number = 1; %\label{lemma:bbatchnumberdefine}%
    last_message: {Id: (Integer, Message)};


%\StartMultiline%upon <al.Deliver | %$\client$%, [Submission, assignment, message, sequence_number, signature, evidence)]>:%\EndMultiline% %\label{line:bdeliversubmit}%
    %\dirin%.import(assignment); %\label{line:bimportsubmit}%
    
    if %\StartMultiline%%$\client$%.verify(signature, [Message, message, sequence_number])
      and verify_plurality(evidence.proof, [Completion, evidence.root, evidence.number])
      and sequence_number <= evidence.number:%\EndMultiline%%\label{line:bverifyclientsignature}%
        let id = %\dirin%[%$\client$%];
        %\StartMultiline%pending[id].push_back(Submission {sequence_number, message, signature, evidence}); %\EndMultiline%%\label{line:bpushback}%

    if %\StartMultiline%%$\client$%.verify(signature, [Message, message, sequence_number]) and sequnce_number = 0: %\EndMultiline%%\label{line:bbypasseveidenceifsequiszero}%
        let id = %\dirin%[%$\client$%];
        pending[id].push_back(Submission {sequence_number, message, signature}); %\label{line:bupdatesumbissionifsequiszero}%


upon exists id in pending such that pending[id] != [] and id not in pool: %\label{line:bexistsinpanding}%
    submission = pending[id].pop_front(); %\label{line:bpandingupdate}%
    pool[id] = submission; %\label{line:bpoolupdate}%


upon pool != {} and collecting = false: %\label{line:bpoolcollectingcheck}%
    collecting = true; %\label{line:bcollectingtrue}%
    trigger <timer.Set | [Flush], 2>; %\label{line:btimerset}%


upon <timer.Ring | [Flush]>: %\label{line:btimerring}%
    collecting = false; %\label{line:bcollectingfalse}%
    
    submissions = pool; %\label{line:bsubmissionsequalpool}%
    pool = {}; %\label{line:bemptypool}%
    
    max_sequence_number = max(submissions.sequence_number); %\label{line:bcomputemaxseq}%
    max_evidence_number = max(submissions.evidence.number);
    max_evidence = submissions.evidence[max_evidence_number];
    
    %\StartMultiline%leaves = [(id, max_sequence_number, message)
      for (id, Submission {.. , message, .., ..}) in submissions];%\EndMultiline% %\label{line:bleavescostruction}%
      
    tree = merkle_tree(leaves); %\label{line:btreecostruction}%
    root = tree.root(); %\label{line:brootcostruction}%
    
    for (id, Submission {.. , message, .., ..}) in submissions:
        %$\client$% = %\dirin%[id];
        proof = tree.prove((id, max_sequence_number, message)); %\label{line:binclusionproof}%
        %\StartMultiline%trigger <al.Send | %$\client$%, [Inclusion, message, root, proof, max_sequence_number, max_evidence]>;%\EndMultiline%
        
    %\StartMultiline%payloads = {id: (max_sequence_number, message)
      for (id, Submission {.., message, .., ..}) in submissions};%\EndMultiline% %\label{line:bpayloaddefine}%
    %\StartMultiline%stragglers = {id: sequence_number
      for (id, Submission {sequence_number, .., .., ..}) in submissions};%\EndMultiline% %\label{line:bstragglersupdate}%
    %\StartMultiline%signatures = {id: signature
      for (id, Submission {.., .., signature, ..}) in submissions}; %\EndMultiline%%\label{line:bsignaturedefine}%
    
    %\StartMultiline%batches[root] = Reducing {max_sequence_number, payloads, signatures, reductions: {}, stragglers}; %\EndMultiline% %\label{line:bbatchesupdate}%    
    trigger <timer.Set | [Reduce, root], 2>; %\label{line:btimersetreducing}%


upon <al.Deliver | %$\client$%, [Reduction, root, multisignature]>: %\label{line:brecivereduction}%
    %\StartMultiline%if let batch alias batches[root] and batch is Reducing 
      and let id = %\dirin%[%$\client$%] and id in batch.payloads:%\EndMultiline% %\label{line:bcheckreducingid}%
        if %$\client$%.multiverify(multisignature, [Reduction, root]): %\label{line:bcheckmultisignereduction}%
            batch.reductions[id] = multisignature; %\label{line:bupdatereduction}%
            batch.signatures.remove(id); %\label{line:bupdateremoveid}%
            batch.stragglers.remove(id); %\label{line:bstragglersremoveid}%


upon <timer.Ring | [Reduce, root]>: %\label{line:breducering}%
    batch alias batches[root]; %\label{line:breducefetchbatch}%
    
    leaves = []; %\label{line:bleavesdefinition}%
    for all id in batch.payloads:
        message = batch.payloads[1]; 
        if id in batch.stragglers:
            leaves.add((id, stragglers[id], message)); %\label{line:baddsleavesfromstraggler}%
        else:
            leaves.add((id, payloads[id])); %\label{line:baddsleavesfrompayloads}%
            
    tree = merkle_tree(leaves); %\label{line:btreecostruction1}%
    new_root = tree.root(); %\label{line:brootcostruction1}%
    batches[new_root] = batches[root]; %\label{line:baddnewroottobatches}%
    batches.remove(root);

    batch alias batches[new_root];
    ids = batch.payloads.keys(); %\label{line:bcompressids}%
    payloads = batch.payloads.values() %\label{line:bpayloads}%

    for %$\server$% in %$\servers$%: %\label{line:bserverinserver}%
        %\StartMultiline%trigger <al.Send | [Batch, batch_number, ids, payloads, stragglers]>; %\EndMultiline%%\label{line:bsendbatch}%

    batch = Witnessing {batch_number, witnesses: {}}; %\label{line:bbatchinwitnessing}%
    batch_number = batch_number + 1; %\label{line:supdatebatchnumberaddingone}%


upon <al.Deliver | %$\server$%, [BatchAcquired, root, unknowns]>: %\label{line:bdeliverbatchacquired}%
    if let batch = batches[root] and batch is Witnessing: %\label{line:bbatchacquiredgetbatch}%
        if exists unknown in unknowns such that unknown not in %\dirin%: %\label{line:bbatchacquiredcheckids}%
            return;
        
        assignments = %\dirin%.export(unknowns..); %\label{line:bbatchacquiredexportassignments}%        
        multisignature = aggregate(batch.reductions.values()); %\label{line:bcomputethemultisignature}%
        signatures = batch.signatures; %\label{line:bcomputethesignature}%
        batch_number = batch.batch_number;
        %\StartMultiline%trigger <al.Send | %$\server$%, 
          [Signatures, root, batch_number, assignments, multisignature, signatures]>;%\EndMultiline% %\label{line:bsendingsignatureormultisignature}%


upon <al.Deliver | %$\server$%, [WitnessShard, root, shard]>: %\label{line:bdeliverwitnessshard}%
    if let batch alias batches[root] and batch is Witnessing: %\label{line:brootandcommitable}%
        batch_number = batch.batch_number;
        if %$\server$%.multiverify(shard, [Witness, root, self, batch_number]): %\label{line:bwitnessshardchecksignature}%
            batch.witnesses[%$\server$%] = shard; %\label{line:bstorewitnessshard}%
            
            
%\StartMultiline%upon exists root in batches such that batches[root] alias batch is Witnessing 
  and |batch.witnesses| >= f + 1 :%\EndMultiline% %\label{line:binwitnessing}%
    certificate = aggregate(batch.witnesses); %\label{line:baggregatewitnesses}%
    batch_number = batch.batch_number;
    
    for %$\server$% in %$\servers$%: %\label{line:bbroadcastwitnessesloop}%
        trigger <al.Send | [Witness, root, batch_number, certificate]>; %\label{line:bbroadcastwitnesses}%
    batch = Completing {completion: {}}; %\label{line:bbatchcompleting}%


%\StartMultiline%upon <al.Deliver | %$\server$%, [CompletionShard, root, completion_root,
  delivery_counter, shard, exceptions]>: %\EndMultiline%%\label{line:bsignedcompletioncertificate}%
    if let batch alias batches[root] and batch in Completing: %\label{line:bcheckbathccompletingreally}%
        if %$\server$%.multiverify(shard, [Completion, completion_root, deliver_counter]): %\label{line:bchecks}%
            completions[(root, delivery_counter, exceptions)].add(shard); %\label{line:baddingsignaturetocomplations}%


%\StartMultiline%upon exists root in batches such that batches[root] alias batch is Completing
  and exists (root, delivery_counter, exceptions) in batch.completions
  such that |batch.completions[(root, delivery_counter, exceptions)]| >= f + 1:%\EndMultiline% %\label{line:bbatchcompleting1}%
    %\StartMultiline%certificate = aggregate(batch.completions[(root, delivery_counter, exceptions)]; %\EndMultiline%%\label{line:screatecompletioncertificate}%
    
    delivery_payload: {Id: (Integer, Message)} = {};
    for all id in batch.ids
        if id not in exceptions:
            if id not in straggler:
                (sequence_number, message) = batch.payloads[id];
            else:
                sequence_number = batch.straggler[id];
                message = batch.payloads[id];
        else 
            (.., message) = batch.payloads[id];
            sequence_number = exceptions[id];
        new_payload[id] = (sequence_number, message);
            
    %\StartMultiline%delivery_leaves = [(id, sequence_number, message) for
     (id, delivery_payloads[id]) in delivery_payloads];%\EndMultiline%
    delivery_tree = merkle_tree(delivery_leaves);
    delivery_root = delivery_tree.root();
    
    for id in batch.ids such that exceptions[id] not %$\bot$%:
        %$\client$% = %\dirin%[id];
        proof = tree.prove((id, delivery_payload[id]));
        sequence_number = delivery_payload[id][0];
        message = delivery_payload[id][1];
        
        %\StartMultiline%trigger <al.Send | %$\client$%, [Completion, delivery_root, message, 
         sequence_number, proof, delivery_counter, certificate]>;%\EndMultiline%%\label{line:bsendbackcompletioncertificate}%

    batches.remove(root); %\label{line:bremoverootfrombatches}%
\end{lstlisting}

\subsubsection{Chop Chop Server}
\label{subsection:pseudocode.csbalserver}
\begin{lstlisting}
implements:
    %\srab%, instance %\srin%


uses:
    %\dirab%, instance %\dirin%
    %AuthenticatedPointToPointLinks%, instance %al%
    %\stobprefix%Server, instance %\stobi%


struct Batch:
    ids: [Id],
    payloads: [(Integer, Message)],
    stragglers: {Id: Integer},
    delivered: Boolean


upon event <%\srin%.Init>:
    batches: {(Root, Broker, Integer): Batch} = {}; %\label{line:sinitbatches}%
    pool: Queue = {}; %\label{line:sinitpoll}%
    expected_batch: {Broker: (Integer: 1, Root)} = {}; %\label{line:sinitexpectedbatch}%
    last_message: {Id: (sequence_number, message)} = {}; %\label{line:slastmessageinit}%
    delivery_counter = 1; %\label{line:sdelivercounterinit}%
    delivery_counter_broker: {Broker: Integer} = {};
    collection: {(Root, Broker, Integer): {Server}} = {}; %\label{line:sdefinecollection}%


procedure join(ids, payloads):
    %\StartMultiline%return [(id, context, message) 
      for (id, (context, message)) in zip(ids, payloads)]; %\EndMultiline%


procedure handle_batch(%$\broker$%, batch_number, ids, payloads, stragglers): %\label{line:shandlebatchprocedure}%
    leaves = join(ids, payloads); %\label{line:scomputeleaveshandle}%
    ids = sorted(ids); %\label{line:serverssortid}%
    
    for all id in leaves:
        if id in stragglers: %\label{line:idinstragglers}%
            leaves[index_of[id]][1] = stragglers[id]; %\label{line:stakestragglersid}%
            
    tree = merkle_tree(leaves); %\label{line:scomputetrehandle}%
    root = tree.root();  %\label{line:scomputeroothandle}%
    key alias (root, %$\broker$%, batch_number); %\label{line:sdefinekeyinhandlebatche}%
    
    if key in batches: %\label{line:sverifiykeyinbatcheshandleprocedure}%
        return %$\bot$%; %\label{line:sverifiykeyinbatcheshandleprocedurereturn}%
    if batch_number != expected_batch[%$\broker$%][0]: %\label{line:sexceptedbatch}%
         return %$\bot$%;
    if expected_batch[%$\broker$%][1] not empty: %\label{line:sexpectedbatchnotempty}%
        return %$\bot$%; %\label{line:sexpectedbatchnotemptyreturn}%
    if ids.has_duplicates() or |ids| != |payloads|: %\label{line:sbatchcheck}%
        return %$\bot$%; %\label{line:sbatchcheckfail}%
    
    unknowns = {id for id in ids such that id not in %\dirin%}; %\label{line:saskforunknows}%
    batches[key] = Batch {ids, payloads, stragglers, delivered: False}; %\label{line:sbatchesupdate}%
    expected_batch[%$\broker$%][1] = root; %\label{line:saddroottoexpected}%
    return [BatchAcquired, root, unknowns]; %\label{line:sbatchacquired}%


%\StartMultiline%upon <al.Deliver | %$\broker$%, [Batch, batch_number, ids, payloads, stragglers]>: %\EndMultiline%%\label{line:sbatchrecive}%
    %\StartMultiline%response = handle_batch(%$\broker$%, batch_number, ids, payloads, stragglers); %\EndMultiline%%\label{line:sinvokeshandlebatchprocedure}%
    if response != %$\bot$%: %\label{line:sbatchreceivecheckresponse}%
        trigger <al.Send | %$\broker$%, response>; %\label{line:sbatchreceivesendresponse}%


%\StartMultiline%procedure handle_signatures(%$\broker$%, root, batch_number, assignments, multisignature, signatures):%\EndMultiline%
    %\dirin%.import(assignments..); %\label{line:shandlesignaturesimportassignments}%
    
    if root not in expected_batch[%$\broker$%]: return %$\bot$%; %\label{line:srootinexbatchofb}%
    if batch_number != expected_batch[%$\broker$%][0]: return %$\bot$%; %\label{line:schcekexpectedbatchinhandlesignature}%
    
    key alias (root, %$\broker$%, batch_number) %\label{line:scomputekeyfrohandlesignature}%
    Batch {ids, payloads, stragglers, delivered} = batches[key]; %\label{line:sexstractbatch}%
    
    if exists id in ids such that id not in %\dirin%: return %$\bot$%; %\label{line:sdircheck}%
    if |signatures| != |stragglers|: return %$\bot$%;
    
    for (id, signature) in signatures: %\label{line:sscrollsignatures}%
        if id not in ids: return %$\bot$%; %\label{line:sidinidsforsignature}%
        if id not in stragglers: return %$\bot$%; %\label{line:sidnotinstragglers}%
            
        (.., message) = payloads[ids.index_of(id)];%\label{line:stakepayloads}%
        sequence_number = stragglers[id];
        
        if not %\dirin%[id].verify(signature, [Message, message, sequence_number]): return %$\bot$%; %\label{line:sverifyclientsignature}%
            
    multisigners = ids \ signatures.keys(); %\label{line:sdeterminemultisigners}%
    
    leaves = join(ids, payloads); %\label{line:scleaves}%
    tree = merkle_tree(leaves); %\label{line:scmerkletree}%
    reduction_root = tree.root(); %\label{line:scroot}%
    
    %\StartMultiline%if not %\dirin%[multisigners..].multiverify(multisignature, [Reduction, reduction_root]): return %$\bot$%; %\EndMultiline%%\label{line:smultiverifyreduction}%
    
    shard = multisign([Witness, root, %$\broker$%, batch_number]); %\label{line:ssignwitness}%
    return [WitnessShard, root, shard]; %\label{line:sreturnwitnessshard}%


%\StartMultiline%upon <al.Deliver | %$\broker$%, [Signatures, root, batch_number, assignments, multisignature, signatures]>: %\label{line:sdeliversignatures}%%\EndMultiline%
    %\StartMultiline%response = handle_signatures(%$\broker$%, root, batch_number, assignments, multisignature, signatures); %\EndMultiline%%\label{line:binvokehandlebatchprocedure}%
    if response != %$\bot$%: %\label{line:shandlesignaturescheckresponse}%
        trigger <al.Send | %$\broker$%, response>; %\label{line:shandlesignaturessendresponse}%


upon <al.Deliver | %$\broker$%, [Witness, root, batch_number, certificate]>: %\label{line:sdeliverwitness}%
    if root not in expected_batch[%$\broker$%][1]: return; %\label{line:srootinexpectedbatchwhendeliverwitness}%
    if batch_number != expected_batch[%$\broker$%][0]: return; %\label{line:scheckexpectedbatcheuqlanumw}%
    
    key alias (root, %$\broker$%, batch_number)
    if !verify_plurality(certificate, [Witness, key]): return; %\label{line:sverifieswkeyistruebefirepropose}%
    
    trigger <stob.Broadcast | %$\broker$%, root, certificate>;


upon <stob.Deliver | %$\broker$%, root, certificate> %\label{line:sdeliverfromconsensus}%
    batch_number = expected_batch[%$\broker$%][0]; %\label{line:scomputenum}%
    key alias (root, %$\broker$%, batch_number) %\label{line:sdefinekeyindeliverstb}%
    if !verify_plurality(certificate, [Witness, key]): return; %\label{line:switnesssignaturecheck}%
    
    expected_batch[%$\broker$%][0] = expected_batch[%$\broker$%][0] + 1; %\label{line:supdatesnum}%
    if expected_batch[%$\broker$%][1] !=  root: %\label{line:sexpectedbatchinroot}%
        for %$\server$% in %$\servers$%:
            trigger <al.Send | %$\server$%, [RequestBatch, key]; %\label{line:srequestbatch}%
            
    pool.push(key); %\label{line:sinsertelementinpool}%
    expected_batch[%$\broker$%][1].remove(root); %\label{line:sremoveroot}%


upon <al.Deliver | %$\server$%, [RequestBatch, key]>: %\label{line:sdeliverarequestbatch}%
    if key in batches: %\label{line:skyeinbatchesrequestbatchdeliver}%
        Batch {ids, payloads, stragglers, delivered} = batches[key];
        assignments = %\dirin%.export(ids..);
        %\StartMultiline%trigger <al.Send | %$\server$%, [AcceptRequestBatch, key, ids, assignments, payloads, stragglers]>; %\label{line:ssendacceptrequestbatch}% %\EndMultiline%


%\StartMultiline%upon <al.deliver | %$\server$%, [AcceptRequestBatch, key, ids, assignments, payloads, stragglers]>: %\EndMultiline%%\label{line:sdelivereacceptbatchkey}%
    if key in batches: return; %\label{line:skeyinbatchesacceptbachdeliver}%
    if key not in pool: return; %\label{line:skeyinpoolacceptebatch}%
    
    %\dirin%.export(assignments..);
    ids = sorted(ids); %\label{line:serverssortid1}%
    
    leaves = [];
    for all id in payloads:
        if id not in %\dirin%: return;
        if id in stragglers:
            leaves = leaves.add((id, stragglers[id]));
        else:
            leaves = leaves.add((id, payloads[id]));
            
    tree = merkle_tree(leaves); %\label{line:scomputesthetreeaccept}%
    check_root = tree.root(); %\label{line:scomputesthecheckingrootaccept}%
    (root, %$\broker$%, batch_number) = key;
    if check_root = root: %\label{line:schecksrootacceptbacth}%
        batches[key] = Batch{ids, payloads, stragglers, delivered:False}; %\label{line:ssetbatchafterskforthetotality}%


upon exists key in pool.peek() such that key in batches: %\label{line:skeyinpoolcheck}%
    key = pool.pop(); %\label{line:skeydequefrompool}%
    completion_batch(key); %\label{line:scallprocedurecompletion}%


procedure completion_batch(key): %\label{line:sinvokesprodecurecompletion}%
    Batch {ids, payloads, stragglers, delivered} = batches[key];
    
    completion_payloads : {Id: (Integer, Message}} = {}; %\label{line:scompletionpayloads}%
    exceptions = {Id : Integer} = {};
    
    for all id in ids: %\label{line:scycleofallids}%
        if id not in straggler:
            (sequence_number, message) = payloads[id]; %\label{line:sextractfrompayload1}%
        else:
            (.., message) = payloads[id]; %\label{line:sextractfrompayload2}%
            sequence_number = straggler[id];
            
        if sequence_number < last_message[id][0]: %\label{line:sifiamaldradyupdate}%
            exceptions[id] = %$\bot$%;
            completion_payload[id] = %$\bot$%;
        else
            if message = last_message[id][1]: %\label{line:sconditionequalmessage}%
                exceptions[id] = last_message[id][0];
                completion_payload[id] = last_message[id]; %\label{line:scompletionpayloadupdate}%
            %\StartMultiline%else if sequence_number > last_message[id][0] and message != last_message[id][1]:%\EndMultiline%%\label{line:sseqbiggerthenlastmessage}%
                last_message[id] = (sequence_number, message); %$\label{line:supdatelastmessage}$%
                completion_payload[id] =last_message[id]; %\label{line:saddmessagetocompletionpayloads}%
                keycard = dir[id];
                trigger <sr.Deliver | (keycard, message)>; %$\label{line:sdelivermessage}$%
            
    completion_leaves = join(completion_payloads.key(), completion_payloads); %\label{line:scomputecompletionleaves}%
    completion_tree = merkle_tree(completion_leaves); %\label{line:scomputecompletiontree}%
    completion_root = tree.completion_root; %\label{line:scomputecompletionroot}%
    
    (root, %$\broker$%, batch_number) = key; %\label{line:skeyincompletionbatch}%
    
    shard = multisign[Completion, completion_root, delivery_counter] %\label{line:ssigncompletionmessage}%
    %\StartMultiline%trigger <al.Send | %$\broker$%, [CompletionShard, root, completion_root,
      delivery_counter, shard, exceptions]>; %\EndMultiline%%\label{line:ssendbackcompletion}%
    
    delivery_counter = delivery_counter + 1; %\label{line:sdelivercounterupdate}%
    batches[key].delivered = True; %\label{line:ssetdelivredbatchestrue}%
    
    if %$\broker$% %$\in$% delivery_counter_broker: %\label{line:sfirsttimecompletionprocedureforb}%
        delivery_counter_broker[%$\broker$%] = delivery_counter_broker[%$\broker$%] + 1; %\label{line:sincresedelivercounterbroker}%
    else:
        delivery_counter_broker[%$\broker$%] = 1; %\label{line:setdelivercounterbroker}%


upon exists key in batches such that batches[key].delivered = True: %\label{line:skeyinbatchesdelivered}%
    collection[key] = {}; %\label{line:saddskeytocollection}%
    for %$\forall$% %$\server$% in %$\servers$%:
        trigger <al.Send | %$\server$%, [Collection, key]>; %\label{line:sbrodcastcollectionmessagefoekey}%


upon <al.Deliverer | %$\server$%, [Collection, key]>:
    (root, %$\broker$%, batch_number) = key;
    if batch_number <= delivery_counter_broker[%$\broker$%]: %\label{line:schecksificandeltekey}%
        trigger <al.Send | %$\server$%, [CollectionAccept, key]>; %\label{line:ssendcollectionacceptmessage}%


upon <al.Deliver | %$\server$%, [CollectionAccept, key]>: %\label{line:sdeliverrcollectionaccept}%
    if key %$\in$% collection:
        collection[key].add(%$\server$%); %\label{line:saddselementtocollectionofkey}%


upon exists key in collection such that |collection[key]| = 3f + 1; %\label{line:scollectionkeymaximumcardinality}%
    batches.remove(key); %\label{line:sremovekeyfrombatches}%
    collection.remove(key);
\end{lstlisting}

\clearpage
\subsection{Correctness}
\label{subsection:chop.correctness}

In this section, we prove to the fullest extent of formal detail that \system implements a \cstobprefix system.

\subsubsection{No Duplication}

In this section, we prove that \system satisfies no duplication.

\begin{notation}[Tuple]
Let $T$ be a tuple. If some element of $T$ is indicated with $\_$, thus the element's value could be any possible value. This notation is used when, for the purpose of proving a statement, we are not interested in the specific value of that element.
For example, if we define $T = (a, \_, c)$, then the second element of $T$ could be any possible value without affecting the correctness of the statement.
\end{notation}

\begin{lemma}
\label{lemma:duplication1}
Let $\server$ be a correct server, let $\client$ be a correct client. Let $m$ be a message broadcast by $\client$ with $seq$ the sequence number associated with $m$. If $m$ is the last message delivered by $\server$ for $\client$ then $(seq, m) \in last\_message[\client]$.
\end{lemma}
\begin{proof}
Upon initialization $last\_message$ is empty at $\server$ (\cref{line:slastmessageinit}). Moreover, $\server$ could only deliver messages by executing \cref{line:sdelivermessage}. Immediately before, $\server$ sets $last\_message[\client] = (seq, m)$ (\cref{line:supdatelastmessage}). We underline that $\server$ updates $last\_message$ only by executing \cref{line:supdatelastmessage}, thus the lemma is proven.
\end{proof}

\begin{lemma}
\label{lemma:duplication2}
Let $\server$ be a correct server, let $\client$ be a correct client. Let $m$ be a message broadcast by $\client$ with $seq$ the sequence number associated with $m$. Let $seq'$ be the sequence number of the last message delivered. If $seq <= seq'$, then $m$ will never be delivered.
\end{lemma}
\begin{proof}
We start by noting that according to \cref{lemma:duplication1}, $(seq', \_) \in last\_message[\client]$. 
Moreover, $\server$ can only deliver messages by executing \cref{line:sdelivermessage}. We underline that $\server$ delivers message $m$ only if the condition $seq > seq'$ (\cref{line:sseqbiggerthenlastmessage}) is satisfied, this proves the lemma.
\end{proof}

\begin{lemma}
\label{lemma:duplication3}
Let $\server$ be a correct server, let $\client$ be a correct client. Let $m$ be a message broadcast by $\client$ with $seq$ the sequence number associated with $m$. Let respectively $(seq', m')$ be the last message and sequence number delivered at $\server$. If $m = m'$ then $m$ is never delivered at $\server$.
\end{lemma}
\begin{proof}
We start by noting that according to \cref{lemma:duplication1}, $(seq', m') \in last\_message[\client]$ at $\server$. Moreover, $\server$ can only deliver messages by executing \cref{line:sdelivermessage}. We underline that $\server$ does only so if $m \neq m'$ (\cref{line:sseqbiggerthenlastmessage}). Thus, the lemma is proven.
\end{proof}

\begin{lemma}
\label{lemma:duplication4}
Let $\client$ be a correct client. Let $m$ be a message broadcast by $\client$ and $l[m]$ be a list of all sequence numbers associated with $m$. Let $m'$ be the last message broadcast by $\client$ and $l[m']$ a list of all sequence numbers associated with $m'$. We have that $max(l[m]) < min(l[m'])$ at $\client$.
\end{lemma}
\begin{proof}
Let $seq$ be a local counter at $\client$. We start by noting that $seq$, $l[m]$ and $l[m']$ are initialized as empty (\cref{line:csubmissionsinit,line:cupdatesumissions}). In the following, we neglect to analyze the event delivery of the completion certificate (Lines from \ref{line:cverifyplurality} to the end of the pseudocode). We underline that $\client$ adds elements to $l[m]$ only executes \cref{line:cupdatesumbission2} and never deletes them.

Moreover, $\client$ can only update $seq$ by executing \cref{line:cseqnumupdate}, which requires the new value of $seq$ to be bigger or equal to the previous one (\cref{line:cchcekcondition}), thus $seq$ can only increase.
We also notice that $\client$, upon updating $seq$, adds $seq$ to $l[m]$ (\cref{line:creturnsequencenumber,line:cupdatesumbission2}). This implies that at every point in time $seq = max(l[m])$.

According to client rule~\ref{rule:client:waitcompletion} (see \cref{rules:Rule}), $\client$ never broadcasts a new message $m'$ before it receives a completion certificate for $m$. When $\client$ receives a completion certificate for $m$, it updates $seq$ (\cref{line:csequpadte}), thus $seq > l[m])$. We underline that this value of $seq$ will be the first element added to $l[m']$ (\cref{line:csubmitinvoke,line:cproceduresubmit,line:caddinitsequncenumbertosub}) and since $seq$ can only grow, this implies that $max(l[m]) < min(l[m'])$. Thus proving the lemma. 
\end{proof}

\begin{theorem}[No duplication]\label{thm:nodup}
\system satisfies no duplication.
\end{theorem}
\begin{proof}
Let $\server$ be a correct server, let $\client$ be a correct client. Let $m$ be a message broadcast by $\client$ with $seq$ the sequence number associated with $m$. Let $m'$ be the last message delivered by $\server$ for $\client$. According to \cref{lemma:duplication1}, $(seq', m') \in last\_message[\client]$ at $\server$. We underline that if $m = m'$, $m$ will never be delivered by $\server$ (\cref{lemma:duplication3}).

Let $(seq'', m'')$ a message with its associated sequence number $seq''$ broadcast by $\client$ after message $m$. According to \cref{lemma:duplication4}, $seq'' > seq$. We distinguish two cases:
\begin{itemize}
    \item If $m'' = m'$, $m''$ will never be delivered by $\server$ (\cref{lemma:duplication3}).
    \item If $m'' \neq m'$ and $m''$ will be delivered, $m'' \in last\_message$ according to \cref{lemma:duplication1}. No message $m$ will never be delivered because $seq < seq''$ (\cref{lemma:duplication2}).
\end{itemize}

To summarize, after message $m'$ is delivered, no duplicate message $m = m'$ broadcast by $\client$ will be delivered. Thus, the theorem is proven.
\end{proof}

\subsubsection{Integrity}

In this section, we prove that \system satisfies integrity.

\begin{definition}[Payload and batch]
Let $\client$ be a client. Let $m$ be a message sent by $\client$ and let $seq$ be an integer representing the sequence number associated with $m$. Let $id$ the $\client$ identifier such that $id = dir[\client]$. We define $l$ the payload of $\client$ such that $l = (id, seq, m)$. We define $b$ a batch such that $b = [l]$. We define the Merkle tree $t$ and its root $r$ built upon $b$ such that $t = merkle\_tree(b)$ and $r = root(t)$.
\end{definition}

\begin{definition}[Key]
Let $\server$ be a correct server. Let $b$ be a batch and $r$ be the root of $b$. Let $num$ be an integer. Let $\broker$ be a broker that broadcasts $(b, num)$. We define $key$ at $\server$ such that $key = (r, \broker, num)$, where $num$ is a sequence number that identifies all batches broadcast by $\broker$.
\end{definition}

\begin{definition}[Witness certificate]\label{def:witness}
Let $f$ be the maximum number of Byzantine servers in the system.
We define the witness certificate as an aggregation of $f+1$ correct multi-signatures of the message $[Witness, key]$. We denote $W[key]$ the witness certificate for $key$.
\end{definition}

\begin{lemma}
\label{lemma:keyinpool}
Let $\server$ be a correct server.
Let $m$ be a message with $seq$ its associated sequence number broadcast by a client $\client$ with $id = dir[\client]$.
Let $b$ be a batch such that $b = [(id, seq, m)]$, $r$ be the root of $b$ such that $r = root(merkle\_tree(b))$, and let $key$ such that $key = (r, \_, \_)$.
If $\server$ delivers $(id, seq, m) \in b$, we have previously $key \in pool$ at $\server$. 
\end{lemma}
\begin{proof}
We start by noting that $\server$ delivers $m$ only by executing \cref{line:sdelivermessage} in the procedure $completion\_batch(key)$ (\cref{line:sinvokesprodecurecompletion}). We underline that $\server$ invokes the procedure $completion\_batch(key)$ only by executing \cref{line:scallprocedurecompletion}, upon doing so $\server$ checks that $key \in pool$ and removes $key$ from $pool$ (\cref{line:skeyinpoolcheck,line:skeydequefrompool}). Thus the lemma is proven. 
\end{proof}

\begin{lemma}
\label{lemma:poolsoexistswitness}
Let $\server$ be a correct server and let $key$ be a key. If $key \in pool$, then there exists $W[key]$.
\end{lemma}
\begin{proof}
We start by noting that upon initialization $pool$ is empty (\cref{line:sinitpoll}). Moreover, $\server$ adds elements to $pool$ only by executing \cref{line:sinsertelementinpool}. Upon doing so, $\server$ verifies that $W[key]$ exists (\cref{line:switnesssignaturecheck}). Thus, the lemma is proven.
\end{proof}

\begin{lemma}
\label{lemma:everyonesignormultisign}
Let $m$ be a message with $seq$ its associated sequence number broadcast by a client $\client$ with $id = dir[\client]$.
Let $b$ be a batch such that $b = [(id, seq, m)]$, $r$ be the root of $b$ such that $r = root(merkle\_tree(b))$, and let $key$ such that $key = (r, \_, \_)$.
Let $W[key]$ be a witness certificate for $key$. We have that:
\begin{gather*}
    \forall index \in b:\\
        id, seq, m = b[index]\\
        \client.verify(signature, [Message, m, s])\\
        \lor\\
        \client.multiverify(multisignature, [Reduction, r]).
\end{gather*}
\end{lemma}
\begin{proof}
According to \cref{def:witness}, if $W[key]$ exists then $f + 1$ servers have signed a $[Witness, key]$ message. Since $f$ is the maximum number of Byzantine servers in the system, this implies that at least one correct server $\server$ has signed the $[Witness, key]$ message.

Let $ids$ be the set of all $id \in b$. Let $signatures$ be the set of client signatures. We start by noting that $\server$ signs the $[Witness, key]$ message only by executing \cref{line:ssignwitness}. Upon doing so, $\server$ checks that $\forall id \in (signatures \land ids): \client.verify(signature, [Message, m, s])$ (\cref{line:sscrollsignatures,line:sidinidsforsignature,line:sverifyclientsignature}). If this is the case, $\server$ removes all $id \in signatures$ from $ids$ (\cref{line:sdeterminemultisigners}) and checks if $\forall id \in (ids \backslash signatures):\client.multiverify(multisignature, [Reduction, root]) = True$.
Only if this is the case, $\server$ multi-signs $[Witness, key]$ (\cref{line:ssignwitness}). We underline that $ids = (signatures\lor ids) + (ids \backslash signatures)$, thus the lemma is proven.
\end{proof}

\begin{lemma}
\label{lemma:sendsignatureaftertriggerbrodcast}
Let $\client$ be a correct client, let $m$ be a message and let $s$ be the signature of $\client$ for $m$. If $\client$ broadcasts $s$ we have that $\client$ had previously triggered the Broadcast event for $m$.
\end{lemma}
\begin{proof}
We first underline that $\client$, after computing $s = sign(m)$ (\cref{line:csignaturecompute}), broadcasts $s$ only by executing \cref{line:csendsubmit} in the procedure $submit(message)$ (\cref{line:cproceduresubmit}). Upon $\client$ triggering the event Broadcast for $m$, it invokes the procedure $submit(m)$ (\cref{line:csubmitinvoke}).

We underline that $\client$ can also invoke the procedure $submit(m)$ executing \cref{line:cinvokesubmitring}, which could only happen when the timer $Submit$ expires (\cref{line:ctimerring}). Moreover, the timer $Submit$ is set only inside the procedure $submit(m)$ (\cref{line:ctimerset}). This implies that $\client$ will never invoke the procedure $submit(m)$ before triggering the event Broadcast for $m$. Thus, the lemma is proven. 
\end{proof}

\begin{lemma}
\label{lemma:sendmultisignatureafterbrodcast}
Let $\client$ be a correct client, let $m$ be a message and let $ms$ be the multi-signature of $\client$ for $m$. If $\client$ broadcasts $ms$ then $\client$ had previously triggered the Broadcast event for $m$.
\end{lemma}
\begin{proof}
We start by noting that $\client$ broadcasts $ms$ for the message $m$ only by executing \cref{line:csendreduction}. Before doing so, $\client$ checks if $m \in submissions$ (\cref{line:cexistsubmissionforinclusion}). We underline that upon initialization $submissions$ is empty (\cref{line:csubmissionsinit}) and $\client$ adds $m$ to $submissions$ only by executing \cref{line:cupdatesumissions}, this happens immediately after triggering the Broadcast event for $m$. Thus, the lemma is proven.
\end{proof}

\begin{lemma}
\label{lemma:idsnodupilicate}
Let $b$ be a batch, and let $r$ be the root of $b$. Let $key$ be a key such that $key = (r, \_, \_)$. Let $ids$ be all $id \in b$. If a witness certificate $W[key]$ exists, then $ids = \cup(\left\{ id \forall id \in ids \right\})$.
\end{lemma}
\begin{proof}
We start by noting that, according to \cref{def:witness}, if there exists a witness certificate $W[key]$ at at least one correct server $\server$, then $\server$ has signed a $[Witness, key]$ message. We underline that the number of Byzantine servers in the system could be at most $f$.

We notice that $\server$ signs a $[Witness, key]$ message only by executing \cref{line:ssignwitness}. Immediately before doing this, $\server$ verifies that $r \in expected\_batch$ (\cref{line:srootinexbatchofb,line:srootinexbatchofb}).
Moreover, $\server$ adds root elements to $expected\_batch$ only by executing \cref{line:saddroottoexpected}. Before doing so, $\server$ first computes the root $r$ such that $l = [(id, \_, \_) \forall id \in ids]$ (\cref{line:scomputeleaveshandle}), $t = merkle\_tree(l)$ (\cref{line:scomputetrehandle}), and $r = root(t)$ (\cref{line:scomputeroothandle}).
Moreover, $\server$ checks if $ids$ contains the same identifier duplicated multiple times otherwise returns (\cref{line:sbatchcheck,line:sbatchcheckfail}. This implies that, if $\server$ adds $r$ to $expected\_batch$, then $ids = \cup(\left\{ id \forall id \in ids \right\})$. Thus, the lemma is proven.
\end{proof}

\begin{theorem}[Integrity]\label{thm:integrity}
\system satisfies integrity.
\end{theorem}
\begin{proof}
Let $\server$ be a correct server. Let $\client$ be a correct client and $id$ the client identifier $id = dir[\client]$. Let $m$ be a message broadcast by $\client$ with $seq$ its associated sequence number and $b$ a batch such that $(id, seq, m) \in b$.

According to \cref{lemma:keyinpool}, if $m$ is delivered by $\server$ then there exists a $key$ such that $key \in pool$. Moreover, according to \cref{lemma:poolsoexistswitness}, if $key \in pool$ then a witness certificate exists for $key$ $W[key]$ at $\server$.
We underline that, according to \cref{lemma:everyonesignormultisign}:
\begin{gather*}
    \forall \; index \in b:\\
        id, seq, m = b[index]\\
        \client.verify(signature, [Message, m, seq]) \\
        \lor \\
        \client.multiverify(multisignature, [Reduction, r]).
\end{gather*}
We notice that, according to \cref{lemma:idsnodupilicate}, if there exists $W[key]$ then $ids = \cup(\left\{ id \forall id \in ids \right\})$.
To conclude, if a witness certificate $W[key]$ exists, then $\client$ has signed its own message $m$ or multi-signed $r$ the root of the batch $b$.

According to \cref{lemma:idsnodupilicate}, it is impossible that $id \in b$ more than once. Moreover, $\client$ broadcasts its own signature or multi-signature only if $\client$ has previously triggered the Broadcast event for $m$ (\cref{lemma:sendsignatureaftertriggerbrodcast,lemma:sendmultisignatureafterbrodcast}). This implies that $m$ could be delivered by $\server$ only if $m$ has previously been broadcast by $\client$. Thus, the integrity property holds. The theorem is proven.
\end{proof}

\subsubsection{Agreement}

In this section, we prove that \system satisfies agreement.

\begin{definition}[Deliver from \stobprefix]
Let $\server$ be a correct server, let $\broker$ be a broker. Let $r$ be the root of a batch $b$ and let $W$ be a witness certificate. We define the delivery from \stob (\stobprefix), instance $\stobi$, when $\server$ triggers the event $<stob.Deliver | \broker, r, W>$.
We note this event $D[\broker, r, W]$.
\end{definition}

\begin{definition}[Broadcast to \stobprefix]
\label{def:brodcasttostb}
Let $\server$ be a correct server, let $\broker$ be a broker. Let $r$ be the root of a batch $b$ and let $W$ be a witness certificate. We define broadcasting to \stobprefix, instance $\stobi$, when $\server$ triggers the event $<stob.Broadcast | \broker, root, W>$.
We note this event $B[\broker, r, W]$.
\end{definition}

\begin{lemma}
\label{lemma:keyinbatcheswhencompletioninvokes}
Let $\server$ be a correct server and let $key$ be a key. If $\server$ invokes the $completion\_batch(key)$ procedure then $key \in batches$ at $\server$.
\end{lemma}
\begin{proof}
We start by noting that $\server$ can invoke the $completion\_batch$ procedure only by executing \cref{line:scallprocedurecompletion}. Before doing this, $\server$ checks that $key \in batches$ (\cref{line:skeyinpoolcheck}). Thus the lemma is proven.
\end{proof}

\begin{lemma}
\label{lemma:delivredtruewasinpool}
Let $\server$ be a correct server, and let $key$ be a key such that $key \in batches$. If $batches[key].delivered = True$, then $key \in pool$ at $\server$.
\end{lemma}
\begin{proof}
We start by noting that, upon initialization, $batches[key].delivered = False$ (\cref{line:sbatchesupdate}). We notice that $\server$ sets $batches[key].delivered = True$ only by executing \cref{line:ssetdelivredbatchestrue} in the procedure $completion\_batch(key)$ (\cref{line:sinvokesprodecurecompletion}).
We underline that $\server$ invokes the procedure $completion\_batch(key)$ only by executing \cref{line:scallprocedurecompletion}, immediately before $server$ checks that $key \in pool$ and removes $key$ from pool. Thus the lemma is proven.
\end{proof}

\begin{lemma}
\label{lemma:keyinpooluponddeliver}
Let $\server$ be a correct server, let $\broker$ be a broker. Let $b$ be a batch and let $r$ be the root of $b$. Let $key$ be a key such that $key = (r, \broker, \_)$.
If and only if the event $D[\broker, r, W]$ happens, we have that $key \in pool$ at some time $t$ at $\server$.
\end{lemma}
\begin{proof}
We start by noting that upon initialization $pool$ is empty at $\server$ (\cref{line:sinitpoll}). Upon $D[\broker, r, W]$ happening (\cref{line:sdeliverfromconsensus}), $\server$ adds $key$ to pool (\cref{line:switnesssignaturecheck,line:sinsertelementinpool}). We underline that before doing this, $\server$ computes $key$ such that $key = (r, \broker, \_)$ (\cref{line:sdefinekeyindeliverstb}) and checks that $W$ is a witness certificate for $key$ (\cref{line:switnesssignaturecheck}).
This proves that $key \in pool$ at some time $t$ at $\server$.
We notice that $\server$ updates $pool$ only by executing \cref{line:sinsertelementinpool} when $D[\broker, r, W]$ happens.
Thus the lemma is proven.
\end{proof}

\begin{lemma}
\label{lemma:completiononlyifeceventhappens}
Let $\server$ be a correct server, let $\broker$ a broker. Let $b$ be a batch and let $r$ be the root of $b$. Let $W$ be a witness certificate. Let $key$ be a key such that $key = (r, \broker, \_)$.
If $\server$ invokes the $completion\_batch(key)$ procedure, then the event $D[\broker, r, W]$ happens at $\server$.
\end{lemma}
\begin{proof}
We start by noting that if $key \in pool$ then the event $D[\broker, r, W]$ happens at $\server$ (\cref{lemma:keyinpooluponddeliver}).
We underline that $\server$ invokes the $completion\_batch$ procedure only by executing \cref{line:scallprocedurecompletion}. Immediately before doing this, $\server$ checks if $key \in pool$ (\cref{line:skeyinpoolcheck}). Thus the lemma is proven.
\end{proof}

\begin{lemma}
\label{lemma:deleterootfromexceptedbatch}
Let $\server$ be a correct server, let $\broker$ be a broker. Let $b$ be a batch and let $r$ be the root of $b$. Let $W$ be a witness certificate. Let $key$ be a key such that $key = (r, \broker, \_)$.
If the event $D[\broker, r, W]$ happens, then $r \notin expected\_batch[\broker]$.
\end{lemma}
\begin{proof}
Upon $D[\broker, r, W]$ happening (\cref{line:sdeliverfromconsensus}), $\server$ removes $r$ to $expected\_batch[\broker]$(\cref{line:sremoveroot}). We underline that before doing this, $\server$ computes $key$ such that $key = (r, \broker, \_)$ (\cref{line:sdefinekeyindeliverstb}) and checks that $W$ is a witness certificate for $key$ (\cref{line:switnesssignaturecheck}). Thus the lemma is proven.
\end{proof}

\begin{lemma}
\label{lemma:deletekeyfrombatchescondition}
Let $\server$ be a correct server, and let $key$ be a key such that $key \in batches$.
$\server$ removes $key$ from batches only if $batches[key].delivered = True$.
\end{lemma}
\begin{proof}
We start by noting that $\server$ removes elements from $batches$ only by executing \cref{line:sremovekeyfrombatches}. Immediately before doing this, $\server$ checks if $|collection[key]| = 3f + 1$.
We underline that upon initialization, $collection$ is empty at $\server$ (\cref{line:sdefinecollection}). Moreover, $\server$ adds $key$ in $collection$ only by executing \cref{line:saddskeytocollection}. Immediately before doing this, $\server$ checks if $batches[key].delivered = True$ (\cref{line:skeyinbatchesdelivered}). We also notice that $\server$ adds elements to $collection[key]$ only by executing \cref{line:saddselementtocollectionofkey}. Immediately before, $\server$ verifies that $key \in collection$.

To summarize, $\server$ adds $key$ to $collection$ only if $batches[key].delivered = True$, adds an element to $collection[key]$ only if $key \in collection$ and in order to delete $key$ checks that $collection[key]$ has enough elements, thus the lemma is proven.
\end{proof}

\begin{lemma}
\label{lemma:rexcpectedkeiinbatches}
Let $\server$ be a correct server and let $\broker$ be a broker. Let $b$ be a batch and let $r$ be the root of the batch. Let $key$ be a key such that $key = (r, \broker, \_)$. If $r \in expected\_batch[\broker]$, we have that $key \in batches$ at $\server$.
\end{lemma}
\begin{proof}
We start by noting that upon initialization $expected\_batch$ and $batches$ are empty (\cref{line:sinitexpectedbatch,line:sinitbatches}). Moreover, $\server$ adds elements to $expected\_batch$ only by executing \cref{line:saddroottoexpected}. Upon $\server$ adding $r$ to $expected\_batch[\broker]$, $\server$ immediately before adds $key$ to batches (\cref{line:saddroottoexpected,line:sbatchesupdate}). We underline that by construction $key = (r, \broker, \_)$ (\cref{line:sdefinekeyinhandlebatche}).

According to \cref{lemma:deletekeyfrombatchescondition}, $\server$ removes $key$ from $batches$ only if $batches[key].delivered = True$.
Moreover, if $batches[key].delivered = True$ then previously $key \in pool$ at $\server$ (\cref{lemma:delivredtruewasinpool}).
According to \cref{lemma:keyinpooluponddeliver}, if $key \in pool$ this implies that $D[\broker, r, W]$ event has happened. We conclude that if the event $D[\broker, r, W[r]]$ has happened, $r \notin expected\_batch[\broker]$ (\cref{lemma:deleterootfromexceptedbatch}).

To summarize, we have $key \in batches$ at $\server$ when $\server$ sets $r \in excepted\_batch[\broker]$ and if $server$ removes $key$ from batches then $r \notin excepted\_batch[\broker]$.
This implies that if $r \in excepted\_batch[\broker]$ then $key \in batches$ at $\server$, thus the lemma is proven.
\end{proof}

\begin{lemma}
\label{lemma:keyinbatchesifsignwitness}
Let $\server$ be a correct server. Let $key$ be a key. If $\server$ signs a $[Witness, key]$ message for $key$ at time $t$, we have that $key \in batches$ at time $t'$ with $t' < t$ at $\server$.
\end{lemma}
\begin{proof}
We start by noting that upon initialization $batches$ is empty at $\server$ (\cref{line:sinitbatches}). Moreover, $\server$ signs a $[Witness, key]$ message only by executing \cref{line:ssignwitness}. We underline that $\server$ computes $key$ such that $key = (r, \_, \_)$ (\cref{line:scomputekeyfrohandlesignature}). 
Immediately before signing the $[Witness, key]$ message, $\server$ checks if $r \in expected\_batch[\broker]$ and returns without signing if this is not the case (\cref{line:srootinexbatchofb}).
According to \cref{lemma:rexcpectedkeiinbatches}, if $r \in expected\_batch[\broker]$ we have that $key \in batches$ at $\server$.
This implies that $key \in batches$ at time $t' < t$, thus the lemma is proven.
\end{proof}

\begin{lemma}
\label{lemma: keysameandsameorder}
Let $\server$ and $\server'$ be any two correct servers. Let $\broker$ be a broker, let $b$ be a batch and $r$ the root of $b$. Let $W$ be a witness certificate. Let $keys$ and $keys'$ two sorted sets of keys.
Upon an event $D[\broker, r, W]$ happening, $\server$ computes $key$ such that $key = (r, \broker, \_)$, and adds $key$ to the set $keys$.
Similarly, upon an event $D[\broker', r', W']$ happening, $\server'$ computes $key'$ such that $key' = (r', \broker', \_)$ and adds $key'$ to the set $keys'$.
Let $l$ the minimum length between $keys$ and $keys'$ such that $l = min(|keys|, |keys'|)$.
Let $keys|_{l}$ the set of the first $l$ elements of $keys$, let $keys'|_{l}$ the set of the first $l$ elements of $keys'$.
At every point in time, we have $keys|_{l} = keys'|_{l}$ and $keys[i] = keys'[i] \forall i < l$.
\end{lemma}
\begin{proof}
Let $keys|_{final}$ the total set of keys delivered by $\server$ until the end of the system. Let $keys'|_{final}$ the total set of keys delivered by $\server'$ until the end of the system.

We start by noting that, for the agreement property of \stobprefix, if the event $D[\broker, r, W]$ happens at $\server$, eventually the event $D[\broker, r, W]$ will happen at $\server'$, and vice versa. Let $key = (r, \broker, \_)$, $\server$ executes $keys.add(key)$, and $\server'$ executes $keys'.add(key)$.
So at time $t = final$ we have that $keys|_{final} = keys'|_{final}$.
For the total order property underlying \stobprefix, if the event $D[\broker, r, W]$ happens at $\server$ and then event $D[\broker', r', W']$ happens at $\server$, it is not possible that at $\server'$ the event $D[\broker', r', W']$ happens before the event $D[\broker, r, W]$. 
To summarize, for the agreement property of \stobprefix, $\server$ and $\server'$ adds to $keys$ and $keys'$ the same sets of key and for the total order property underlying \stobprefix, $\server$ and $\server'$ add to $keys$ and $keys'$ the keys in the same order.

We underline that due to the no duplication property of \stobprefix, $\server$ and $\server'$ cannot add the same key twice.
This implies that $\server$ and $\server'$ add to $keys$ and $keys'$ the same set of keys in the same order.
So in conclusion, we have that at any point in time $keys|_{l} = keys'|_{l}$ and $keys[i] = keys'[i] \forall i < l$.
Thus, the lemma is proven.
\end{proof}

\begin{lemma}
\label{lemma:callcomplitionbatchinthesameorder}
Let $\server$ and $\server'$ be any two correct servers. Let $keys$ and $keys'$ be two sorted sets of keys with $|keys| = |keys'|$ and $keys[i] = keys'[i] \forall i < keys.length$.
At $\server$, let $\forall key \in keys \rightarrow key \in batches$.
At $\server'$, let $\forall key \in keys' \rightarrow key \in batches$.
Let $completion\_batch()_{i}$ be the $i$-th time that a server invokes the procedure $completion\_batch()$.
If $\server$ invokes $completion\_batch(key)_{i}$ with $key \in keys$, then $\server'$ will eventually invoke $completion\_batch(key')_{j}$ with $key' \in keys' \And i = j$ and we have that $key = key' \forall i < keys.length$.
\end{lemma}
\begin{proof}
We start by noting that, upon initialization, $pool$ and $pool'$ are empty at $\server$ and $\server'$ (\cref{line:sinitpoll}).
Let $\broker$ be a broker, let $b$ be a batch and let $r$ be the root of $b$. Upon an event $D[\broker, r, W]$ happening at $\server$ (\cref{line:sdeliverfromconsensus}), $\server$ computes $key = (r, \broker, \_)$ (\cref{line:sdefinekeyindeliverstb}) and adds $key$ to $pool$ (\cref{line:sinsertelementinpool}).
Similarly upon an event $D[\broker', r', W']$ happening at $\server'$ (\cref{line:sdeliverfromconsensus}), $\server'$ computes $key' = (r', \broker', \_)$ (\cref{line:sdefinekeyindeliverstb}) and adds $key'$ to $pool$ (\cref{line:sinsertelementinpool}).
We underline that, according to \cref{lemma: keysameandsameorder}, $\server$ and $\server'$ respectively add to $pool$ and $pool'$ the same keys in the same order.
This implies that $key = key' \forall i$ such that $key = pool.pop()_{i}$, the $i$-th time that $\server$ extracts the first element from $pool$, and $key' = pool'.pop()_{i}$ the $i$-th time that $\server'$ extracts the first element from $pool'$.
We remark that for assumption $\forall key \in pool \rightarrow key \in batches$ at $\server$ and $\forall key \in pool' \rightarrow key \in batches$ at $\server'$.

To conclude, every times that $\server$ executes $key = pool.pop()$ it also invokes $completion\_batch(key)$ (\cref{line:skeydequefrompool,line:scallprocedurecompletion}).
This implies that if $\server$ invokes $completion\_batch(key)_{i}$ with $key = pool.pop()_{i}$, then $\server'$ will eventually invoke $completion\_batch(key')_{i}$ with $key' = pool.pop()_{i}$.
And we have $key = key' \forall i$.
We underline that if $\server$ adds $key$ to $pool$, eventually $\server'$ adds $key$ to $pool'$ for the agreement property of \stobprefix. Thus, the lemma is proven. 
\end{proof}

\begin{lemma}
\label{lemma:collectionmessagesoicallcompletionprecedure}
Let $\server$ be a correct server and let $key$ be a key. If $\server$ broadcasts a $[Collection, key]$ message for $key$, then previously $\server$ invokes $completion\_batch(key)$.
\end{lemma}
\begin{proof}
We start by noting that $\server$ broadcasts a $[Collection, key]$ message only by executing \cref{line:sbrodcastcollectionmessagefoekey}. Immediately before, $\server$ checks if $batches[key].delivered = True$ (\cref{line:skeyinbatchesdelivered}).
We underline that upon $\server$ inserting $key$ in $batches$, we have $batches[key].delivered = False$ (\cref{line:sbatchesupdate,line:ssetbatchafterskforthetotality})
Moreover, $\server$ sets $batches[key].delivered = True$ only by executing \cref{line:ssetdelivredbatchestrue} in the procedure $completion\_batch(key)$ (\cref{line:sinvokesprodecurecompletion}). Thus, the lemma is proven.
\end{proof}

\begin{lemma}
\label{lemma:completionbatchequaldelivertimes}
Let $\server$ be a correct server. Let $keys$ be a set of key such that $keys = \left\{ key: key = (\_, \broker, \_) \right\}$.
Let $|completion\_batch(key)|$ the number of times that $\server$ invokes the $completion\_batch()$ procedure for $key \in keys$.
We have that at every time $|completion\_batch(key)| = delivery[\broker]$
at $\server$.
\end{lemma}
\begin{proof}
We notice that upon initialization, $delivery[\broker]$ is empty and $\server$ updates it only by executing \cref{line:sincresedelivercounterbroker,line:setdelivercounterbroker} in the procedure $completion\_batch()$ (\cref{line:sinvokesprodecurecompletion}). 

Let $keys$ a set of keys such that $keys = \left\{ key: key = (\_, \broker, \_) \right\}$.
We underline that, every time that $\server$ invokes the procedure $completion\_batch(key)$ with $key \in keys$, $\server$ increases the $delivery[\broker]$ by one (\cref{line:sincresedelivercounterbroker}). We remark that the first time that $\server$ invokes the $completion\_batch(key)$, $\server$ sets $delivery[\broker] = 1$ (\cref{line:sfirsttimecompletionprocedureforb,line:setdelivercounterbroker}).
This implies that at every point in time $|completion\_batch(key)| = delivery[\broker]$ at $\server$. Thus, the lemma is proven.
\end{proof}

\begin{lemma}
Let $\server$ be a correct server. Let $\broker$ be a broker. Let $|D[\broker, \_, \_]|$ the number of $D[\broker, \_, \_]$ events happens at $\server$ for the same $\broker$.
Let $num$ be a number such that $num = |D[\broker, \_, \_]|$.
We have that at every point in time at $\server$ $num \ge delivery[\broker]$.
\end{lemma}
\begin{proof}
We start by noting that when an event $D[\broker, \_, \_]$ happens, $\server$ computes $key$ such that $key = (\_, \broker, num)$ with $num = expected\_batch[\broker]$ (\cref{line:sdeliverfromconsensus,line:scomputenum}).
We underline that upon initialization, $expected\_batch[\broker] = 0$ (\cref{line:sinitexpectedbatch}). Moreover, $\server$ updates $expected\_batch[\broker]$, adding one, only executing \cref{line:supdatesnum} when an event $D[\_, \broker, \_]$ happens (\cref{line:sdeliverfromconsensus}). 
Let $keys$ a set of keys such that $keys = \left\{ key: key = (\_, \broker, \_) \right\}$.
Let $|completion\_batch(key)|$ the number of times that $\server$ has invoked the $completion\_batch()$ procedure for $key \in keys$.
We remark that at every point in time $|completion\_batch(key)| = delivery[\broker]$ at $\server$ (\cref{lemma:completionbatchequaldelivertimes}).
Moreover, according to \cref{lemma:completiononlyifeceventhappens}, $\server$ invokes the $completion\_batch(key)$ procedure with $key = (r, \broker, \_)$ only if the event $D[r, \broker, W[r]]$ happens. This implies that at every point in time $num \ge delivery[\broker]$ at $\server$.
Thus the lemma is proven.
\end{proof}

\begin{lemma}
\label{lemma:deliverikeyi}
Let $\server$ be a correct server and let $\broker$ be a broker. Let $D[\_, \broker, \_]_{i}$ be the $i$-th time that an event $|D[\_, \broker', \_]|$ happened at $\server$ with $\broker = \broker'$.
Let $key$ be the key computes by $\server$ when the event $D[\_, \broker, \_]_{i}$ happens. We have that $key = (\_, \broker, i)$.
\end{lemma}
\begin{proof}
We start by noting that $\server$ delivers from \stobprefix only by executing \cref{line:sdeliverfromconsensus}. We underline that $expected\_batch[\broker]$ is initialized to $1$ at $\server$ (\cref{line:sinitexpectedbatch}). We also notice that $\server$ updates $expected\_batch[\broker]$, adding $1$, only when an event $D[\_, \broker, \_]$ happens (\cref{line:sdeliverfromconsensus,line:supdatesnum}). 
Moreover, when an event $D[\_, \broker, \_]$ happens, $\server$ computes $key$ such that $key = (\_, \broker, n)$ with $n = expected\_batch[\broker]$ (\cref{line:scomputenum}).
To summarize the first time that an event $D[\_, \broker, \_]$ happens (we call this event $D[\_, \broker, \_]_{1}$), we have that $\server$ computes $key$ such that $key = (\_, \broker, 1)$ and $\server$ adds $1$ to $expected\_batch[\broker]$.
This implies that at every point in time if the $D[\_, \broker, \_]_{i}$ event happens then $key = (\_, \broker, i)$.
Thus the lemma is proven.
\end{proof}

\begin{lemma}
\label{lemma:calldeliverandproceduresameorder}
Let $\server$ be a correct server and let $\broker$ be a broker. 
Let $keys$ be a set of keys such that $keys = \left\{ key: key = (\_, \broker, \_) \right\}$.
Let $D[\_, \broker, \_]$ an event that happens at $\server$ before the event $D'[\_, \broker, \_]$. Let $key \in keys$ and $key' \in keys$ the keys computed by $\server$ when the event $D[\_, \broker, \_]$ and $D'[\_, \broker, \_]$ happens. 
We have that $\server$ cannot invoke $completion\_batch(key')$ procedure without first invoking $completion\_batch(key)$.
\end{lemma}
\begin{proof}
We start by noting that when an event $D[\_, \broker, \_]$ at $\server$ (\cref{line:sdeliverfromconsensus}), $\server$ computes $key$ such that $key = (\_, \broker, \_)$ (\cref{line:sdefinekeyindeliverstb}). Moreover, $\server$ adds $key$ to $pool$ (\cref{line:sinsertelementinpool}).
We underline that upon initialization, $pool$ is empty at $\server$ (\cref{line:sinitpoll}) and $\server$ adds elements to the back of queue $pool$ only by executing \cref{line:sinsertelementinpool}. 
We notice that $\server$ invokes the $completion\_batch()$ procedure only by executing \cref{line:scallprocedurecompletion}. Immediately before doing this, $\server$ peeks the first element of $pool$ and removes it from $pool$ (\cref{line:skeyinpoolcheck,line:skeydequefrompool}).
This implies that, if $\server$ inserts $key$ in $pool$ before inserting $key'$, then $\server$ will invoke the $completion\_procedure(key)$ before $completion\_procedure(key')$. Thus, the lemma is proven.
\end{proof}

\begin{lemma}
\label{lemma:nsomeordercompletionbatch}
Let $\server$ be a correct server and let $\broker$ be a broker. 
Let $keys$ be a set of keys such that $keys = \left\{ key: key = (\_, \broker, \_) \right\}$.
Let $D[\_, \broker, \_]_{n}$ be the $n$-th event for $\broker$ happens at $\server$. Let $key$ the key computed by $\server$ when the event $D[\_, \broker, \_]_{n}$ happens.
Let $completion\_batch()_{i}$ be the $i$-th time that $\server$ invokes a procedure $completion\_batch$ for some $key \in keys$. If $\server$ invokes $completion\_batch(key')_{n}$, then $key' = key$ and $key = (\_, \broker, n)$.
\end{lemma}
\begin{proof}
We start by noting that when an event $D[\_, \broker, \_]_{n}$ happens, $\server$ computes $key$ as $key = (\_, \broker, n)$ (\cref{lemma:deliverikeyi}).
According to \cref{lemma:calldeliverandproceduresameorder}, $\server$ invokes the $completion\_batch()$ procedure for $key$ in the same order that the associated event $D[\_, \broker, \_]$ happens.
To summarize, if $\server$ invokes $completion\_batch(key')_{n}$ then $key' = key$ and $key = (\_, \broker, n)$.
Thus, the lemma is proven. 
\end{proof}

\begin{lemma}
\label{lemma:collectionacceptcompletionbatch}
Let $\server$ be and $\server'$ two correct servers. Let $n$ be a number and let $key$ be a key such that $key = (\_, \broker, n)$.
Let $\server$ broadcasts a $[Collection, key]$ message, if $\server'$ replays with a message $[CollectionAccept, key]$, then $\server'$ has previously invoked $completion\_batch(key)$.
\end{lemma}
\begin{proof}
We start by noting that $\server'$ sends a $[CollectionAccept, key]$ message only by executing \cref{line:ssendcollectionacceptmessage}. Immediately before doing this, $\server'$ checks if $n \le delivery[\broker]$ (\cref{line:schecksificandeltekey}).
We notice that if $\server$ broadcasts a $[Collection, key]$ message for $key$, $\server$ has previously invoked the $completion\_batch(key)$ procedure. (\cref{lemma:collectionmessagesoicallcompletionprecedure}).
Let $keys$ be the set of key such that $keys = \left\{ key: key = (\_, \broker, \_) \right\}$.
Let $completion\_batch()_{i}$ be the $i$-th time that $\server$ invokes the $completion\_batch()$ for some $key \in keys$.
According to \cref{lemma:nsomeordercompletionbatch}, we have at $\server$ that $completion\_batch(key)_{n}$ and $key = (\_, \broker, n)$.

Moreover, according to \cref{lemma:callcomplitionbatchinthesameorder}, $\server$ and $\server'$ invoke the $completion\_batch()$ procedure for $key \in keys$ in the same order. This implies that $key = key'$ when $completion\_batch(key)_{n}$ is invoked by $\server$ and $completion\_batch(key')_{n}$ by $\server'$.
Let $|completion\_batch(key)|$ be the number of times that $\server'$ invokes the $completion\_batch()$ procedure for $key \in keys$.
We remark that at every point in time $|completion\_batch(key)| = delivery[\broker]$ at $\server'$ (\cref{lemma:completionbatchequaldelivertimes}).
Let $delivery[\broker] = i$. This implies that $\server'$ has already invoked $i$ times the procedure $completion\_batch()$ for $key \in keys$.
To conclude, if $n \le i$, then $\server'$ had already invoked $completion\_batch(key')_{n}$ and $key' = key = (\_, \broker, n)$.
Thus, the lemma is proven.
\end{proof}

\begin{lemma}
\label{lemma:notrenovekeyfrombatchesuntilecompletionbatch}
Let $\server$ be a correct server, let $key$ be a key such that $key \in batches$ at $\server$. If $\server$ removes $key$ from $batches$, then all correct servers have previously invoked the procedure $completion\_batch(key)$.
\end{lemma}
\begin{proof}
We start by noting that $\server$ removes $key$ from $batches$ only by executing \cref{line:sremovekeyfrombatches}. This happens only if $|collection[key]| = 3f + 1$ (\cref{line:scollectionkeymaximumcardinality}). We remember that, for hypothesis, the total number of servers in the system is $3f + 1$.
We underline that upon initialization $collection[key]$ is empty (\cref{line:sdefinecollection}) and $\server$ add elements to it only by executing \cref{line:saddselementtocollectionofkey}. This implies that $\server$ removes $key$ from $batches$ only after receiving the $[CollectionAccept, key]$ message from all the servers in the system (\cref{line:sdeliverrcollectionaccept}).
We remark that according to \cref{lemma:collectionacceptcompletionbatch}, if a correct server $\server'$ broadcasts a $[CollectionAccept, key]$ message this implies that $\server'$ had already invoked the $completion\_batch()$ procedure for $key$.
To summarize, $\server$ removes $key$ from batches only if all correct servers had already invoked the $completion\_batch()$ procedure for $key$. Thus, the lemma is proven.
\end{proof}

\begin{lemma}
\label{lemma:requestbatchbeforecompletionbatch}
Let $\server$ be a correct server. Let $key$ be a key. If $\server$ sends a $[RequestBatch, key]$ message at time $t$, $\server$ has never invoked the $completion\_batch(key)$ procedure at time $t'$ with $t' \leq t$.
\end{lemma}
\begin{proof}
We start by noting that $\server$ sends a $[RequestBatch, key]$ message with $key = (r, \broker, \_)$ only if an event $D[r, \broker, \_]$ happens (\cref{line:sdeliverfromconsensus,line:srequestbatch}). We call time $t$ the moment at which $\server$ sends a $[RequestBatch, key]$ message.
After doing this, $\server$ adds $key$ to $pool$ (\cref{line:sinsertelementinpool}), we identify with $t'$ the moment at which this happens. We underline that sending a $[RequestBatch, key]$ message and adding $key$ to $pool$ are sequential operations, this implies that $t \le t'$.
We remark that, according to \cref{lemma:deliverikeyi}, $\server$ never computes the same $key$ twice. 
This implies that this is the first and the only time that $\server$ adds $key$ to $pool$.
We underline that $\server$ invokes the $completion\_batch(key)$ procedure only by executing \cref{line:scallprocedurecompletion} and immediately $\server$ checks that $key \in pool$ (\cref{line:skeyinpoolcheck}). This implies that if we call $t''$ the moment when $\server$ invokes the $completion\_batch(key)$ then $t \le t' \leq t''$.
Thus, the lemma is proven.
\end{proof}

\begin{lemma}
\label{lemma:keyeventuallinbatchesifyouask}
Let $\server$ and $\server'$ be two correct servers. Let $key$ be a key such that $key \in batches$ at $\server'$. If $\server$ broadcasts a $[RequestBatch, key]$, then eventually $key \in batches$ at $\server$.
\end{lemma}
\begin{proof}
Let $\broker$ be a broker, let $b$ be a batch and let $r$ be the root of $b$. Let $W$ be a witness certificate. Let $D[r, \broker, W]$ be an event happens at $\server$ and let $key = (r, \broker, \_)$ the $key$ computes by $\server$ upon the event happening. 
We start by noting that $\server$ sends a $[RequestBatch, key]$ (\cref{line:srequestbatch}) only if the event $D[r, \broker, W]$ happens (\cref{line:sdeliverfromconsensus}) and $W$ is a witness certificate for $key$ (\cref{line:switnesssignaturecheck}).

We underline that, according to \cref{lemma:requestbatchbeforecompletionbatch}, when $\server$ sends a $[RequestBatch, key]$ message, then $\server$ has not yet invoked the procedure $completion\_batch(key)$.
Thus, if $\server$ sends a $[RequestBatch, key]$ message at time $t$ we have two possible cases:
\begin{itemize}
    \item At time $t'$ with $t' \ge t$, $\server$ invokes the procedure $completion\_batch(key)$, implying that $key \in batches$ which proves the lemma.
    \item $\server$ receives an $[AcceptRequestBatch, key]$ message from $\server'$ because, according to \cref{lemma:notrenovekeyfrombatchesuntilecompletionbatch}, it is impossible that $\server'$ has removed $key$ from batches because this could happen only if all correct servers have previously invoked the $completion\_batch(key)$ procedure. But in this case, $\server$ has not yet invoked it.
\end{itemize}

To summarize, $\server'$ eventually receives the $[RequestBatch, key]$ message from $\server$ (\cref{line:sdeliverarequestbatch}) and replays the $[AcceptRequestBatch, key]$ message (\cref{line:ssendacceptrequestbatch}). Moreover, $\server$ eventually delivers the $[AcceptRequestBatch, key]$ message (\cref{line:sdelivereacceptbatchkey}). Upon receiving this, $\server$ checks:
\begin{itemize}
    \item If $key \in batches$: if this is the case the lemma is already proven (\cref{line:skeyinbatchesacceptbachdeliver}).
    \item If $key \in pool$: this is satisfied because we assume that an event $D[\broker, r, \_]$ happens and, according to \cref{lemma:keyinpooluponddeliver}, the respective key is in pool (\cref{line:skeyinpoolacceptebatch}).
\end{itemize}
Also, $\server$ computes the root $r'$ of the batch as follows: $l = (id, seq, m)$, $b = [l]$, $t = merkle\_tree(b)$, and $r' = root(t)$ (\cref{line:scomputesthetreeaccept,line:scomputesthecheckingrootaccept}).
In conclusion $\server$ checks if $r = r'$ (\cref{line:schecksrootacceptbacth}) and adds $key$ to $batches$ (\cref{line:ssetbatchafterskforthetotality}). Thus, the lemma is proven.
\end{proof}

\begin{lemma}
\label{lemma:ifeeventeventuallykeyinbatches}
Let $\server$ be a correct server, let $\broker$ be a broker. Let $b$ be a batch and let $r$ be the root of $b$. Let $W$ be a witness certificate. Let $key$ be a key such that $key = (r, \broker, \_)$.
If $D[\broker, r, W]$ event happens, we have eventually $key \in batches$ at $\server$.
\end{lemma}
\begin{proof}
When an event $D[\broker, r, W]$ happens at $\server$ (\cref{line:sdeliverfromconsensus}), $\server$ computes $key$ such that $key = (r, \broker, \_)$ and $\server$ checks that $W$ is a witness certificate for $key$
(\cref{line:sdefinekeyindeliverstb,line:switnesssignaturecheck}).
Moreover, $\server$ checks if $r \in expected\_batch[\broker]$ (\cref{line:sexpectedbatchinroot}).
We remark that, according to \cref{lemma:rexcpectedkeiinbatches}, if $r \in expected\_batch[\broker]$, then $key \in batches$ at $\server$ and the lemma is satisfied. 

If this is not the case, $\server$ sends to all servers a $[RequestBatch, key]$ message, we say that this happens at time $t$ (\cref{line:srequestbatch}).
We remark that $W$ is a witness certificate thus we have $f + 1$ servers that have signed a $[Witness, key]$ message in which $f$ is the maximum number of Byzantine servers in the system (\cref{def:witness}). We also underline that there is at least one correct server $\server'$ that signs the $[Witness, key]$ message and, according to \cref{lemma:rexcpectedkeiinbatches}, $key \in batches$ at $\server'$ at least until $\server'$ signs the message.
To summarize, we have that $\server'$ signs $[Witness, key]$ at time $t''$ and that $key \in batches$ at time $t'$ at $\server'$ with $t' \leq t''$.

We notice that when $\server$ sends the $[RequestBatch, key]$ message at $t$, the witness certificate $W$ already exists, this could happen only if $f + 1$ servers already signed a $[Witness, key]$ message. So we have that $t'' < t$.

We underline that $\server'$ removes $key$ from $batches$ only if all correct servers have invoked the $completion\_batch(key)$ procedure (\cref{lemma:notrenovekeyfrombatchesuntilecompletionbatch}). Also, we notice that according to \cref{lemma:keyinbatcheswhencompletioninvokes} a correct server can invoke the $completion\_batch()$ procedure only if $key \in batches$ at $\server$. This implies that we can have two possible scenarios:
\begin{itemize}
    \item If $key \in batches$ at $\server'$ when $\server'$ receives the $[RequestBatch, key]$ message from $\server$, we have that eventually $key \in batches$ at $\server$ (\cref{lemma:keyeventuallinbatchesifyouask}).
    \item If $key \notin batches$ at $\server'$ when $\server'$ receives the $[RequestBatch, key]$ message from $\server$, we have that $\server'$ removes $key$ from $batches$. This implies that all correct servers, including $\server$, have already invoked the $completion\_batch(key)$ which is only possible if $key \in batches$. So we have $key \in batches$ at $\server$.
\end{itemize}
Thus the lemma is proven.
\end{proof}

\begin{lemma}
\label{lemma:idsdeterministsorted}
Let $\server$ be a correct server. Let $key$ be a key such that $key \in batches$ at $\server$. Let $ids$ be the set of identifiers such that $ids = batches[key].ids$.
We have that $ids$ is deterministically sorted at $\server$. 
\end{lemma}
\begin{proof}
We start by noting that $\server$ adds $key$ to $batches$ only by executing \cref{line:sbatchesupdate,line:ssetbatchafterskforthetotality}. Immediately before doing this, $\server$ orders the $ids$ set (\cref{line:serverssortid,line:serverssortid1}). Thus, the lemma is proven.
\end{proof}

\begin{lemma}
\label{lemma:delivermessageinordergivenakey}
Let $\server$ and $\server'$ be two correct servers. Let $m$ be a message with $seq$ its associated sequence number broadcast by a client $\client$ with $id = dir[\client]$. Let $b$ be a batch such that $b = [(id, seq, m)]$, $r$ a root with $r = root(merkle\_tree(b))$, and $key$ be a key such that $key = (r, \_, \_)$.
Let $\server$ and $\server'$ invoke $completion\_batch(key)$ procedure.
Let $d$ be the elements of $b$ delivered by $\server$ and let $d'$ the elements of $b$ delivered by $\server'$, such that $d \subset b \And d' \subset b $.
Let $d[i]$ be the $i$-th element of $d$ and $d'[i]$ the $i$-th element of $d'$.
We have that $d = d'$ and $d[i] = d'[i] \forall i \in d.length $.
\end{lemma}
\begin{proof}
Let $completion\_batch()_{i}$ be the $i$-th time that $\server$ invokes the $completion\_batch()$ procedure and let $completion\_batch()_{j}$ the $j$-th time that $\server'$ invokes $completion\_batch()$ procedure. If $\server$ invokes $completion\_batch(key)_{i}$ and $\server'$ invokes $completion\_batch(key')_{j}$ with $i = j$, then $key = key'$ (\cref{lemma:callcomplitionbatchinthesameorder}).

We underline that when $\server$ invokes the $completion\_batch(key)$ procedure, $\server$ chooses whether to deliver an element of $b$ based only on the value of $last\_message[\client]$ (\cref{line:sifiamaldradyupdate,line:sconditionequalmessage,line:sseqbiggerthenlastmessage}). We remark that upon initialization $last\_message[\client]$ is empty (\cref{line:slastmessageinit}) and $\server$ adds elements to it only by executing \cref{line:supdatelastmessage} in the procedure $completion\_batch()$. 
To summarize, $\server$ and $\server'$ invoke the $completion\_batch(keys)$ procedure for the same set of $keys$ in the same order. Moreover, $\server$ and $\server'$ process the elements of $b$ in the same order (\cref{line:scycleofallids}) because $ids$ is sorted in a deterministic and equal way at $\server$ and $\server'$ (\cref{lemma:idsdeterministsorted}).

We underline that upon $\server$ and $\server'$ invoking the $completion\_batch(key)$ procedure we have $last\_message[\client]$ at $\server$ is equal to $last\_message[\client]$ at $\server'$, and we remark that $\server$ and $\server'$ process elements of $b$ in the same order. Thus $\server$ and $\server'$ modify the $last\_message[\client]$ in the same way. This implies that if $\server$ does not deliver an element of $b$, $\server'$ will also not deliver it. 

To conclude, we prove that $\server$ and $\server'$ deliver the same elements of $b$ in the same order.
So let $d$ the elements of $b$ delivered by $\server$ and let $d'$ the elements of $b$ delivered by $\server'$, such that $d \subset b$ and $d' \subset b$.
Let $d[i]$ be the $i$-th element of $d$ and $d'[i]$ the $i$-th element of $d'$.
We have that $d = d'$ and $d[i] = d'[i] \forall i \in d.length$.
Thus the lemma is proven.
\end{proof}

\begin{lemma}
\label{lemma:keyinpoolinbatchescompletionbatchkey}
Let $\server$ be a correct server and let $key$ be a key. If $key \in pool$ such that $key = pool.peek()$ and $key \in batches$, then $\server$ invokes the $completion\_batch(key)$ procedure.
\end{lemma}
\begin{proof}
We start by noting that $\server$ invokes the $completion\_batch(key)$ procedure only by executing \cref{line:scallprocedurecompletion}. Upon doing this, $\server$ checks that $key$ is the first element of $pool$ and $key \in batches$ (\cref{line:skeyinpoolcheck}).
We underline that if $\server$ removes $key$ from $batches$ this implies that all correct servers and in particular $\server$ had already invoked $completion\_batch(key)$ (\cref{lemma:notrenovekeyfrombatchesuntilecompletionbatch}).
We also remark that if $\server$ removes $key$ from $pool$ immediately after $\server$ invokes the $completion\_batch(key)$.
Thus, the lemma is proven.
\end{proof}

\begin{theorem}[Agreement]\label{thm:agreement}
\system satisfies agreement.
\end{theorem}
\begin{proof}
Let $\server$ be a correct server. 
Let $m$ be a message with $seq$ its associated sequence number broadcast by a client $\client$ with $id = \dir[\client]$. Let $b$ be a batch such that $b = [(id, seq, m)]$, $r$ a root such that $r = root(merkle\_tree(b))$, and $key$ be a key such that $key = (r, \_, \_)$.
Let $\server$ deliver $(id, seq, m)$.

We underline that $\server$ can only deliver messages by executing \cref{line:sdelivermessage} in the $completion\_batch(key)$ procedure. Let $W$ be a witness certificate. According to \cref{lemma:completiononlyifeceventhappens}, if $\server$ invokes $completion\_batch(key)$ then the event $D[\_, r, W]$ happens at $\server$.
By the agreement property of \stobprefix, the event $D[\_, r, W]$ eventually happens at every correct server. Let $\server'$ be a correct server. We remark that when the event $D[\_, r, W]$ happens at $\server'$ then $key \in pool$ (\cref{lemma:keyinpooluponddeliver}) and that eventually $key \in batches$ (\cref{lemma:ifeeventeventuallykeyinbatches}) at $\server'$.
If $key \in pool$ and $key \in batches$, $\server'$ eventually invokes the $completion\_batch(key)$ procedure (\cref{lemma:keyinpoolinbatchescompletionbatchkey}).

Let $completion\_batch()_{i}$ be the $i$-th time that $\server$ invokes $completion\_batch()$ procedure, let $completion\_batch()_{j}$ the $j$-th time that $\server$ invokes $completion\_batch()$ procedure. 
If $\server$ invokes $completion\_batch(key)_{i}$ and $\server'$ invokes $completion\_batch(key')_{j}$ with $i = j$ then $key = key'$.
This implies that $\server$ and $\server'$ invoke $completion\_batch()$ for the same set of $keys$ in the same order (\cref{lemma:callcomplitionbatchinthesameorder}).
Let $d$ be the subset of $b$ delivered by $\server$ when it invokes $completion\_batch(key)$, let $d'$ the subset of $b$ delivered by $\server'$ when it invokes $completion\_batch(key)$.
Let $(id, seq, m) \subset d$, then we have that $(id, seq, m) \subset d'$ because $d = d'$ according to \cref{lemma:delivermessageinordergivenakey}. Thus the theorem is proven.
\end{proof}

\subsubsection{Total Order}

In this section, we prove that \system satisfies total order.

\begin{theorem}[Total order]\label{thm:totalorder}
\system satisfies total order.
\end{theorem}
\begin{proof}
Let $\server$ and $\server'$ be two correct servers. Let $m$ be a message with $seq$ its associated sequence number broadcast by a client $\client$ with $id = dir[\client]$. Let $b$ be a batch such that $b = [(id, seq, m)]$, $r$ be a root such that $r = root(merkle\_tree(b))$, and let $key$ be a key such that $key = (r, \_, \_)$.

Let $completion\_batch()_{i}$ be the $i$-th time that $\server$ invokes $completion\_batch()$ procedure, let $completion\_batch()_{j}$ the $j$-th time that $\server$ invokes $completion\_batch()$ procedure. 
If $\server$ invokes $completion\_batch(key)_{i}$ and $\server'$ invokes $completion\_batch(key')_{j}$ with $i = j$ then $key = key'$.
This implies that $\server$ and $\server'$ invoke $completion\_batch()$ for the same set of $keys$ in the same order. (\cref{lemma:callcomplitionbatchinthesameorder}).
Let $d \subset b$ the elements delivered by $\server$ if $\server$ invokes $completion\_batch(key)$ procedure. Let $d' \subset b$ be the elements delivered by $\server'$ if $\server'$ invokes $completion\_batch(key)$ procedure. Let $d[i]$ be the $i$-th element of $d$ and $d'[i]$ be the $i$-th element of $d'$. We have that $d = d'$ and $d[i] =d'[i]$ (\cref{lemma:delivermessageinordergivenakey}).

To summarize, the agreement property of \system ensures that if $\server$ delivers a $(id, seq, m)$ then every correct $\server$ eventually delivers it.
From the above observations:
\begin{itemize}
    \item All correct servers invoke the $completion\_batch()$ procedure in the same order for the same set of $keys$ (\cref{lemma:callcomplitionbatchinthesameorder}).
    \item All correct servers, upon invoking the $completion\_batch(key)$ procedure, deliver the same subsets of $b$ in the same order (\cref{lemma:delivermessageinordergivenakey}).
\end{itemize}
This implies that all correct servers deliver the same sets of messages in the same order.

To conclude, if $\server$ delivers $(id, seq, m)$ before $(id', seq', m')$. We have that all the other correct servers that have delivered $(id', seq', m')$ must have previously delivered $(id, seq, m)$. Thus, the theorem is proven.
\end{proof}

\subsubsection{Validity}

In this section, we prove that \system satisfies validity.

\begin{notation}[$\oplus$ operation]
Let $t$ and $t'$ be tuples. We define the $\oplus$ operation such that
\begin{gather*}
    x = t 
    \; \oplus \;
    x = t'
\end{gather*}
in which the variable $x$ could be set only to $t$ or $t'$ and it is not possible to change the assigned value to $x$.
\end{notation}

\begin{definition}[Root of message]\label{def:rootofm}
Let $m$ be a message with $seq$ its associated sequence number broadcast by a client $\client$ with $id = \dir[\client]$. Let $b$ be a batch such that $(id, seq, m) \in b$ and let $r$ be the root of $b$. We say that $r$ is root of $m$ which we note as $r[m]$.
\end{definition}

\begin{definition}[Completion certificate]\label{def:completioncertificate}
We define $C[m]$ a completion certificate for a message $m$, broadcast by a client $\client$ with $seq$ its associated sequence number, if $f + 1$ servers have signed $[Completion, r[m], counter]$ in which $counter$ is an integer and $r[m]$ is the root of $m$. We call $C[m]$ completion certificate for $m$.
\end{definition}

\begin{definition}[Submission of message]\label{def:submission}
Let $\client$ be a correct client. Let $m$ the message broadcast by $\client$ and let $seq$ the sequence number associated with $m$. Let $id = \dirin[\client]$ be the identifier of $\client$. Let $s$ be the signature of $\client$ for the following message $[Message, m, seq]$. Let $C[\_]$ a generic completion certificate such that $C[\_].counter \geq seq$.
We define Submission of $m$ ($sub[m]$) the tuple $sub[m] = (id, seq, m, s, C[\_])$.
\end{definition}

\begin{definition}[Core of message]\label{def:coreofmessage}
Let $\client$ be a correct client and let $sub[m]$ be a submission such that $sub[m] = (id, seq, m, s, C[\_])$.
Let $max\_seq$ be an integer such that $max\_seq \geq seq$. Let $r[m]$ be a root of $m$ for $(id, max\_seq, m)$. Let $ms$ be the multi-signature of $\client$ for the message $[Reduction, r[m]]$.
We define $c[m]$ as follows:
\begin{gather*}
    c[m] = (id, seq, m, s)
    \; \oplus \;
    c[m] = (id, max\_seq, m, ms).
\end{gather*}
\end{definition}

\begin{definition}[Existence of witness certificate]\label{notation:existawitnesscertificatatserver}
Let $\server$ be a correct server. Let $\broker$ be a broker, and let $r$ be a root. Let $key$ be a key such that $key = (r, \broker, \_)$.
Let $W[key]$ be a witness certificate for $key$.
We define that $\exists W[key]$ at $\server$ when an event $D[r, \broker, W[key]]$ happens.
\end{definition}

\begin{lemma}
\label{lemma:deliversigncompletion}
Let $\server$ be a correct server. Let $m$ be a message from client $\client$. If $\server$ delivers $m$ from $\client$, $\server$ signs $[Completion, r[m], \_]$.
\end{lemma}
\begin{proof}
We start by noting that $\server$ delivers $m$ only by executing \cref{line:sdelivermessage}. Immediately before doing this, $\server$ adds $m$ to $completion\_payloads$ (\cref{line:saddmessagetocompletionpayloads}). Moreover, $\server$ computes $r$ based on the element of $completion\_payloads$ (\cref{line:scomputecompletionleaves,line:scomputecompletiontree,line:scomputecompletionroot}). This implies that $r = r[m]$.
Finally, $\server$ signs $[Completion, r[m], \_]$ (\cref{line:ssigncompletionmessage}). Thus the lemma is proven.
\end{proof}

\begin{lemma}
\label{lemma:lastmessagedelievred}
Let $\server$ be a correct server. Let $\client$ be a client with $id = dir[\client]$. Let $m$ be the last message from $\client$ delivered from $\server$. We have that at every point in time $last\_message[id] = m$ at $\server$.
\end{lemma}
\begin{proof}
We start by noting that $last\_message[id]$ is empty upon initialization at $\server$. Moreover, $\server$ updates it only by executing \cref{line:supdatelastmessage}. We remark that in an atomic procedure $\server$ executes $last\_message[id] = m$ and $< Deliver m >$ (\cref{line:supdatelastmessage,line:sdelivermessage}).
This implies that $last\_message[id]$ is equal to the last message delivered at every point in time. Thus the lemma is proven.
\end{proof}

\begin{lemma}
\label{lemma:ifsigncomplsodelivered}
Let $\server$ be a correct server. Let $m$ be the message from client $\client$. If $\server$ signs $[Completion, r[m], \_]$, then $\server$ delivers $m$ from $\client$.
\end{lemma}
\begin{proof}
We start by noting that $\server$ signs $[Completion, r[m], \_]$ only by executing \cref{line:ssigncompletionmessage}. We underline that $\server$ signs $[Completion, r[m'], \_] \forall m' \in completion\_leaves$ (\cref{line:scomputecompletionleaves,line:scomputecompletiontree,line:scomputecompletionroot}).
Moreover, $\server$ adds to $completion\_leaves$ all the elements $(seq, m) \in completion\_payloads$ (\cref{line:scomputecompletionleaves}) in which $seq$ is the sequence number associated with $m$. We underline that $completion\_payloads$ is initialized empty (\cref{line:scompletionpayloads}) and $\server$ updates it executing \cref{line:scompletionpayloadupdate} or \ref{line:saddmessagetocompletionpayloads}.
In both cases $\server$ adds to $completion\_payloads$ the value of $last\_message[id]$. Let $m$ be the last message from $\client$ delivered by $\server$. According to \cref{lemma:lastmessagedelievred}, $last\_message[id] = m$ which implies that $\server$ adds to $completion\_payloads$ only message that had already delivered or that is delivering. Thus, the lemma is proven.
\end{proof}

\begin{lemma}
\label{lemma:minsubmissionsatclient}
Let $\client$ be a correct client. Let $C[m]$ be a completion certificate. If $\client$ broadcasts a message $m$ with $seq$ its associated sequence number then $m \in submissions \lor \exists C[m]$ at $\client$.
\end{lemma}
\begin{proof}
We start by noting that upon initialization $submissions$ is empty at $\client$ (\cref{line:csubmissionsinit}).
Moreover, $\client$ broadcasts $m$ only by executing \cref{line:cbrodcastmessage}. Immediately after doing this, $\client$ adds $m$ to $submissions$ (\cref{line:cupdatesumissions}).
We underline that $\client$ removes elements from $submissions$ only by executing \cref{line:sremoveelementfromsumissions}. Before doing that, $\client$ verifies that $C[m]$ is a completion certificate for $m$ by checking that $f + 1$ servers sign $[Completion, r, \_ ]$ (\cref{line:cverifyplurality}) and checking that $r = r[m]$ (\cref{line:cproverrdim}).
This implies that if $\client$ had broadcast $m$ with $seq$ its associated sequence number and $m \notin submissions$, it is because a completion certificate $C[m]$ exists at $\client$. Thus the lemma is proven.
\end{proof}

\begin{lemma}
\label{lemma:onecorrectserverdeliversifexistacompletioncertificate}
Let $\client$ be a correct client, let $m$ be a message and let $seq$ be a sequence number. If there exists a completion certificate $C[m]$, then at least one correct server $\server$ has delivered $m$ from $\client$.
\end{lemma}
\begin{proof}
If a completion certificate $C[m]$ exists, according to \cref{def:completioncertificate}, $f + 1$ servers have signed $[Completion, r[m], \_ ]$.
We remark that at most $f$ servers in the system could be Byzantine. Thus, at least one correct server $\server$ had signed $[Completion, r[m], \_ ]$.
According to \cref{lemma:ifsigncomplsodelivered}, if $\server$ signs $[Completion, r[m], \_ ]$, then $\server$ has delivered $m$. Thus, the lemma is proven.
\end{proof}

\begin{lemma}
\label{lemma:poolwillbeempty}
Let $\broker$ be a correct broker. If $pool \neq \varnothing$, we eventually have that $pool = \varnothing$ at $\broker$.
\end{lemma}
\begin{proof}
We start by noting that upon initialization, $pool$ is empty at $\broker$ (\cref{line:bpoolinit}) and $collecting$ is false (\cref{line:bcollectinginit}). We underline that $\broker$ adds elements to $pool$ only by executing \cref{line:bpoolupdate}.

If $pool \neq \varnothing$, we have two possible scenarios: $collecting$ is either set to false or to true.
In the first case, $\broker$ will update $collecting$ to true and sets a timer $Flush$ (\cref{line:bpoolcollectingcheck,line:bcollectingtrue,line:btimerset}). Eventually the timer expires (\cref{line:btimerring}), at this point $\broker$ sets $collecting = false$ and $pool = \varnothing$ (\cref{line:bcollectingfalse,line:bemptypool}).
Instead, if $collecting = true$ this implies that $\broker$ had already sets the timer $Flush$, because $\broker$ only sets $collecting = true$ immediately before sets $Flash$ (\cref{line:bcollectingtrue}). Moreover, the timer has not yet expired otherwise $collecting$ will be equal to $false$ (\cref{line:bcollectingfalse}). We underline that eventually, the timer will expire and $\broker$ can as before reset $pool = \varnothing$ and update $collecting = true$. Thus the lemma is proven.
\end{proof}

\begin{lemma}
\label{lemma:subinsubmissions}
Let $\broker$ be a correct broker. Let $m$ be a message and let $sub[m]$ be a submission of $m$. If $\broker$ delivers $sub[m]$ from a client $\client$ then $sub[m] \in submissions$ at $\broker$.
\end{lemma}
\begin{proof}
We start by remarking that according to \cref{def:submission} $sub[m]$ is defined such that $sub[m] = (id, seq, m, s, C[\_])$.
We also notice that $\broker$ can deliver a $sub[m]$ only by executing \cref{line:bdeliversubmit}. If this happens, $\broker$ first checks that $sub[m]$ is a submission for $m$ by checking that $id = \dirin[\client]$ (\cref{line:bimportsubmit}), $s = \client.sign([Message, m, seq])$ (\cref{line:bverifyclientsignature}), $C[\_]$ is a completion certificate (\cref{line:bverifyclientsignature}), and $C[\_].counter \geq seq$ (\cref{line:bverifyclientsignature}).
If this is the case, $\broker$ adds $sub[m]$ to $pending[id]$ (\cref{line:bpushback}).

We underline that upon initialization both $pending$ and $pool$ are empty at $\broker$ (\cref{line:bpandingdefinition,line:bpoolinit}). We notice that eventually $\broker$ will set $pool[id] = pending[id]$ if $id \notin pool$, $\broker$ does this immediately (\cref{line:bexistsinpanding,line:bpandingupdate,line:bpoolupdate}). Otherwise, according to \cref{lemma:poolwillbeempty} we know that eventually, $pool$ will be empty again so the condition $id \notin pool$ will be satisfied. Thus, $\broker$ can eventually set $pool[id] = pending[id]$. 

According to the above observation, we have now that $pool[id]$ is not empty at $\broker$. We analyze the two following scenarios:
\begin{itemize}
    \item If $collecting = false$: then $\broker$ sets a $Flush$ timer that eventually expires and $\broker$ can finally add $pool[id]$ to $submissions$ (\cref{line:bpoolcollectingcheck,line:btimerset,line:btimerring,line:bsubmissionsequalpool}).
    \item If $collecting = true$: this implies that $\broker$ has already set the $Flush$ that has not yet expired. We remark that eventually the timer $Flush$ will expire and the $\broker$ can, as before, add $pool[id]$ to $submissions$ (\cref{line:bsubmissionsequalpool}).
\end{itemize}

To conclude, we have that $\broker$ stores $sub[m]$ in $pending$, then $pending$ in $pool$ and finally $pool$ in $submissions$, so at the end $sub[m] \in submissions$.
Thus the lemma is proven.
\end{proof}

\begin{lemma}
\label{lemma:subminsumissionsatbroker}
Let $\client$ be a correct. Let $m$ be a message and let $m \in submissions$ at $\client$. Let $sub[m]$ a submission.
We eventually have $sub[m] \in submissions$ at some correct broker $\broker$.
\end{lemma}
\begin{proof}
We start by noting that upon initialization $submissions$ is empty at $\client$ (\cref{line:csubmissionsinit}). Moreover, $\client$ updates it only by executing \cref{line:cupdatesumissions}, immediately after $\client$ invokes atomically the $submit(m)$ procedure (\cref{line:csubmitinvoke}).

The first time that $\client$ invokes the $submit(m)$ procedure, $\client$ randomly selects a broker $\broker$ (\cref{line:csubmissioncheck}) and deletes $\broker$ from the list of future available brokers (\cref{line:caddinitsequncenumbertosub}). Moreover, $\client$ computes $sub[m]$ as follows $sub[m] = (id, seq, m, s, C[\_])$ with $id = \dirin[\client]$ and $s = sign([Message, m, seq])$ (\cref{line:cdirexport,line:csignaturecompute}).
Thus $\client$ sends to $\broker$ the submission $sub[m]$ (\cref{line:csendsubmit}).
If $\broker$ is correct, according to \cref{lemma:subinsubmissions}, $sub[m] \in submissions$ at $\broker$ which proves the lemma.
If $\broker$ is not correct, $\client$ will send the $sub[m]$ to another broker. Indeed, every time that $\client$ invokes the procedure $submit(m)$ for the same message $m$, $\client$ selects a new broker from the list of available ones, updates the list and sets a timer $Submit$ (\cref{line:csubmissioncheck,line:caddinitsequncenumbertosub,line:ctimerset}). Eventually the timer $Submit$ expires and $\client$ invokes the $submit(m)$ procedure again (\cref{line:ctimerring,line:cinvokesubmitring}). For the above considerations, we deduce that the maximum number of times that $\client$ can invoke the $submit(m)$ procedure for the same $m$ is equal to total number of brokers in the system.

To summarize, we remark that for assumption there is at least one correct broker $\broker'$ in the system. This implies that $\client$ eventually sends $sub[m]$ to $\broker'$. Moreover, according to \cref{lemma:subinsubmissions} we have $sub[m] \in submissions$ at $\broker'$ that is correct, thus the lemma is proven.
\end{proof}

\begin{lemma}
\label{lemma:newroot}
Let $\broker$ be a correct broker, let $r[m]$ be a root of $m$, let $r'$ be a root. If $\broker$ insert $r'$ in $batches$ such that $batches[r'] = batches[r[m]]$ then $r' = r[m]$.
\end{lemma}
\begin{proof}
We start by noting that $\broker$ inserts a new root $r'$ to $batches$ and initializes it with another value of $batches$ only by executing \cref{line:brootcostruction}.
Let $b$ be a batch and $r$ be the root of $b$ such that $(id, seq, m) \in b$ and $r = r[m]$. We underline that $\forall id \in b$, $\broker$ computes the following: batch $b' = [(id, \_, m)]$, tree $t' = merkle\_tree(b')$, and root $r' = merkle\_root(t')$ (\cref{line:bleavesdefinition,line:baddsleavesfromstraggler,line:baddsleavesfrompayloads,line:btreecostruction1,line:brootcostruction1}).
This implies that, by \cref{def:rootofm}, $r' = r[m]$ thus the lemma is proven.
\end{proof}

\begin{lemma}
\label{lemma:coreinbatchsattimet}
Let $\broker$ be a correct broker and let $m$ be a message. Let $sub[m]$ be a submission, $c[m]$ be a core for $m$ and $r[m]$ be a root of $m$. Let $t$ be the time in which $\broker$ adds a new batch $b$ to $batches$. If $sub[m] \in submissions$, then $r[m] \in batches$ and $c[m] \in batches[r[m]]$ at $\broker$ at $t$.
\end{lemma}
\begin{proof}
We start by noting that upon initialization $submissions$ is empty (\cref{line:binitsubmissions}) and $\broker$ adds elements to it only by executing \cref{line:bsubmissionsequalpool}. Immediately after doing this, $\broker$ computes the $max\_seq$ such that $max(sub[m].seq) \forall sub[m] \in submissions$ (\cref{line:bcomputemaxseq}).

We remark that, according to \cref{def:submission}, $sub[m]$ is defined such that $sub[m] = (id, seq, m, s, C[\_])$.
Moreover, $\broker$ after computes $max\_seq$ also calculates a root $r$ as follows: batch $b = [(id, max\_seq, m)] \forall sub[m] \in submissions$, tree $t = merkle\_tree(b)$, and root $r = merkle\_root(t)$ (\cref{line:bleavescostruction,line:btreecostruction,line:brootcostruction}).
This implies that $r = r[m]$.

We also remark that according to \cref{def:coreofmessage}, $c[m]$ is defined such that $c[m] = (id, seq, m, s)$.
We underline that $\broker$ computes $c[m]$. Indeed, $\forall sub[m] \in submissions$ $\broker$ updates the variables $payloads$, $stragglers$, $signatures$ such that $payloads[id] = (max\_seq, sub[m].m)$ (\cref{line:bpayloaddefine}), $stragglers[id] = sub[m].seq$ (\cref{line:bstragglersupdate}), and $signatures[id] = sub[m].seq$ (\cref{line:bsignaturedefine}).
We remark that upon initialization, $batches$ is empty at $\broker$ (\cref{line:bemptybatches}), moreover, $\broker$ adds roots to $batches$ and creates a new batch (\cref{line:bbatchesupdate}).

Let $t$ be the moment in which $\broker$ adds $r[m]$ to $batches$. We notice also that $\broker$ at the same time $t$ adds to $batches[r[m]]$ the variables $payloads$, $stragglers$, $signatures$ and we remark that we $\broker$ can build $c[m]$ as follows:
\begin{gather*}
    id = id \in payloads \\
    seq = stragglers[id] \\
    m = payloads[id] \\
    s = signatures[id] \\
    c[m] = (id, seq, m, s) 
\end{gather*}
(\cref{line:bbatchesupdate}). 
To conclude, we have that $c[m] \in batches[r[m]]$ at time $t$ at $\broker$, thus the lemma is proven.
\end{proof}

\begin{lemma}
\label{lemma:coreinbatchesforever}
Let $\broker$ be a correct broker. Let $r[m]$ be a root of $m$. Let $c[m]$ a core. Let $t$ be the time in which $\broker$ creates a new batch $b$. If $\broker$ adds $r[m] \in batches$ at $t$, then $c[m] \in batches[r[m]]$.
\end{lemma}
\begin{proof}
We start by noting that $\broker$ can create a new batch and add $r[m]$ to $batches$ at the same time $t$ only by executing \cref{line:bbatchesupdate}. We underline that, according to \cref{lemma:coreinbatchsattimet}, we have $c[m] \in batches[r[m]]$ at $t$ at $\broker$.

We underline that $\broker$ can modify $batches[r[m]]$ only by executing \cref{line:bupdatereduction,line:bupdateremoveid,line:bstragglersremoveid}. This happens only if $\broker$ delivers from client $\client$ a $[Reduction, r, ms]$ message, in which $r$ is a root and $ms$ is a multi-signature.
Upon receiving this message, $\broker$ checks if $r \in batches$, $batches[r]$ is reduced, and if $id = \dirin[\client] \in batches[r]$ (\cref{line:bcheckreducingid}).
We first remark that if $id \in batches[r]$, let $(id, seq, m) \in batches[r]$ we have that $r = r[m]$ according to \cref{def:rootofm}.
After the root verification, $\broker$ checks that $ms$ is a valid $\client$ multi-signature for $[Reduction, r[m]]$. Moreover, $\broker$ adds $ms$ to $reductions$ and removes $id$ from $signature$ and $straggler$ (\cref{line:bupdatereduction,line:bupdateremoveid,line:bstragglersremoveid}).
Despite this modification to $batches[r[m]]$, $c[m]$ is still in $batches[m]$. Indeed, $\broker$ can compute $c[m]$ as follows:
\begin{gather*}
    id = id \in payloads \\
    max\_seq = payloads[id] \\
    m = payloads[id] \\
    ms = reductions[i] \\
    c[m] = (id, max\_seq, m, ms).
\end{gather*}
This implies that if $r[m]$ is inserted in $batches$ when the batch is created, then $c[m] \in batches[r[m]]$ at every point in time, until of course $\broker$ removes $r[m]$ from $batches$.

We also remark that if $id \notin signatures$ or $id \notin stragglers$, we cannot build $c[m] = (id, seq, m, s)$ anymore. Moreover, we argue that if $id$ was in $signatures$ and then $\broker$ removes $id$ from $signatures$ we have $id \in reductions$. This is because $\broker$ removes all $id$ from $signatures$ only by executing \cref{line:bupdateremoveid} and immediately before adding $id \in reductions$ (\cref{line:bupdatereduction}). The same reasoning holds for $stragglers$ (\cref{line:bstragglersremoveid}).

To conclude we have that at every point in time that $c[m] \in batches[r[m]]$ could be
\begin{gather*}
    c[m] = (id, seq, m, s)
    \; \oplus \;
    c[m] = (id, max\_seq, m, ms)
\end{gather*}
which satisfies the \cref{def:coreofmessage} of $c[m]$, thus the lemma is proven. 
\end{proof}

\begin{lemma}
\label{lemma:cinbatchesevenwithreplay}
Let $\broker$ be a correct broker. Let $r[m]$ be a root of $m$ and $c[m]$ be a core of $m$. Let $r'$ be a root. If $\broker$ insert $r'$ in $batches$ such that $batches[r'] = batches[r[m]]$, then $c[m] \in batches[r'[m]]$.
\end{lemma}
\begin{proof}
We start by noting that according to \cref{lemma:newroot}, $r' = r'[m]$.
We notice that $\broker$ adds a new root $r'$ to $batches$ without creating a new batch only by executing \cref{line:baddnewroottobatches}. We underline that $\broker$ inserts $r'$ to $batches$ and initializes it with the value $batches[r[m]]$. Moreover, $\broker$ does this only when the timer $[Reduce, r]$ expires (\cref{line:breducering}), the timer is set by executing \cref{line:btimersetreducing} immediately after that $\broker$ adds $r[m]$ to $batches$ creating a new batch.
We underline that according to \cref{lemma:coreinbatchesforever}, we have $c[m] \in batches[r[m]]$, because for what we prove above we are in the case in which $r[m]$ is added to $batches$ and a new batch is creating in the same time $t$. To summarize we have:
\begin{gather*}
    c[m] \in batches[r[m]] \\
    batches[r'] = batches[r[m]] \\
    r' = r'[m]
\end{gather*}
which implies that $c[m] \in batches[r'[m]]$.
Thus the lemma is proven.
\end{proof}

\begin{lemma}
\label{lemma:cminbatchesnomatterthecase}
Let $\broker$ be a correct broker. Let $r[m]$ be a root of $m$ and let $c[m]$ be a core of $m$. If $r[m] \in batches$, then $c[m] \in batches[r[m]]$ at $\broker$.
\end{lemma}
\begin{proof}
We start by noting that $\broker$ adds elements to $batches$ only by executing \cref{line:bbatchesupdate,line:baddnewroottobatches}.
Let t be the time in which $\broker$ creates a new batch. In the first case, $\broker$ also adds $r[m]$ to $batches$ at time $t$. According to \cref{lemma:coreinbatchesforever} we have that $c[m] \in batches[r[m]]$ at $\broker$.
In the second case, we have that $\broker$ adds a $r[m]$ to $batches$ and initializes with a value of an existing batch. According to \cref{lemma:cinbatchesevenwithreplay}, we still have that $c[m] \in batches[r[m]]$ at $\broker$.
This implies that upon $\broker$ adding a new $r[m]$ to $batches$ the lemma holds. We remark that, according to both \cref{lemma:cinbatchesevenwithreplay,lemma:cinbatchesevenwithreplay}, $c[m] \in batches[r[m]]$ in every moment even if $\broker$ modifies $batches[r[m]]$. Thus the lemma is proven.
\end{proof}

\begin{lemma}
\label{lemma:brokersendssubreductiontoserver}
Let $\broker$ be a correct broker. Let $m$ be a message. Let $r[m]$ be a root of $m$ and let $c[m]$ be a core of $m$ such that:
\begin{gather*}
    c[m] = (id, seq, m, s) 
    \; \oplus \;
    c[m] = (id, max\_seq, m, ms).
\end{gather*}
If $r[m] \in batches$ at $\broker$, then $\broker$ sends to all servers:
\begin{gather*}
    (id, seq, m)
    \; \oplus \;
    (id, max\_seq, m).
\end{gather*}
\end{lemma}
\begin{proof}
We start by noting that, according to \cref{lemma:cminbatchesnomatterthecase}, if $r[m] \in batches$ at $\broker$ then $c[m] \in batches[r[m]]$ at $\broker$.
Moreover, immediately after, $\broker$ adds $r[m]$ to $batches$ (\cref{line:bbatchesupdate}) then sets a timer $[Reduce, r[m]]$ (\cref{line:btimersetreducing}). When the timer expires (\cref{line:breducering}), $\broker$ computes $r'$, with $r' = r'[m]$ (\cref{lemma:newroot}) and adds $r'[m]$ to $batches$ such that $batches[r'[m]] = batches[r[m]]$ (\cref{line:baddnewroottobatches}).
Finally $\broker$ sends $(id, max\_seq, m)$ and in case $id \in stragglers$ also $seq$ to all servers (\cref{line:bcompressids,line:bpayloads,line:bsendbatch}). We underline that for construction $id = c[m].id$ and $seq = c[m].seq$, thus the lemma is proven.
\end{proof}

\begin{lemma}
\label{lemma:batchiswitnessatbroker}
Let $\broker$ be a correct broker. Let $r[m]$ be a root of message $m$ and let $c[m]$ be the core of $m$. Let $r[m] \in batches$ at $\broker$. Upon $\broker$ sending $c[m]$ to all servers, we have that $batches[r[m]]$ is $Witnessing$ at $\broker$.
\end{lemma}
\begin{proof}
We start by noting that for \cref{def:coreofmessage} $c[m]$ is defined as follows:
\begin{gather*}
    c[m] = (id, seq, m, s)
    \; \oplus \;
    c[m] = (id, max\_seq, m, ms).
\end{gather*}
We underline that, according to \cref{lemma:brokersendssubreductiontoserver}, if $r[m] \in batches$ then $\broker$ sends to all servers:
\begin{gather*}
    (id, seq, m)
    \; \oplus \;
    (id, max\_seq, m).
\end{gather*}
We notice that $\broker$ does this only by executing \cref{line:bsendbatch}. Immediately after, $\broker$ sets $Witnessing$ the state of $baches[r[m]]$ (\cref{line:bbatchinwitnessing}).
We underline that $\broker$, before sending $s$ or $ms$ (\cref{line:bsendingsignatureormultisignature}) in order to complete the sending of $c[m]$, checks that $batches[r[m]]$ is $Witnessing$ (\cref{line:bbatchacquiredgetbatch}). We remark that $\broker$ immediately after sending $c[m]$ does not change the state of $batches[r[m]]$. To conclude, when $\broker$ sends to all servers $c[m]$ we have that $batches[r[m]]$ is $Witnessing$ at $\broker$. Thus the lemma is proven.
\end{proof}

\begin{lemma}
\label{lemma:ifkeyinbatchesafteracceptexistwitness}
Let $\server$ be a correct server. Let $key$ be a key. Let $W[key]$ be a witness certificate for $key$. If $\server$ adds $key$ to $batches$ when receives a $[AcceptRequestBatch, key]$ message, we have that $\exists W[key]$.
\end{lemma}
\begin{proof}
We start by noting that $\server$ when receives a $[AcceptRequestBatch, key]$ message (\cref{line:sdelivereacceptbatchkey}), before adding $key \in batches$ (\cref{line:ssetbatchafterskforthetotality}, $\server$ checks that $key \in pool$ (\cref{line:skeyinpoolacceptebatch}).
Moreover, $\server$ adds elements to $pool$ only by executing \cref{line:sinsertelementinpool}. Immediately before doing this, $\server$ checks that $W[key]$ is a witness certificate for $key$ (\cref{line:switnesssignaturecheck}). This implies that $\exists W[key]$, thus the lemma is proven.
\end{proof}

\begin{lemma}
\label{lemma:keyinbatcheswitnessorbatchacquire}
Let $\server$ be a correct server. Let $\broker$ be a broker, and let $r$ be a root. Let $key$ be a key such that $key = (r, \broker, \_)$.
Let $W[key]$ be a witness certificate for $key$.
If $key \in batches$ we have that $\server$ sends $[BatchAcquired, key]$ or $\exists W[key]$.
\end{lemma}
\begin{proof}
We start by noting that upon initialization $batches$ is empty at $\server$ (\cref{line:sinitbatches}). Moreover, $\server$ adds elements to $batches$ only by executing \cref{line:sbatchesupdate,line:ssetbatchafterskforthetotality}.
In the first case, immediately after adding $key \in batches$, $\server$ returns a $[BatchAcquired, key]$ message to the procedure $handle\_batch$, and sends it to the respective broker $\broker$ (\cref{line:sbatchacquired,line:sbatchreceivecheckresponse,line:sbatchreceivesendresponse}).
In the second case, $\server$ adds $key$ to $batches$ only after receiving the message $[AcceptRequestBatch, key]$, this implies that according to \cref{lemma:ifkeyinbatchesafteracceptexistwitness}, exists a $W[key]$. Thus the lemma is proven.
\end{proof}

\begin{lemma}
\label{lemma:cmstoreinbatchesatserver}
Let $r[m]$ be a root of message $m$, let $c[m]$ be a core of $m$. Let $key$ be a key such that $key = (r[m], \_, \_)$.
If a witness certificate exists for $key$ $W[key]$, then there is at least one correct server $\server$ such that $c[m] \in batches[key]$ at $\server$.
\end{lemma}
\begin{proof}
We start by noting that according to \cref{def:witness}, if there exists $W[key]$, then at least $f + 1$ servers have multi-signed the message $[Witness, key]$.
Let $\server$ be a correct server that multi-signed $[Witness, key]$. We remark that according to \cref{def:coreofmessage}, $c[m]$ is defined as follows:
\begin{gather*}
    c[m] = (id, seq, m, s)
    \; \oplus \;
    c[m] = (id, max\_seq, m, ms).
\end{gather*}
We underline that, according to \cref{lemma:keyinbatchesifsignwitness}, we have that $key \in batches$ at $\server$ before $\server$ multi-signed $[Witness, key]$. This implies that $c[m] \in batches[key]$ except for $s$ or $ms$. Indeed, $\server$ can access to $batches[key]$ and have the following information:
\begin{gather*}
    id = id \in payloads \\
    seq = stragglers[id] \; \oplus \; max\_seq = payloads[id] \\
    m = payloads[id]
\end{gather*}
(\cref{line:sexstractbatch}).
In addition, $\server$ computes $r[m]$ (\cref{line:scleaves,line:scmerkletree,line:scroot}).
Moreover, $\server$ checks if $id \in straggler$ and that $s = sign[Message, m, seq]$ (\cref{line:sidnotinstragglers,line:sverifyclientsignature}), and, if not, that $ms = multisign[Reduction, r[m]]$ (\cref{line:sdeterminemultisigners,line:smultiverifyreduction}).
Finally, $\server$ can multi-sign $[Witness, key]$ (\cref{line:ssignwitness}). 
To summarize, if there exists a $W[key]$, then at least one correct server has signed $[Witness, key]$. And if a server $\server$ has signed $[Witness, key]$ then $c[m] \in batches[key]$ at $\server$. Thus the lemma is proven.
\end{proof}

\begin{lemma}
\label{lemma:noduplicatedid}
Let $\broker$ be a correct broker. Let $r[m]$ be a root of the message $m$ and $c[m]$ be a core of $m$ such that $c[m] = (id, \_, m, \_)$.
Let $c'[m']$ be a core of $m'$ such that $c'[m'] = (id, \_, m', \_)$.
If $c[m] \in batches[r[m]]$ at $\broker$. We cannot have $c'[m'] \in batches[r[m]]$.
\end{lemma}
\begin{proof}
We start by noting that $\broker$ adds elements to $batches[r[m]]$ only by executing \cref{line:bbatchesupdate}. Moreover, we argue that $\broker$ does this only for the elements that were in $pool$ (\cref{line:bsubmissionsequalpool}). 
Let $s[m]$ be the respective submission of $c[m]$ such that $s[m] = (id, \_, m, \_, \_)$.
We underline that $\broker$ adds $s[m]$ to $pool$ only by executing \cref{line:bpoolupdate}. Immediately before doing this, $\broker$ checks that $id \notin pool$ (\cref{line:bexistsinpanding}) and only if this is the case $\broker$ adds $s[m]$ to $pool$. This implies that if $s[m] \in pool$, $\broker$ never adds to $pool$ any $s'[m']$ such that $s'[m'] = (id, \_, m', \_, \_)$.
For what we prove above, if $c[m] \in batches[r[m]]$ at $\broker$ we cannot have $c'[m'] \in batches[r[m]]$. Thus the lemma is proven.
\end{proof}

\begin{lemma}
\label{lemma:expectyedwhatiobtain}
Let $\broker$ be a correct broker and let $\server$ be a correct server. Let $keys$ the set of $key$ that $\server$ receives from $\broker$. Let $batch\_number$ the sequence number for $keys$ such that $key = (\_, \broker, batch\_number) \forall key \in keys$.
Let $D[\_, \broker, \_]_{tot}$ be the total number of times that an event $D[\_, \broker', \_]$ with $\broker' = \broker$ happens at $\server$. Let $expected\_batch[\broker] = D[\_, \broker, \_]_{tot} + 1$. We have that $batch\_number = expected\_batch[\broker]$ at $\server$.
\end{lemma}
\begin{proof}
We prove this statement by induction. We start by analyzing the initial case. Upon initialization, $batch\_number$ is equal to $1$ at $\broker$ (\cref{lemma:bbatchnumberdefine}). We also notice that $\broker$ updates, adding one, $batch\_number$ only by executing \cref{line:supdatebatchnumberaddingone}. Immediately before doing this, $\broker$ sends to all servers a key $key$ such that $key = (\_, \broker, batch\_number)$.
This implies that the first $key$ that $\broker$ sends has $batch\_number = 1$. We argue that $batch\_number$ is a sequence number for all keys that $\broker$ sends.

We underline that upon initialization, $expected\_batch[\broker]$ is also equal to $1$ at $\server$. Moreover, $\server$ updates $ expected\_batch[\broker]$ only by executing \cref{line:supdatesnum}. This occurs only when an event $D[\_, \broker, \_]$ happens (\cref{line:sdeliverfromconsensus}). Thus implying that at every point in time $expected\_batch[\broker]$ is equal to the total number of events $D[\_, \broker, \_]$ happened for $\broker$ plus 1.

To summarize, upon initialization, we have that the first $key$ that $\server$ receives from $\broker$ has $batch\_number = 1$ and $expected\_batch[\broker] = 1$ thus the base case is satisfied indeed $batch\_number = expected\_batch[\broker]$ at $\server$.

Let $n$ be an integer. We assume that $batch\_number = expected\_batch[\broker] = n$ at $\server$. This implies that $\broker$ sent $n$ different $keys$ to $\server$ and the last $key$ sent is defined such that $key = (\_, \broker, n)$.
We underline that according to broker rule~\ref{rule:broker:waitdelivery} (see \cref{rules:Rule}), $\broker$ waits for $key$ to be delivered from \stobprefix before sending another one.
When $key$ is delivered from \stobprefix, $\server$ sets $expected\_batch[\broker] = n + 1$, because for what we argue above $expected\_batch[\broker]$ is equal to the total number plus $1$ of events $D[\_, \broker, \_]$ happen for $\broker$ at $\server$.

To conclude, let $t$ the time in which $\server$ sets $expected\_batch[\broker] = n + 1$, $\broker$ sends the new key $key$ such that $key = (\_, \broker, batch\_number)$ and $batch\_number = n + 1$ only at time $t'$ with $t' > t$.
This implies that when $\server$ receives $key$, $batch\_number = expected\_batch[\broker] = n + 1$.
Thus the lemma is proven.
\end{proof}

\begin{lemma}
\label{lemma:batchstate}
Let $\broker$ be a correct broker. Let $r[m]$ be a root of message $m$ such that $r[m] \in batches$. We have that the following statements on the states of $batches[r[m]]$ hold:
\begin{gather*}
    \text{if } r[m] \notin batches \rightarrow batches[r[m]] \text{ was } Completing \\
    \text{if } batches[r[m]] \text{ is } Completing \rightarrow batches[r[m]] \text{ was } Witnessing
\end{gather*}
\end{lemma}
\begin{proof}
We start by noting that $\broker$ removes $r[m]$ from $batches$ only by executing \cref{line:bremoverootfrombatches}. Immediately before doing this, $\broker$ checks that $batches[r[m]]$ is $Completing$ (\cref{line:bbatchcompleting1}).
We underline that $\broker$ updates the state of $batches[r[m]]$ to $Completing$ only by executing \cref{line:bbatchcompleting}. Immediately before doing this, $\broker$ checks that $batches[r[m]]$ was $Witnessing$ (\cref{line:binwitnessing}).
Thus the lemma is proven.
\end{proof}

\begin{lemma}
\label{lemma:signsendbackawitnessmessage}
Let $\broker$ be a correct broker. Let $\client$ be a client. Let $r[m]$ be a root of message $m$ and $c[m]$ be a core of $m$.
Let $key$ be a key such that $key = (r[m], \broker, \_)$.
If $r[m] \in batches$ at $\broker$ then there is at least $f+1$ servers that have signed and sent back to $\broker$ a $[Witness, key]$ message.
\end{lemma}
\begin{proof}
We start by noting that if $r[m] \in batches$ at $broker$, $\broker$ sends to all servers:
\begin{gather*}
    (id, seq, m)
    \; \oplus \;
    (id, max\_seq, m)
\end{gather*}
(\cref{lemma:brokersendssubreductiontoserver}).
Let $\server$ be a correct server.
We underline that, $\server$ receives $(id, seq, m)$ or $(id, max\_seq, m)$ only by executing \cref{line:sbatchrecive}. Immediately after, $\server$ invokes the procedure $handle\_batch$ (\cref{line:sinvokeshandlebatchprocedure}).

At the beginning of the $handle\_batch$ procedure, $\server$ computes $r[m]$ (\cref{line:scomputetrehandle,line:scomputeroothandle}). We remark that if $\server$ receives $(id, max\_seq, m)$ then $\server$ computes $r[m]$ on top of it (\cref{line:scomputeleaveshandle}), otherwise if $id \in stragglers$ then $\server$ computes $r[m]$ for $(id, seq, m)$ (\cref{line:idinstragglers,line:stakestragglersid}).
Moreover, $\server$ computes $key = (r[m], \broker, batch\_number)$ (\cref{line:sdefinekeyinhandlebatche}) and returns in case $key \in batches$ (\cref{line:sverifiykeyinbatcheshandleprocedure,line:sverifiykeyinbatcheshandleprocedurereturn}). We underline that if $key \in batches$ at $\server$ then sends $[BatchAcquired, key]$ or $\exists W[key]$ at $\server$ (\cref{lemma:keyinbatcheswitnessorbatchacquire}).
We remark that if $\exists W[key]$, we have according to \cref{def:witness} that at least a $f + 1$ servers have signed $[Witness, key]$.
Thus in this case the lemma is proven.

We analyze now the others two cases at $\server$:
\begin{enumerate}
    \item $key \in batches \rightarrow \server$ already sent $[BatchAcquired, key]$.
    \item $key \notin batches \rightarrow \server$ adds $key \in batches$ and sends $[BatchAcquired, key]$.
\end{enumerate}

The first case is proven by \cref{lemma:keyinbatcheswitnessorbatchacquire}.
For the second case, we remark that if $r[m] \in expected\_batch(\broker) \rightarrow key \in batches$ (\cref{lemma:rexcpectedkeiinbatches}).
This implies that we are in the first case or that a $W[key]$ exists. Indeed, we notice that if $r[m] \in expected\_batch(\broker)$ at $\server$, then $\server$ returns from the procedure (\cref{line:sexpectedbatchnotempty,line:sexpectedbatchnotemptyreturn}).
Moreover, let $c[m]$ and $c'[m']$ two cores such that:
\begin{gather*}
     c[m] = (id, \_, m, \_) \\
     c'[m'] = (id, \_, m', \_) \\
     r[m] = r[m'].
\end{gather*}
According to \cref{lemma:noduplicatedid}, if $c[m] \in batches[r[m]]$ at $\broker$ then $c'[m'] \notin batches[r[m]]$ at $\broker$. We also remark that if we assume that $\broker$ is correct, this implies that $\forall (id, id') \text{ such that } id, id' \in r[m]: id \neq id'$.
Thus the checks of duplicate identifiers is satisfied at $\server$ (\cref{line:sbatchcheck,line:sbatchcheckfail}).
We also underline that, according to \cref{lemma:expectyedwhatiobtain}, we have that $batch\_number = expected\_batch[\broker]$ at $\server$, thus the check at \cref{line:sexceptedbatch} is verified.
To conclude, if $key \notin batches$ at $\server$, $\server$ adds $key$ to $batches$, adds $r[m] \in expected\_batch[\broker]$ and sends a $[BatchAcquired, key]$ message to $\broker$ (\cref{line:sbatchesupdate,line:saddroottoexpected,line:sbatchacquired}).

If $\server$ sends a $[BatchAcquired, key]$ message to $\broker$, $\broker$ eventually delivers it (\cref{line:bdeliverbatchacquired}). Let $key$ be defined such that $key = (r[m], \broker, \_)$.
We remark that according to \cref{lemma:batchiswitnessatbroker}, $batches[r[m]]$ is $Witnessing$ because $\broker$ is sending $c[m]$. Moreover, $r[m]$ is still in $batches$ because in order to remove it from $batches$, the state of $batches[r[m]]$ must be $Completing$ and it is impossible that $batches[r[m]]$ was $Completing$ before the beginning of $Witnessing$ (\cref{lemma:batchstate}).
Moreover, upon receiving a $[BatchAcquired, key]$ message, $\broker$ sends back to $\server$ the signature $s$ or the multi-signature $ms$ associated to $c[m]$ in a $[Signature, key]$ message (\cref{line:bcomputethemultisignature,line:bcomputethesignature,line:bsendingsignatureormultisignature}). We underline that \cref{def:coreofmessage} and \cref{lemma:cminbatchesnomatterthecase} prove that $c[m] \in batches[r[m]]$ and that we have $c[m].signature$ is equal to $s$ or $ms$ but not both.

We also notice that in case $id \notin dir[\client]$ at $\server$ asks to $\broker$ the $id$ in the $[BatchAcquired, key]$ message (\cref{line:saskforunknows,line:sbatchacquired}). In this case, $\broker$ sends back to $\server$ with $s$ or $ms$ also the $id$ (\cref{line:bbatchacquiredexportassignments,line:bsendingsignatureormultisignature}).
Upon $\server$ receiving the $[Signature, key]$ message from $\broker$ invokes the procedure $handle\_signature$ (\cref{line:sdeliversignatures,line:binvokehandlebatchprocedure}). We notice in the procedure $handle\_signature$ (\cref{line:shandlesignaturesimportassignments}), $\server$ verifies that $r[m] \in expected\_batch[\broker]$ (\cref{line:srootinexbatchofb}) this is the case because $\server$ adds $r[m]$ to $expected\_batch[\broker]$ immediately before sending the $[BatchAcquired, key]$ message. Moreover, according to \cref{lemma:expectyedwhatiobtain}, we have that $batch\_number = expected\_batch[\broker]$ at $\server$, thus the check at \cref{line:schcekexpectedbatchinhandlesignature} is verified. 

Let $c[m]$ the core of $m$ that $\server$ receives from $\broker$, we remark that
\begin{gather*}
    c[m] = (id, seq, m, s)
    \; \oplus \;
    c[m] = (id, max\_seq, m, ms).
\end{gather*}
Finally, $\server$ verifies $s$ or $ms$. In particular, $\server$ checks that $\client = \dirin[id]$ correct signs $[Message, m, seq]$ (\cref{line:sverifyclientsignature}) or multi-signs $[Reduction, r[m]]$ with $r[m]$ builds on top of $(id, max\_seq, m)$ (\cref{line:smultiverifyreduction}).
We underline that, $\server$ immediately after checking the signature, multi-signs $[Witness, key]$ (\cref{line:ssignwitness}) with $key$ such that $key = (r[m], \broker, batch\_number)$ (\cref{line:scomputekeyfrohandlesignature}).
Moreover, $\server$ sends back to $\broker$ the $[Witness, key]$ message (\cref{line:sreturnwitnessshard,line:binvokehandlebatchprocedure,line:shandlesignaturessendresponse}).

To conclude, we remark that for assumption at least $2f + 1$ servers in the system are correct. We also remember that $\broker$ sends $c[m]$ to all servers, this implies that $\broker$ will eventually receive $f + 1$ $[Witness, key]$ messages or that there already exists a witness certificate $W[key]$. Thus the lemma is proven.
\end{proof}

\begin{lemma}
\label{lemma:batchiswitnessingorcompletingaatb}
Let $\broker$ be a correct broker. Let $\server$ be a correct server. Let $r[m]$ be a root of message $m$, such that $r[m] \in batches$ at $\broker$. Let $key$ be a key such that $key = (r[m], \broker, \_)$.
If $\server$ sends a $[Witness, key]$ message, then $batches[r[m]]$ is $Witnessing$ or $batches[r[m]]$ is $Completing$ at $\broker$.
\end{lemma}
\begin{proof}
We start by noting that $\server$ can only send to $\broker$ a $[Witness, key]$ message by executing \cref{line:sreturnwitnessshard}, immediately before doing this $\broker$ checks that $r[m] \in expected\_batch(\broker)$. We underline that $\server$ updates $expected\_batch$ only by executing \cref{line:saddroottoexpected} in the $handle\_batch$ procedure (\cref{line:shandlebatchprocedure}). Moreover, $\server$ invokes the $handle\_batch$ procedure only if receives from $\broker$:
\begin{gather*}
    (id, seq, m)
    \; \oplus \;
    (id, max\_seq, m) 
\end{gather*}
(\cref{line:sbatchrecive,line:sinvokeshandlebatchprocedure}).
According to \cref{lemma:batchiswitnessatbroker}, upon $\broker$ sending $(id, \_, m)$ to $\server$ we have that $batches[r[m]]$ is $Witnessing$ at $\broker$.
We remark that, according to \cref{lemma:batchstate}, the state of $batches$ could be $Completing$ only if it was $Witnessing$ before. Thus the lemma is proven. 
\end{proof}

\begin{lemma}
\label{lemma:existsawitnessidcompleting}
Let $\broker$ be a correct broker. Let $r[m]$ be a root of $m$ such that $r[m] \in batches$ at $\broker$. If $batches[r[m]]$ is $Completing$, then $\exists W[key]$ at $\broker$.
\end{lemma}
\begin{proof}
We start by noting that $\broker$ updates the state of $batches[r[m]]$ to $Completing$ only by executing \cref{line:bbatchcompleting}. We underline that $\broker$ immediately before creates a witness certificate $W[key]$ for $key$ (\cref{line:baggregatewitnesses}). Thus, the lemma is proven.
\end{proof}

\begin{lemma}
\label{lemma:existaawitnessecrtificatetobroker}
Let $\broker$ be a correct broker. Let $r[m]$ be a root of message $m$, such that $r[m] \in batches$ at $\broker$. Let $key$ be a key such that $key = (r[m], \broker, \_)$.
If $\broker$ receives at least $f + 1$ correctly signed $[Witness, key]$ messages, then $\exists W[key]$ and $batches[r[m]]$ is $Completing$ at $\broker$.
\end{lemma}
\begin{proof}
We start by noting that $\broker$ delivers $[Witness, key]$ only by executing \cref{line:bdeliverwitnessshard}. Immediately before, $\broker$ checks that $r[m] \in batches$ and that $batches[r[m]]$ is $Witnessing$ (\cref{line:brootandcommitable}). We remark that among the $f + 1$ servers that sent a $[Witness, key]$ message to $\broker$, at least one was correct. Thus, according to \cref{lemma:batchiswitnessingorcompletingaatb}, $batches[r[m]]$ is $Witnessing$ or $Completing$. We underline that if $batches[r[m]]$ is $Completing$, then $\exists W[key]$ (\cref{lemma:existsawitnessidcompleting}), thus the lemma is proven.

This implies that we can assume for the rest of the proof that $batches[r[m]]$ is $Witnessing$.
Moreover, $\broker$ verifies that the server $\server$ correctly multi-signed $[Witness, key]$ (\cref{line:bwitnessshardchecksignature}).
Let $ms$ the multi-signature of $\server$ for the message $[Witness, key]$. We notice that $\broker$ adds $ms$ to $batches[r[m]].witnesses$ (\cref{line:bstorewitnessshard}).
We underline that, upon initialization, $batches[r[m]].witnesses$ is empty (\cref{line:bbatchinwitnessing}).
Moreover, $\broker$ continues doing these operations for the next $f + 1$ $[Witness, key]$ incoming messages.

Upon $\broker$ collecting $f + 1$ multi-signatures for $[Witness, key]$ message (\cref{line:bdeliverwitnessshard}), $\broker$ aggregates all the multi-signatures in one (\cref{line:baggregatewitnesses}) and sends to all servers the witness certificate $W[key]$ (\cref{line:bbroadcastwitnesses}). Immediately after, $\broker$ updates also the state of $batches[r[m]]$ to $Completing$ (\cref{line:bbatchcompleting}). Thus the lemma is proven.
\end{proof}

\begin{lemma}
\label{lemma:conditiononsignwitnessmessage}
Let $\server$ be a correct server. Let $\broker$ be a correct broker. Let $r[m]$ be a root of message $m$. Let $key$ be a key such that $key = (r[m], \broker, \_)$.
If $\server$ signs a $[Witness, key]$ message then $r[m] \in expected\_batch[\broker]$ or $\exists W[key]$ at $\server$.
\end{lemma}
\begin{proof}
We start by noting that $\server$ signs a $[Witness, key]$ message only by executing \cref{line:ssignwitness}, immediately before $\broker$ computes $key$ such that $key = (r[m], \broker, \_)$ (\cref{line:scomputekeyfrohandlesignature}) and checks that $r[m] \in expected\_batch[\broker]$ (\cref{line:srootinexbatchofb}).
We underline that $\server$ removes $r[m]$ from $expected\_batch[\broker]$ only by executing \cref{line:sremoveroot} when an event $D[\broker, r[m], W]$ happens (\cref{line:sdeliverfromconsensus}). Before doing this, $\server$ computes $key$ such that $key = (r[m], \broker, \_) $ (\cref{line:sdefinekeyindeliverstb}) and $\server$ verifies that $W[key]$ is a witness certificate for $key$ (\cref{line:switnesssignaturecheck}). Thus the lemma is proven.
\end{proof}

\begin{lemma}
\label{line:expectedbatchorwitnesscertificate}
Let $\server$ be a correct server. Let $\broker$ be a correct broker. Let $r[m]$ be a root of message $m$. Let $num$ be a sequence number. Let $key$ be a key such that $key = (r[m], \broker, num)$.
If $\server$ signs a $[Witness, key]$ message then $expected\_batch[\broker] = num$ or $\exists W[key]$ at $\server$.
\end{lemma}
\begin{proof}
We start by noting that $\server$ signs a $[Witness, key]$ message only by executing \cref{line:ssignwitness}. Immediately before, $\server$ checks that $expected\_batch[\broker] = num$ (\cref{line:schcekexpectedbatchinhandlesignature,line:scheckexpectedbatcheuqlanumw}).
We remark that $\server$ modifies $expected\_batch[\broker]$ only by executing \cref{line:supdatesnum} when an event $D[r[m], \broker, W]$ happens (\cref{line:sdeliverfromconsensus}). Immediately before, $\server$ computes $key$ such that $key = (r[m], \broker, expected\_batch[\broker])$ (\cref{line:scomputenum,line:sdefinekeyindeliverstb}). 
Finally, $\server$ verifies that $W$ is a witness certificate for $key$ (\cref{line:switnesssignaturecheck}). 
To summarize, if $\server$ updates $expected\_batch[\broker]$ according to \cref{notation:existawitnesscertificatatserver}, then $\exists W[key]$ at $\server$. Thus the lemma is proven.
\end{proof}

\begin{lemma}
\label{lemma:ifexistsawitnessecerificateservercalltotalorder}
Let $\broker$ be a correct broker, let $r[m]$ a root of $m$ and let $key$ be a key such that $key = (r[m], \broker, \_)$.
Let $W[key]$ be a witness certificate for $key$. If $\exists W[key]$ at $\broker$, then at least one server $\server$ invokes $B[\broker, r[m], W[key]]$.
\end{lemma}
\begin{proof}
We start by noting that $\broker$ creates a witness certificate only by executing \cref{line:baggregatewitnesses}. Immediately after doing this, $\broker$ sends $W[key]$ to all servers (\cref{line:bbroadcastwitnesses}). We also remark that according to \cref{def:witness}, $W[key]$ exists because at least $f + 1$ servers have signed a $[Witness, key]$ message.
Let $\server$ be a correct server. We underline that by assumption at most $f$ servers could be Byzantine in the system, this implies that if $\exists W[key] \rightarrow \server$ signs $[Witness, key]$.

We underline that $\server$ receives from $\broker$, $W[key]$ only by executing \cref{line:sdeliverwitness}. Moreover, $\server$ checks if $r[m] \in expected\_batch[\broker]$ (\cref{line:srootinexpectedbatchwhendeliverwitness}). We remark that $\server$ signs $[Witness, key]$ message, this implies that according to \cref{lemma:conditiononsignwitnessmessage}, we have that $r[m] \in expected\_batch[\broker]$ or $\exists W[key]$ at $\server$.
We remark that according to \cref{notation:existawitnesscertificatatserver}, if $\exists W[key]$ at $\server$ this implies that an event $D[r[m], \broker, \_]$ happens at $\server$. According to the integrity property of \stobprefix, we have that if an event $D[r[m], \broker, \_]$ happens is because an event $B[r[m], \broker, \_]$ was triggered before. Let $\server'$ the server that invokes $B[r[m], \broker, \_]$, thus the lemma is proven.

We analyze now the case in which $r[m] \in expected\_batch[\broker]$ at $\server$. Upon receiving $W[key]$ with $key$ defined as $key = (r[m], \broker, num)$.
$\server$ checks if $num = expected\_batch[\broker]$ (\cref{line:scheckexpectedbatcheuqlanumw}).
We underline that, according to \cref{line:expectedbatchorwitnesscertificate}, we have that $expected\_batch[\broker] = num$ or $\exists W[key]$
at $\server$ because we remark that $\server$ signed the $[Witness, key]$ message. In case $expected\_batch[\broker] \neq num$ and $\exists W[key]$ the lemma is proven as we argue above.

To conclude, if $r[m] \in expected\_batch[\broker]$ and $expected\_batch[\broker] = num$, then $\server$ first verifies that $W[key]$ is a witness certificate for $key$ (\cref{line:sverifieswkeyistruebefirepropose}).
Second, $\server$ broadcasts to \stobprefix invoking $B[\broker, r[m], W[key]]$ (\cref{def:brodcasttostb}). Thus the lemma is proven.
\end{proof}

\begin{lemma}
\label{lemma:completioninvocationafterpropose}
Let $broker$ a broker. Let $r[m]$ be a root of $m$. Let $key$ be a key such that $key = (r[m], \broker, \_)$.
Let $W[key]$ be a witness certificate for $key$. If an event $B[\broker, r[m], W[key]]$ has been triggered, then there is at least one correct server $\server$ that will invoke $completion\_batch(key)$.
\end{lemma}
\begin{proof}
We start by noting that if $B[\broker, r[m], W[key]]$ has been invoked for the validity property of \stobprefix then eventually all correct servers deliver $D[\broker, r[m], W[key]]$. This implies that eventually $\exists W[key]$ at all correct servers according to \cref{notation:existawitnesscertificatatserver}.

We underline that, according to \cref{def:witness}, there is at least one correct server $\server$ that has signed a $[Witness, key]$ message. Let $t$ the moment in which $\server$ signs $[Witness, key]$, then $key \in batches$ at $\server$ at time $t'$ such that $t' < t$ (\cref{lemma:keyinbatchesifsignwitness}).
We remark that, if $\server$ removes $key$ from $batches$ we have that $\server$ has previously invoked the procedure $completion\_bathc(key)$ (\cref{lemma:notrenovekeyfrombatchesuntilecompletionbatch}) in this case the lemma is proven. 

We also underline that we have $key \in pool$ at $\server$, because the event $D[\broker, r[m], W[key]]$ happens (\cref{lemma:keyinpooluponddeliver}).
To summarize, we have that $key \in pool$ and $key \in batches$ at $\server$, this implies that according to \cref{lemma:keyinpoolinbatchescompletionbatchkey}, $\server$ will invoke the $completion\_batch(key)$ procedure, thus the lemma is proven.
\end{proof}

\begin{lemma}
\label{lemma:withcompletionbatchdeliverthemessage}
Let $\client$ be a correct client with $id = \dirin[\client]$. Let $\server$ be a correct server. Let $r[m]$ be a root of message $m$. Let $key$ be a key such that $key = (r[m], \_, \_)$ if $\server$ invokes $completion\_batch(key)$ procedure, we have that $\server$ delivers $m$.
\end{lemma}
\begin{proof}
Let $c[m]$ be the core of $m$ such that $c[m] = (id, seq, m, \_)$.
We start by noting that $\server$ invokes the $completion\_batch(key)$ only by executing \cref{line:scallprocedurecompletion}, immediately before doing this $\server$ verifies that $key \in batches \rightarrow c[m] \in batches[key]$ (\cref{line:skeyinpoolcheck}).

Let $(seq', m')$ the last message from $\client$ delivered by $\server$. We remark that, according to \cref{lemma:duplication1}, we have that $(seq', m') \in last\_message[\client]$ at $\server$.
Let $l[m]$ be a list of all sequence numbers associated with $m$ and $l[m']$ the list of all sequence numbers associated with $m'$. We also remark that $\client$ is a correct client thus, according to \cref{lemma:duplication4}, we have that $max(l[m]) < min(l[m'])$ at $\client$, this implies that $seq > seq'$.

We underline that according to client rule~\ref{rule:client:noduplicate} (see \cref{rules:Rule}), $\client$ never triggers $\event{\clin}{Broadcast}[message]$ twice in a row with the same message. This implies that $m \neq m'$.
To conclude, we have that the two conditions to deliver a message are satisfied (\cref{line:sseqbiggerthenlastmessage}), thus $\server$ updates $last\_message[\client]$ (\cref{line:supdatelastmessage}) and deliver $m$ (\cref{line:sdelivermessage}). This proves the lemma.
\end{proof}

\begin{lemma}
\label{lemma:completioncertificatesigned}
Let $\server$ be a correct server. Let $m$ be a message delivered by $\server$, let $r[m]$ a root of $m$, we have that $\server$ signs $[Completion, r[m], counter]$.
\end{lemma}
\begin{proof}
We start by noting that $\server$ delivers a message only by executing \cref{line:sdelivermessage}. Immediately before doing this, $\server$ adds the $(seq, m)$ that will be delivered to $completion\_payload[id]$ (\cref{line:supdatelastmessage}). Moreover, $\server$ computes the new $completion\_root$ on top of all the elements of $completion\_payload$ (\cref{line:scomputecompletionleaves,line:scomputecompletiontree,line:scomputecompletionroot}). We notice that, according to \cref{def:completioncertificate} and for what we argue before $completion\_root$ is a root of $m$.
To conclude, $\server$ immediately after delivers $m$, signs $[Completion, r[m], counter]$ (\cref{line:ssigncompletionmessage}). Thus the lemma is proven.
\end{proof}

\begin{lemma}
\label{lemma:batchiscompletingatbroker}
Let $\broker$ be a correct broker. Let $r[m]$be a root of $m$ such that $r[m] \in batches$. Let $key$ be a key such that $key = (r[m], \broker, \_)$.
If one correct server invoke $completion\_batch(key)$ procedure, then $batches[r[m]]$ is $Completing$ at $\broker$.
\end{lemma}
\begin{proof}
We start by noting that $\server$ invokes the procedure $completion\_batch(key)$ only by executing \cref{line:sinvokesprodecurecompletion}. We remark that, according to \cref{lemma:completiononlyifeceventhappens}, if $\server$ invokes $completion\_batch(key)$ this implies that an event $D[\broker, r[m], W]$ happens at $\server$.
We underline that, according to \cref{notation:existawitnesscertificatatserver}, we have that $\exists W[key]$ at $\server$. 
Moreover, $\broker$ computes the witness certificate $W[key]$ only by executing \cref{line:baggregatewitnesses} and immediately after $\broker$ updates the state of $batches[r[m]]$ to $Completing$ (\cref{line:bbatchcompleting}). Thus the lemma is proven.
\end{proof}

\begin{lemma}
\label{lemma:existacompletioncertificatwatbroker}
Let $\broker$ be a correct broker. Let $r[m]$ be a root of $m$. If there is a correct server $\server$ that has signed $[Completion, r[m], counter]$, then $\exists C[m]$ at $\broker$.
\end{lemma}
\begin{proof}
We start by noting that $\server$ signs $[Completion, r[m], counter]$ only by executing \cref{line:ssigncompletionmessage}, immediately after $\server$ sends back to $\broker$ the $[Completion, r[m], counter]$ signed to $\broker$ (\cref{line:ssendbackcompletion}).
Let $key$ be a key such that $key = (r[m], \broker, \_)$.
We remark that $\server$ signs the $[Completion, r[m], counter]$ in the $completion\_batch(key)$ procedure. This implies that, according to \cref{lemma:batchiscompletingatbroker}, $batches[r[m]]$ is $Completing$ at $\broker$.
We underline that if $r[m] \notin batches$, then $\broker$ removes $r[m]$. Moreover, $\broker$ can only do this by executing \cref{line:bremoverootfrombatches}, but immediately before this $\broker$ computes the completion certificate $C[m]$ for $m$ (\cref{line:screatecompletioncertificate}), which implies that in this case the lemma is proven. For the rest of the proof, we assume that $r[m] \in batches$ at $\broker$.
Moreover, when $\broker$ receives a signed $[Completion, r[m], counter]$ message (\cref{line:bsignedcompletioncertificate}), $\broker$ checks that $batches[r[m]]$ is $Completing$ (\cref{line:bcheckbathccompletingreally}), the signature is valid (\cref{line:bchecks}) and adds the signature to $completion$ (\cref{line:baddingsignaturetocomplations}) that upon initialization is empty (\cref{line:bbatchcompleting}).

We remark that for the agreement property of \system, if a correct server $\server$ invokes $completion\_batch(key)$ all correct servers eventually call $completion\_batch(key)$. We remember that the total number of correct servers in the system is at least $2f +1$. This implies that $\broker$ will eventually collect at least $f + 1$ $[Completion, r[m], counter]$ signed messages.
To conclude, $\broker$ upon receiving $f+1$ correct $[Completion, r[m], counter]$ messages (\cref{line:bbatchcompleting1}), $\broker$ aggregates all $f+1$ multi-signatures in just one and creates a completion certificate $C[m]$ for $m$ (\cref{line:screatecompletioncertificate}). Thus the lemma is proven.
\end{proof}

\begin{theorem}[Validity]\label{thm:validity}
\system satisfies validity.
\end{theorem}
\begin{proof}
Let $\client$ be a correct client. Let $m$ be the message that $\client$ has broadcast, and let $seq$ the sequence number associated with $m$. We start by noting that, according to \cref{lemma:minsubmissionsatclient}, we have $m \in submissions$ or $\exists C[m]$ at $\client$, with $C[m]$ a completion certificate for $m$.
If $\exists C[m]$ for $m$, then at least one correct server delivers the message $m$ (\cref{lemma:onecorrectserverdeliversifexistacompletioncertificate}), thus the validity property is satisfied. We now analyze the case in which $m \in submissions$ at $\client$.

Let $sub[m]$ be a submissions for $m$, we remark that according to \cref{def:submission}, $sub[m]$ is defined such that $sub[m] = (id, seq, m, s, C[\_])$ in which $id$ is the client identifier such that $id = \dirin[\client]$, $s$ is the client signature for the $[Message, m, seq]$ message and $C[\_]$ is a generic completion certificate such that $C[\_].counter \geq seq$.
We underline that, according to \cref{lemma:subminsumissionsatbroker}, because $m \in submissions$ at $\client$ we have that $sub[m] \in submissions$ at some correct broker $\broker$.
Let $r[m]$ be a root of $m$ (\cref{def:rootofm}). Let $c[m]$ be a core of $m$ such that, according to \cref{def:coreofmessage},
\begin{gather*}
    c[m] = (id, seq, m, s)
    \; \oplus \;
    c[m] = (id, max\_seq, m, ms)
\end{gather*}
in which $ms$ is the client multi-signature for the message $[Reduction, r[m]]$, and $max\_seq \geq seq$.
We notice that, because $sub[m] \in submissions$ at $\broker$, according to \cref{lemma:coreinbatchsattimet}, we have that $r[m] \in batches$ and $c[m] \in batches[r[m]]$ at $\broker$ at $t$, with $t$ we identify the moment in which $\broker$ adds $r[m]$ to $batches$.
Moreover, because now $r[m] \in batches$ at $\broker$ we have that $c[m] \in batches[r[m]]$ not only at time $t$ but until $\broker$ does not remove $r[m]$ from $batches$ (\cref{lemma:cminbatchesnomatterthecase}).

Let $key$ be a key such that $key = (r[m], \broker, \_)$.
For what we argue above, because $r[m] \in batches$ and $c[m] \in batches[r[m]]$, we have that at least $f+1$ servers, in which $f$ is the maximum number of Byzantine servers on the system, sign and send back to $\broker$ a $[Witness, key]$ message (\cref{lemma:signsendbackawitnessmessage}).
Let $W[key]$ be a witness certificate for $key$. We underline that, according to \cref{lemma:existaawitnessecrtificatetobroker}, because $\broker$ receives at least $f+1$ correctly signed $[Witness, key]$ messages, we have that $\exists W[key]$ and $batches[r[m]]$ is $Completing$ at $\broker$.
Moreover, because $\exists W[key]$ at $\broker$, according to \cref{lemma:ifexistsawitnessecerificateservercalltotalorder}, we have that at least one correct server $\server$ broadcasts $B[\broker, r[m], W[key]]$ to \stobprefix (\cref{def:brodcasttostb}).

Following the above considerations, if an event $B[\broker, r[m], W[key]]$ has been triggered by a correct $server$, then at least one correct $\server'$ invokes $completion\_batch(key)$ (\cref{lemma:completioninvocationafterpropose}).
We remark that $key = (r[m], \broker, \_)$.
To conclude, according to \cref{lemma:withcompletionbatchdeliverthemessage}, because $\server'$ invokes $completion\_batch(key)$, we have that $\server'$ delivers $m$.
Thus we prove that there is a correct server $\server'$ that delivers the message $m$, implying that \system satisfies the validity property.

Moreover, according to client rule~\ref{rule:client:waitcompletion} (see \cref{rules:Rule}), client $\client$ can broadcast a new message $m'$ only after receiving a completion certificate $C[m]$ for the previous message $m$. For the rest of the theorem, we prove that $\client$ eventually obtains $C[m]$ because $m$ is delivered by at least one correct server.
We remark that, according to \cref{lemma:completioncertificatesigned}, $\server'$ signs $[Completion, r[m], counter]$ because $\server'$ has delivered $m$.
This implies, according to \cref{lemma:existacompletioncertificatwatbroker}, that $\exists C[m]$ at $\broker$.
Moreover, we have that $\broker$ sends back $C[m]$ to $\client$ (\cref{line:bsendbackcompletioncertificate}). We underline that $\client$ eventually receives $C[m]$ (\cref{line:crecivecompletioncertifcate}) and can finally remove $m$ from $submissions$ (\cref{line:sremoveelementfromsumissions}). We notice that $\client$ can now broadcast a new message. This implies that the validity property is satisfied when client rule~\ref{rule:client:waitcompletion} (see \cref{rules:Rule}) is observed. Thus the theorem is proven.
\end{proof}

\clearpage
\section{Rank}
\label{appendix:diral}

This section contains the formal definition (\cref{subsection:name.interface}), full pseudocode (\cref{subsection:name.pseudocode}) and full correctness proofs (\cref{subsection:name.correctness}) of \diral.
\diral is an algorithm that implements a \dir~\cite{directory-protocol-beatcs10} in a way that assigns an identifier incrementally to each process $\process$ in the system while ensuring that the set of identifiers remains dense.
Given $|\process|$ the total number of processes, \diral is able to represent each identifier with $\ceil{\log_2\rp{|\process|}}$ bits which corresponds to the information-theoretical minimum number of bits necessary to uniformly represent a set of $|\process|$ processes.
\system uses an instance of \diral to assign dense client identifiers.

\paragraph{Model.}
As in \system, \diral assumes a set of $n$ permissioned servers such that up to $f$ servers may behave arbitrarily, \ie $n \geq 3f+1$~\cite{byz-generals-toplas82}.
The set of processes that want to obtain an identifier is permissionless hence of unbounded and unknown size.
Unlike \system, the only participants in the system executing \diral are \emph{processes} and \emph{servers}.
\diral uses an underlying instance of \stob (\stobprefix) (see \cref{subsection:chop.interface}) to reach agreement between servers.

\paragraph{Algorithm overview.}
A correct process $\process$ sends a signed \emph{signup request} to all servers in order to obtain an identifier.
When a correct server $\server$ delivers the signup request, $\server$ first checks the correctness of the signature for the request then checks if $\process$ already had an assigned identifier.
If this is not the case, $\server$ proposes to the underlying \stobprefix instance the process request.
Upon delivery of the process request by \stobprefix, $\server$ simply assigns to $\process$ its position in the totally ordered list as $\process$'s identifier.
For the agreement and total order properties of \stobprefix, all correct servers eventually assign the same identifier to the same $\process$.
Once the assignment is done, $\server$ sends back to the process a signed \emph{assignment confirmation} with the associated identifier.
The process waits to collect $f + 1$ assignment confirmations for the same identifier and aggregates all the signatures to obtain an \emph{assignment certificate}.
Upon reception and verification of the assignment certificate, every participant in the system can trust the association between the process $\process$ and its identifier since at least one correct server certified it.

\subsection{Directory Interface and Properties}
\label{subsection:name.interface}

\paragraph{Directory interface.}
A \dir interface (instance $\dirin$) exposes the following procedures and events:
\begin{itemize}
    \item \textbf{Request} $\event{\dirin}{Signup}{}$: Requests that an identifier be assigned to the local process.
    \item \textbf{Indication} $\event{\dirin}{SignupComplete}$: Indicates that an identifier has successfully been assigned to the local process.
    \item \textbf{Getter} $\dirin[id]$: Returns the process associated with identifier $id$, if known, otherwise returns $\bot$.
    \item \textbf{Getter} $\dirin[process]$: Returns the identifier associated with process $process$, if known, otherwise returns $\bot$.
    \item \textbf{Getter} $\dirin.export(id)$: Returns the assignment for identifier $id$, if known, otherwise returns $\bot$.
    \item \textbf{Setter} $\dirin.import(assignment)$: Imports assignment $assignment$.
\end{itemize}

\paragraph{Directory properties.}
A \dir satisfies the following properties:
\begin{itemize}
    \item \textbf{Signup integrity}: A correct process never triggers $SignupComplete$ before triggering $Signup$.
    \item \textbf{Signup validity}: If a correct process triggers $Signup$, it eventually triggers $SignupComplete$.
    \item \textbf{No duplication}: No process has more than one identifier $id$ assigned.
    \item \textbf{Self-knowledge}: Upon triggering $SignupComplete$, a correct process knows the identifier $id$ associated to itself.
    \item \textbf{Transferability}: If a correct process $\process$ invokes $\dirin.export(id)$ to obtain an assignment $a$, then any correct process knows the identifier $id$ associated to $\process$ upon invoking $\dirin.import(a)$.
    \item \textbf{Density}: Let $ids$ be the set of all the (integer) identifiers assigned at time $t$. For every time $t'$, $ids$ is dense. This implies that if $id$ is included in $ids$ then $id-1$ is also included in $ids$.
\end{itemize}

\clearpage
\subsection{Pseudocode}
\label{subsection:name.pseudocode}

This section contains the complete code of the \dir implementation \diral.
\diral's client is implemented in \Cref{subsection:pseudocode.diral} and \diral's server in \cref{subsection:pseudocode.diralserver}.

\subsubsection{Rank Client}
\label{subsection:pseudocode.diral}

\begin{lstlisting}
implements:
    %\dirab%, instance %\dirin%


uses:
    %AuthenticatedPointToPointLinks%, instance %al%


struct Assignment:
    id: Integer,
    process: Process,
    certificate: Certificate


enum Status(Outsider, SigningUp, SignedUp)


upon <%\dirin%.Init>:
    association: {Integer: [Multisignature]} = {}; %\label{line:minitassociation}%
    directory: {(Id, Process)} = {}; 
    certificates: {Id: Certificate} = {}; 
    status : Status = Outsider; %\label{line:namestautusinit}%


upon <%\dirin%.Signup>: %\label{line:ntriggersignup}%
    signature = sign([Signup]);
    status = SigningUp; %\label{line:nstatussigningup}%
    for %$\server$% in %$\servers$%:
        trigger <al.Send | %$\server$%, [Signup, signature]>; %\label{line:mbrodcasttoallservers}%


upon <al.Deliver | %$\server$%, [Assignment, %$\process$%, id, certificate]> %\label{line:mdeliveranassigmentcertficate}%
    if not %$\server$%.verify(certificate, [Assignment, self, id]): return; %\label{line:nveriftsigcert}%
    association[id].add(certificate); %\label{line:maddsasssocation}%


%\StartMultiline%upon exist id in association such that |association[id]| %$\geq$% f+1 and status = SigningUp:%\label{line:ncheckthestatus}% %\EndMultiline%
    certificate = aggregate(association[id]); %\label{line:mcreateassigment}%
    status = SignedUp; %\label{line:nstuatusupdate}%
    %\dirin%.import(Assignment{id, process: self, certificate}); %\label{line:nimportinvokes}%
    trigger <%\dirin%.SignupComplete>; %\label{line:ntriggersignupcomplete}%


procedure %\dirin%.import(assignment): 
    %\StartMultiline%if assignment.certificate.verify_quorum(
     [Assignment, assignment.id, assignment.process]):%\EndMultiline% %\label{line:ncertficatecheckimport}%
        directory.add((assignment.id, assignment.process)); %\label{line:naddstodi}%
        certificates[assignment.id] = assignment.certificate; %\label{line:naddtodirin}%


procedure %\dirin%.[](query): %\label{line:dinvokeprocedure}%
    if query is Id:
        if exists process such that (query, process) in directory: %\label{line:dcheckkeycardindirectory}%
            return process; %\label{line:dreturnkycard}%
    else if query is Process:
        if exists id such that (id, query) in directory: %\label{line:dcheckidindirectory}%
            return id; %\label{line:dreturnid}%
    return %$\bot$%;


procedure %\dirin%.export(id): %\label{line:ddprocedureexport}%
    if let certificate = certificates[id]: %\label{line:ddcertificateincerticatesid}%
        process = %\dirin%[id]; %\label{line:ddequationkeycardid}%
        return Assignment {id, process, certificate}; %\label{line:ddreturnassigment}%
    else:
        return %$\bot$%; %\label{line:ddreturnassigmentreturn}%
\end{lstlisting}

\subsubsection{Rank Server}
\label{subsection:pseudocode.diralserver}

\begin{lstlisting}
implements:
    %\dirsrab%, instance %\dirsrin%
    

uses:
    %AuthenticatedPointToPointLinks%, instance %al%
    %\stobprefix%Server, instance %\stobi%


upon <%\dirsrin%.Init>:
    Id: Integer = 0; %\label{line:nidinitialiation}%
    association: {Process : id} = {}; %\label{line:nassocationinit}%


upon <al.Deliver | %$\process$%, [Signup, signature]>: %\label{line:nrecivesignupandsignature}%
    if not %$\process$%.verify(signature, [Signup]): return; %\label{line:nverifysignatureafterconsensus}%
    if %$\process$% in association: return;
    trigger <stob.Broadcast | Signup, %$\process$%, signature>; %\label{line:nproposetoconsensus}%


upon <stob.Deliver | Signup, %$\process$%, signature>: %\label{line:ndeliverfromconsenuss}%
    if not %$\process$%.verify(signature, [Signup]): return; %\label{line:nchecksignaturefromconsenusss}%
    if %$\process$% in asociation: return; %\label{line:nprocesnotassociation}%
        
    certificate = multisign([Assignment, %$\process$%, id]); %\label{line:nnumtisigncertificate}%
    trigger <al.Send | [Assignment, %$\process$%, id, certificate]>; %\label{line:nsendbacktoprocess}%
    
    association[%$\process$%] = id; %\label{line:nassociationupdate}%
    id = id + 1; %\label{line:nupdateid}%
\end{lstlisting}

\clearpage
\subsection{Correctness}
\label{subsection:name.correctness}

In this section, we prove to the fullest extent of formal detail that \diral implements a \dir.

\subsubsection{Signup Integrity}

In this section, we prove that \diral satisfies signup integrity.

\begin{theorem}[Signup integrity]
\diral satisfies signup integrity.
\end{theorem}

\begin{proof}
Let $\process$ be a correct process. We start by noting that upon $\process$ triggering $SignUp$ (\cref{line:ntriggersignup}), $\process$ sets the status to $SigningUp$ (\cref{line:nstatussigningup}). We also remark that upon initialization, status is set to $Outsider$ (\cref{line:namestautusinit}), this implies that if the status is equal to $SigningUp$ it is because $\process$ triggers $SignUp$. 
Moreover, $\process$ triggers $SignupComplete$ only by executing \cref{line:ntriggersignupcomplete}. Immediately before, $\process$ checks that the state is $SigningU$ (\cref{line:nstuatusupdate}). As we argued before, this implies that $\process$ can trigger $SignupComplete$ only after having triggered $SignUp$. Thus the theorem is proven.
\end{proof}

\subsubsection{Signup Validity}

In this section, we prove that \diral satisfies signup validity.

\begin{definition}[Assignment certificate]\label{def:assigmentcertficate}
Let $\process$ be a correct process. Let $id$ be an integer. We define an assignment certificate for $\process$, that we call $A[\process]$, an aggregate of at least $f + 1$ correctly multi-signed $[Assignment, \process, id]$ messages sent by different servers.
\end{definition}

\begin{lemma}
\label{lemma:propsebconsensus}
Let $\process$ be a correct process. Let $s$ be the signature of $\process$ for the $[Signup]$ request. If $\process$ triggers $Signup$ there is at least one correct server that invokes $B[Signup, \process, s]$.
\end{lemma}
\begin{proof}
We start by noting that $\process$ broadcasts to all servers the $[Signup, s]$ message upon triggering $Signup$ (\cref{line:ntriggersignup,line:mbrodcasttoallservers}). Let $\server$ be a correct server. We underline that $\server$ eventually receives the $[Signup, s]$ message from $\process$ (\cref{line:nrecivesignupandsignature}) and invokes $B[Signup, \process, s]$ (\cref{line:nproposetoconsensus}). Thus the lemma is proven.
\end{proof}

\begin{lemma}
\label{lemma:countid}
Let $\server$ be a correct server. Let $|D[\_]|$ be the number of times that $\server$ triggers an event $D[Signup, \process, s]$, with $\process$ a generic process and $s$ a correct signature of $\process$ for $[Signup]$ message. Let $id$ be an integer. We have that at every point in time $|D[\_]| = id$.
\end{lemma}
\begin{proof}
We start by noting that upon initialization, $id = 0$ at $\server$ (\cref{line:nidinitialiation}). Moreover, we notice that $\server$ updates $id$ only by executing \cref{line:nupdateid}. We notice that $\server$ does this only if an event $D[Signup, \process, s]$ happens and $s$ is a valid signature of $\process$ for the message $[Signup]$ (\cref{line:nverifysignatureafterconsensus,line:ndeliverfromconsenuss}). To conclude we remark that an event $D[\_]$ can happen only at \cref{line:ndeliverfromconsenuss}. This implies that if and only if an event $D[\_]$ happens, we have that $\server$ increments $id$ by one. Thus the lemma is proven.
\end{proof}

\begin{lemma}
\label{lemma:createasigmenetcertificate}
Let $\process$ be a correct process and let $s$ be the signature of $\process$ for the $[Signup]$ request.
Let $\server$ be a correct server, if $\server$ invokes $B[Signup, \process, s]$, then $\exists A[\process]$ at $\process$.
\end{lemma}
\begin{proof}
We start by noting that if a correct server invokes $B[Signup, \process, s]$, according to the validity and agreement of \stobprefix, we have that all correct servers in the system deliver from \stobprefix the same message $[Signup, \process, s]$.

Moreover, according to \cref{lemma:countid}, we know that $id$ at every correct server is equal to the number of times that an event $D[Signup, \process, s]$ happens if $s$ is a valid signature of $\process$ for the $[Signup]$ message. We notice that when a correct server triggers a $D[Signup, \process, s]$ event, it checks that $s$ is valid and updates $id$ only if this is the case (\cref{line:nchecksignaturefromconsenusss,line:nupdateid}).

According to the total order property of \stobprefix, all correct servers deliver the messages in the same order. This implies that all correct servers assign the same identifier $id$ to the same $\process$. Thus all correct servers multi-sign the same message $[Assignment, \process, id]$ (\cref{line:nnumtisigncertificate}).
Moreover, after multi-signing the message, all correct servers send it back to $\process$ (\cref{line:nsendbacktoprocess}).

Let $\server$ be a correct server. Upon $\process$ receiving a signed $[Assignment, \process, id]$ message, it checks that the signature is valid (\cref{line:mdeliveranassigmentcertficate,line:nveriftsigcert}) and the signature to $association[id]$ (\cref{line:maddsasssocation}). We remark that upon initialization $association$ is empty (\cref{line:minitassociation}).
Let $f$ be the number of Byzantine servers in the system.
To conclude, we remark that at least $f + 1$ servers have sent to $\process$ a correctly signed $[Assignment, \process, id]$ message. This implies that eventually $\exists id: |association[id]| = f+1$ (\cref{line:ncheckthestatus}).

Thus, $\process$ can aggregate all $f+1$ servers' multi-signatures and create an assignment certificate $A[\process]$ (\cref{line:mcreateassigment}). This proves the lemma.
\end{proof}

\begin{theorem}[Signup validity]
\diral satisfies signup validity.
\end{theorem}
\begin{proof}
Let $\process$ be a correct process. Let $s$ be the signature of $\process$ for $[Signup]$. We start by noting that, according to \cref{lemma:propsebconsensus}, if $\process$ triggers $Signup$ we have that at least a correct server $\server$ invokes $B[Signup, \process, s]$.
Moreover, because $\server$ invokes $B[Signup, \process, s]$ we have that $\exists A[\process]$ at $\process$ (\cref{lemma:createasigmenetcertificate}).
To conclude, we remark that $\process$ assembles an assignment certificate $A[\process]$ only by executing \cref{line:mcreateassigment}. Immediately after, $\process$ triggers $SignupComplete$ and imports the certificate in the \dir (\cref{line:nimportinvokes,line:ntriggersignupcomplete}). Thus the theorem is proven.
\end{proof}

\subsubsection{No duplication}

In this section, we prove that \diral satisfies no duplication.

\begin{lemma}
\label{lemma:unique assigmentcertficate}
Let $\process$ be a process. If an assignment certificate $A[\process]$ exists for $\process$, we have that $A[\process]$ is unique. Thus, if $\exists A[\process], \nexists A'[\process]: \forall A'[\process] \neq A[\process]$.
\end{lemma}
\begin{proof}
We start by noting that according to \cref{def:assigmentcertficate}, if an assignment certificate $A[\process]$ exists, then at least $f + 1$ servers have signed a $[Assignment, \process, id]$ message, among which at least one is correct by definition.
Let $\server$ be a correct server that has signed $[Assignment, \process, id]$.
We underline that $\server$ can sign a $[Assignment, \process, id]$ only by executing \cref{line:nnumtisigncertificate}. Moreover, $\server$ does this only if $\process \notin association$ (\cref{line:nprocesnotassociation}). We notice that upon initialization $association$ is empty (\cref{line:nassocationinit}) and $\server$ adds elements immediately after signing an $[Assignment, \process, id]$ message (\cref{line:nnumtisigncertificate,line:nassociationupdate}). This implies that $\server$ signs just one $[Assignment, \process, id]$ message for $\process$. Thus the lemma is proven.
\end{proof}

\begin{theorem}[No duplication]
\diral satisfies no duplication.
\end{theorem}
\begin{proof}
Let $\process$ be a process. Let $A[\process]$ be an assignment certificate for $\process$. We underline that according to \cref{lemma:unique assigmentcertficate} if $A[\process]$ exists we have that $A[\process]$ is unique for $\process$. We remark that according to \cref{def:assigmentcertficate}, an assignment certificate certifies the relation between $\process$ and an identifier $id$. To summarize, because $A[\process]$ is unique for $\process$, and an assignment certificate maps $\process$ into $id$, the identifier associated with every process is unique. Thus the theorem is proven.
\end{proof}

\subsubsection{Self-knowledge}

In this section, we prove that \diral satisfies self-knowledge.

\begin{definition}[Assignment]\label{def:assigmnet}
Let $\process$ be a correct process. Let $id$ be an integer. Let $A[\process]$ be an assignment certificate. We define an assignment and we call $Z[\process]$ the tuple $Z[\process] = (\process, id, A[\process])$.
\end{definition}

\begin{theorem}[Self-knowledge]
\diral satisfies self-knowledge.
\end{theorem}
\begin{proof}
Let $\process$ be a correct process. We start by noting that $\process$ triggers $SignupComplete$ only by executing \cref{line:ntriggersignupcomplete}. We underline that immediately before doing this, $\process$ imports in the \dir its own assignment $Z[\process]$ such that $Z[\process] = (\process, id, A[\process])$ (\cref{line:nimportinvokes}).
We underline that, according to \cref{def:assigmnet}, $A[\process]$ is an assignment certificate computed at \cref{line:mcreateassigment}. Thus implying that $\process$ knows its identifier $id$ because the assignment associated is stored in the \dir, thus the theorem is proven. 
\end{proof}

\subsubsection{Transferability}

In this section, we prove that \diral satisfies transferability.

\begin{theorem}[Transferability]
\diral satisfies transferability.
\end{theorem}
\begin{proof}
Let $\process$ be a correct process. Let $id$ be an integer. We start by noting that when $\process$ invokes the procedure $export(id)$ if $id \in \dir$, we have that $\process$ obtains an assignment $Z[\process]$ such that $Z[\process] = (\process, id, A[\process])$ (\cref{line:ddreturnassigment}). Let $\process'$ be a correct process. If $\process'$ invokes $import(Z[\process])$, it first checks that $A[\process]$ is an assignment certificate (\cref{line:ncertficatecheckimport}) and then adds $Z[\process]$ in the \dir (\cref{line:naddstodi,line:naddtodirin}). This implies that $\process'$ knows the association between $id$ and $\process$. Thus the theorem is proven. 
\end{proof}

\subsubsection{Density}

In this section, we prove that \diral satisfies density.

\begin{theorem}[Density]
\diral satisfies density.
\end{theorem}
\begin{proof}
Let $\server$ be a correct server. Let $\process$ be a process. Let $id$ be an identifier and let $A[\process]$ be an assignment certificate for $\process$. In order to assign $id$ to $\process$, $\server$ has to sign a $[Assigment, \process, id]$ message. 
We underline that $\server$ does this only by executing \cref{line:nnumtisigncertificate}. Immediately after, $\server$ updates $id$ adding one (\cref{line:nupdateid}). We also remark that this is the only place in which $\server$ modifies $id$. Moreover, according to the no duplication property, $\server$ signs no more than one $[Assigment, \process, id]$ message for each process. This implies that $\server$ signs an $Assigment$ message by associating different processes with an increased $id$ number. To conclude, we remark that for the agreement and total order property of \stobprefix, all correct servers sign the same set of $Assigment$ messages in the same order. Thus they assign the same $id$ to $\process$. For what we argue above, we can conclude that the theorem is proven.
\end{proof}

\end{document}